\documentclass[12pt,english,aps,tightenlinesletterpaper,groupaddress,secnumarabic,nofootinbib]{revtex4}
\usepackage{amsmath}
\usepackage{mathptmx}
\usepackage{newtxmath}

\usepackage[T1]{fontenc}
\usepackage[latin9]{inputenc}
\usepackage[a4paper]{geometry}
\usepackage{babel}
\usepackage{float}
\usepackage{graphicx}
\usepackage[unicode=true,pdfusetitle,
 bookmarks=true,bookmarksnumbered=false,bookmarksopen=false,
 breaklinks=false,pdfborder={0 0 1},backref=false,colorlinks=false]
 {hyperref}

\makeatletter

\providecommand{\tabularnewline}{\\}

\@ifundefined{textcolor}{}
{%
 \definecolor{BLACK}{gray}{0}
 \definecolor{WHITE}{gray}{1}
 \definecolor{RED}{rgb}{1,0,0}
 \definecolor{GREEN}{rgb}{0,1,0}
 \definecolor{BLUE}{rgb}{0,0,1}
 \definecolor{CYAN}{cmyk}{1,0,0,0}
 \definecolor{MAGENTA}{cmyk}{0,1,0,0}
 \definecolor{YELLOW}{cmyk}{0,0,1,0}
}

\usepackage{babel}
\newcommand{\pA}{\cite{Chung:2021lfg}}

\usepackage{xcolor}

\makeatother

\begin{document}
\title{Power Spectrum in the Chaotic Regime of Axionic Blue Isocurvature
Perturbations}
\author{Daniel J. H. Chung}
\email{danielchung@wisc.edu}

\affiliation{Department of Physics, University of Wisconsin-Madison, Madison, WI
53706, USA}
\author{Sai Chaitanya Tadepalli}
\email{stadepalli@wisc.edu}

\affiliation{Department of Physics, University of Wisconsin-Madison, Madison, WI
53706, USA}
\begin{abstract}
Large blue tilted spectral index axionic isocurvature perturbations
can be produced when the axion sector is far out of equilibrium during
inflation through an initial Peccei-Quinn (PQ) symmetry breaking field
displacement along a nearly flat direction in the effective potential.
As a companion to a previous work, we present analytic formulae for
the blue isocurvature spectrum for the case of the kinetic energy
density of the PQ symmetry breaking field being larger than the quartic
power of the final spontaneous PQ symmetry breaking scale. It corresponds
to a regime in which the nonlinearities of the classical potential
become important many times during the formation of the axion isocurvature
quantum perturbations leading to interesting resonant behavior. One
consequence of this nonlinearity-driven resonance is the chaotic nature
of the map that links the underlying Lagrangian parameters to the
isocurvature amplitudes. We point out an accidental duality symmetry
between the perturbation equations and the background field equations
that can be used to understand this. Finally, we present two types
of analytic results. The first relies on a computation utilizing an
effective potential wherein fast time scale fluctuations have been
integrated out. The second is grounded in a functional ansatz, requiring
only a limited set of fitting parameters. Both analytic results should
be useful for carrying out forecasts and fits to the data.
\end{abstract}
\maketitle
\tableofcontents{}

\section{Introduction}

Axions offer a compelling solution to the strong CP problem \citep{Kim:2008hd,Srednicki:2002ww,Peccei:2006as,Peccei:2010ed,DiLuzio:2020wdo}.
Due to their weak interactions with the Standard Model (SM), axions
can also potentially constitute a substantial portion of the cosmological
cold dark matter (CDM) \cite{Abbott:1982af,Preskill:1982cy,Dine:1982ah,Kolb:1990vq,Sikivie:2006ni,Kawasaki:2013ae,Marsh:2015xka}.
Given this dual significance of axions in both particle physics and
cosmology, numerous experiments have been dedicated to their detection
\cite{B_hre_2013,2017,Ouellet_2019,Braine_2020,Gramolin_2020,Kwon_2021,Brubaker_2017,McAllister:2017lkb,Alesini:2017ifp,Alesini:2020vny,ARIADNE:2017tdd,Budker_2014,Brun_2019,Armengaud_2014,Armengaud2019}.
Reviews on direct detection can be found in sources such as \cite{Ringwald:2012hr,Redondo_2011,Ringwald_2014,Stadnik_2017,Carosi:2013rla,Graham_2013,Agarwal_2011,Irastorza_2018}.
In situations where the axions are spectator fields during inflation,
a well-known cosmological observable called CDM-photon isocurvature
perturbations can become detectable if the axions interact sufficiently
weakly and do not thermalize. The generation of isocurvature perturbations
by spectator axions, including its model-specific characteristics
and the related observational limitations, have been extensively investigated
in \cite{Kasuya1997,Kawasaki1995,Nakayama2015,Harigaya2015,Kadota2014,Kitajima2014,Kawasaki2014,Higaki2014,Jeong2013,Kobayashi2013,Hamann2009,Hertzberg2008,Beltran2006,Fox:2004kb,Estevez2016,Kearney2016,Nomura2015,Kadota2015,Hikage2012,Langlois:2003fq,Mollerach1990,Axenides1983,Jo2020,Iso:2021tuf,Bae2018,Visinelli2017,Takeuchi:2013hza,Bucher:2000hy,Lu:2021gso,Sakharov:2021dim,Rosa:2021gbe,Jukko:2021hql,Chen:2021wcf,Jeong:2022kdr,Cicoli:2022fzy,Koutsangelas:2022lte,Kawasaki:2023zpd}.
The isocurvature spectrum studied in these cases is typically flat,
and comparisons with data leads to constraints in the ($H$,~$F_{a}$)
parameter space \cite{Marsh:2015xka} where $H$ is the inflationary
Hubble scale and $F_{a}$ is the axion decay constant in equilibrium.
For PQ symmetry breaking to complete before or during inflation, these
parameters typically need to satisfy $F_{a}>H$. Phenomenologically,
these isocurvature constraints can be naturally relaxed by introducing
blue-tilted isocurvature fluctuations that can be highly suppressed
on large-scales where most of the observational constraints are most
severe.

Interestingly, it has been shown that axionic sector out of equilibrium
dynamics during inflation can generate a large blue spectral tilt
to the quantum isocurvature perturbations \citep{Kasuya2009}. Notably,
the work of \cite{Chung:2015tha} has highlighted that a detectable
isocurvature signal from a linear spectator with spectral index $\gtrsim2.4$
provides a nontrivial evidence of dynamical degrees of freedom with
time-dependent masses during inflation. In a companion paper \pA,
we have analytically and numerically computed the blue-tilted isocurvature
spectrum in the model of \citep{Kasuya2009} in the underdamped parametric
region which produces background classical field dynamics that are
only mildly resonant with the isocurvature quantum fluctuations. In
this work, we focus on analytically capturing the isocurvature perturbations
in strongly resonant situations in which the two Peccei-Quinn (PQ)
symmetry breaking background fields cross each other many times while
undergoing strongly nonlinear classical oscillations.

During these crossings, the axion perturbation amplitudes can become
amplified through an effective negative mass squared effects similar
to the physics of \pA. In our earlier study \pA, we focused on cases
involving a maximum of one crossing after the transition where at
the moment of that one crossing, the dominant force in the system
is the Hubble damping term (even when the nonlinear forces driving
the resonance are significant in magnitude). In this current study,
as we explore situations with higher kinetic energy corresponding
to quartic-potential driven forces dominating over the damping force
during the crossings, we observe that the motion of the two fields
becomes chaotic when the nonlinear forces dominate over the Hubble
expansion rate driven damping force and the harmonic linear forces.
Qualitatively, the quartic interaction within the blue axion system
considered here as well as in \pA~can induce a chaotic behavior
akin to the chaos observed in classical so-called Yang-Mills-like
potentials \citep{Carnegie_1984,Dahlqvist:1990zz,Marcinek:1994}.
Quantitatively, we observe that the field trajectory can become chaotic
when the average quartic interaction energy at transition surpasses
a certain threshold, $\approx O(\alpha_{{\rm Ch}})\left(F_{a}/H\right)^{4}$,
where $\alpha_{{\rm Ch}}$ is an $O(1)$ threshold parameter that
varies mildly with $F_{a}/H$.

This background field dynamics approximately determines the amplitude
of the long wavelength quantum fluctuations because of an accidental
duality between the linearized quantum mode equations and the nonlinear
background field equations. This means that for the rising part of
the isocurvature spectrum and the first few bumps after the resonant
transition occurs, the magnitude of the isocurvature amplitude maps
chaotically to the underlying Lagrangian parameters. In addition to
explaining the dynamics in this strongly resonant situation, this
paper presents two sets of fitting models that can be used for forecasts
and data fits. One set is based on modeling the dynamical axion mass
through a set of approximately square and exponential effective potentials
that can be derived after integrating out the high-frequency fluctuations
of the background fields. The other is a slightly simpler fitting
function designed to directly match the shape and amplitude of the
final isocurvature spectrum and is checked by matching with explicitly
solved numerical examples.

The order of presentation is as follows. In Sec.~\ref{sec:Massive-underdamped-fields}
we provide a review of our axion toy-model introduced by Kasuya-Kawasaki
in \citep{Kasuya2009} and explore the dynamics of the background
fields for massive underdamped fields. Moving to Sec.~\ref{Sec:Estimate-Tzjc},
we expand upon the analysis of \pA ~by considering massive fields
that result in multiple zero-crossings of the background fields before
the transition. We present an analytic expression for estimating the
transition time $T_{c}$ in these cases. Next in Sec.~\ref{sec:Isocurvature-power-spectrum},
we examine the characteristics of the isocurvature power spectrum
in the blue-tilted region. We accomplish this by analyzing the zero-mode
($k=0$) system and establishing appropriate matching conditions (based
on an accidental duality of the mode equations) to reconcile the values
with those of finite $k$ modes. This approach enables us to investigate
the shape and magnitude of the isocurvature power spectrum in greater
detail. We find that for massive background fields with large $O(F_{a}^{4}/H^{4})$
nonlinear interaction, the corresponding zero-mode amplitudes can
show chaotic structure. We end that section by demonstrating how the
duality can be used to understand the chaotic map between the Lagrangian
parameters and the isocurvature amplitudes. In Sec.~\ref{sec:Fitting-functions}
we provide empirical fitting functions of the zero-mode amplitudes
for the non-chaotic cases and a distribution function for the chaotic
cases. In Sec.~\ref{sec:Generic-mass-model}, we revisit the mass-model
first presented in \pA, and expand it by applying it to several cases,
fitting both the blue-tilted and oscillating regions of the spectra.
Inspired by the results of the mass model, we present a simpler 7-parameter
sinusoidal fitting function in Sec.~\ref{sec:sine-model-ftting}
to reduce the complexity of possible future fitting efforts. We conclude
in Sec.~\ref{sec:Conclusions}.

We present some of the finer details of our work in the following
list of appendices. In App.~\ref{sec:-phim_Transient} we estimate
the non-adiabatic effects from zero-crossings and quantify their effects
on the transition of the background fields. In App.~\ref{sec:Approx_theta}
we give an approximate estimation of the phase, $\theta$, of the
zero-mode solution $I_{0}$. App.~\ref{sec:Chaotic-structure} discusses
the chaotic structure of the background fields. We explore one of
the subdominant mass parameter dependence of the isocurvature power
spectrum in App.~\ref{sec:cminus}. Finally, in App.~\ref{sec:Fitted-sine-model-parameters}
we list the best-fit model parameters for the examples discussed in
Sec.~\ref{sec:sine-model-ftting}

\section{\label{sec:Massive-underdamped-fields}Massive underdamped fields}

This paper is concerned with a scenario in which the complex field
sector containing axion is far out of equilibrium. The key non-axion
field degrees of freedom that determine the properties of the axion
are $\phi_{\pm}$ fields where $\phi_{+}$ is initially displaced
far from the minimum of the potential located near $F_{a}$ which
is the axion decay constant. The non-equilibrium dynamics of $\phi_{\pm}$
lead to a rich set of isocurvature power spectra for the axion.

In a previous work \pA, we presented analytic results for axionic
blue isocurvature power spectrum for the resonant underdamped cases
within a specific region of the parametric space. Here, ``underdamped''
refers to the situation when the spectator field has a time-dependent
effective mass $m$ such that $m^{2}/H^{2}>9/4$ for which the perturbation
mode behaves like an underdamped oscillator. Even though $m$ is time-dependent
because of its time-varying background field $\phi_{\pm}$ dependence,
it happens to be approximately constant for some initial time period
(specified more fully below). In \pA, the analysis was strictly limited
to cases where the mass $m$ is minimally greater than $3H/2$ such
that the background fields, $\phi_{\pm}$, cross each other close
to the first zero-crossing of $\phi_{+}$ and the kinetic energy at
the crossing is $\leq O(F_{a}^{4}/H^{4})$. One of the aims of this
paper is to explain the dynamics for the case when $m^{2}\gg H^{2}$
(initially) which will lead to multiple zero-crossings.

In this section, we will first give a brief review of an example axion
model and then discuss background field dynamics for massive fields
with multiple zero-crossings before transition.

\subsection{\label{subsec:Review}A brief review of an example axion model}

In \cite{Kasuya2009}, the authors pointed out that if a PQ charged
SM singlet moves along a flat direction lifted only by gravity-mediated
supersymmetry (SUSY) breaking masses of $O(H)$, then the amplitude
of the isocurvature fluctuations can generically have a strong blue
tilt.

Consider then the chiral superfields $\Phi_{\pm,0}$ from \citep{Kasuya2009}
where the indices indicate the associated $U(1)_{{\rm PQ}}$ global
Peccei-Quinn charges. The resulting effective potential obtained after
summing up $F$-term and Kaehler induced contributions while looking
along the flat direction $\Phi_{0}=0$ is 
\begin{equation}
V\approx h^{2}|\Phi_{+}\Phi_{-}-F_{a}^{2}|^{2}+c_{+}H^{2}|\Phi_{+}|^{2}+c_{-}H^{2}|\Phi_{-}|^{2}.\label{effpotential}
\end{equation}
where $h$ is a coupling coefficient, $c_{\pm}$ are positive constants,
and $H$ is the inflationary Hubble scale. The parameter $c_{+}$
dominantly controls the blue isocurvature spectral index, $n_{{\rm I}}-1=3-2\sqrt{9/4-c_{+}}$
since $\Phi_{+}$ is initially displaced hierarchically larger than
$\Phi_{-}$ and $F_{a}$ along the flat direction $\Phi_{+}\Phi_{-}-F_{a}^{2}=0$.
Soft SUSY-breaking mass terms (${\rm TeV}$ range) are neglected as
they are assumed to be much smaller than the inflationary Hubble scale
$H$. This setup (even generalized away from this SUSY example) implicitly
assumes that the inflation sector can be arranged to have $H\ll F_{a}$
and that the flat directions are only lifted by the quadratic terms
at renormalizable level.

During inflation, the $U(1)_{{\rm PQ}}$ is broken and the $\Phi_{+}$
field rolls down along the flat direction from an initial large displacement.
The magnitude of the initial displacement will eventually determine
the $k$ interval over which the blue spectrum persists, and the maximum
displacement of the field is of $O(M_{P})$ to have the effective
field theory be under control. Such large displacements can be generically
induced through supergravity induced effects from a UV completion
of the theory. The Nambu-Goldstone boson associated with a linear
combination of the phases of the two fields is recognized as the axion.
In particular, with the parameterization 
\begin{equation}
\Phi_{\pm}\equiv\frac{\varphi_{\pm}}{\sqrt{2}}\exp\left(i\frac{a_{\pm}}{\sqrt{2}\varphi_{\pm}}\right)\label{eq:angularparam}
\end{equation}
where $\varphi_{\pm}$ and $a_{\pm}$ are real, the axion is 
\begin{equation}
a=\frac{\varphi_{+}}{\sqrt{\varphi_{+}^{2}+\varphi_{-}^{2}}}a_{+}-\frac{\varphi_{-}}{\sqrt{\varphi_{+}^{2}+\varphi_{-}^{2}}}a_{-}\label{eq:axion}
\end{equation}
while the heavier partner is ignored as it is not dynamically important.
Using the scalings defined as
\begin{equation}
\phi_{\pm}\equiv\Phi_{\pm}\frac{h}{H}
\end{equation}
\begin{equation}
F=hF_{a}/H\label{eq:Fscale}
\end{equation}
\begin{equation}
\xi(\phi_{+},\phi_{-})\equiv\phi_{+}\phi_{-}-F^{2}\label{eq:xidef}
\end{equation}
and 
\begin{equation}
T\equiv tH,
\end{equation}
the background equations of motion with the interaction force $\xi\phi_{\pm}$
can be written as 
\begin{align}
\ddot{\phi}_{+}(T)+3\dot{\phi}_{+}(T)+c_{+}\phi_{+}+\xi(\phi_{+},\phi_{-})\phi_{-} & =0\label{eq:backgroundeom0}\\
\ddot{\phi}_{-}(T)+3\dot{\phi}_{-}(T)+c_{-}\phi_{-}+\xi(\phi_{+},\phi_{-})\phi_{+} & =0\label{eq:backgroundeom}
\end{align}
for motions of $\Phi_{\pm}$ where the background $\Phi_{\pm}$ does
not change its phase. The associated mode equation for the fluctuations
$I_{\text{\ensuremath{\pm}}}\equiv\delta a_{\pm}/2$ (see \cite{Chung:2016wvv}
for details) is 
\begin{equation}
\left(\partial_{T}^{2}+3\partial_{T}\right)I+\left(\frac{Ka(0)}{a(T)}\right)^{2}I+\tilde{M}^{2}I=0\label{eq:modeeq}
\end{equation}
where 
\begin{equation}
K\equiv\frac{k}{a(0)H}\label{eq:KDEF}
\end{equation}
is the scaled physical wave vector at the initial time of $\phi_{+}$
rolling defined as $T=0$, the vector $I=(I_{+},I_{-})$ contains
the quantum axion fluctuation information, and the mass matrix is
\begin{equation}
\tilde{M}^{2}(T)\equiv\left(\begin{array}{cc}
c_{+} & F^{2}\\
F^{2} & c_{-}
\end{array}\right)+\left(\begin{array}{cc}
\phi_{-}^{2}(T) & 0\\
0 & \phi_{+}^{2}(T)
\end{array}\right).\label{eq:massmat}
\end{equation}
Hence, at the Lagrangian level, the effective set of parameters governing
the background and linearized perturbation dynamics is $\{c_{+},c_{-},F\equiv hF_{a}/H\}$.
As will be detailed in Sec~\ref{subsec:Resonance}, the initial condition
of the background fields will be restricted to a two parameter family
$\left(\phi_{+}(0),\dot{\phi}_{+}(0)\right)$ with the $\phi_{-}(0)$
and $\dot{\phi}_{-}(0)$ fixed according to the constraint that the
fields are sitting on the flat direction. The boundary conditions
for the mode functions will be Bunch-Davies (BD).

The expression for the isocurvature fluctuations during inflation
can be written as 
\begin{equation}
\Delta_{s}^{2}(t,\vec{k})\approx4\omega_{a}^{2}\frac{k^{3}}{2\pi^{2}}I^{\dagger}\left(\begin{array}{cc}
r_{+}^{2} & 0\\
0 & r_{-}^{2}
\end{array}\right)I\label{eqspec}
\end{equation}
for 
\begin{equation}
r_{\pm}\equiv\sqrt{\frac{\phi_{\pm}^{2}(t)}{(\phi_{+}^{2}(t)+\phi_{-}^{2}(t))^{2}\theta_{+}^{2}(t_{i})}}
\end{equation}
where $\omega_{a}$ is the ratio of axion energy density to the dark
matter fraction today and $\theta_{+}(t_{i})$ is the initial axion
angle.\footnote{Without loss of significant generality in the current scenario, we
are assuming $\Phi_{+}$ has all of the initial axion angle.} The quantity $\omega_{a}$ is sensitive to the assumption of whether
or not the axion contained in $I$ is the QCD axion. For specificity,
the reader can assume that this is the QCD axion and refer to the
formula of the misalignment scenario dark matter fraction given in
\cite{Chung:2016wvv}, although this paper is largely insensitive
to this assumption. Hence, as far as the CDM-photon isocurvature is
concerned, there is one more parameter of $\omega_{a}^{2}/\theta_{+}^{2}(t_{i})$.
In summary, as far as the power spectrum is concerned, what we will
focus in on this paper is a 3 (Lagrangian) + 2 (initial conditions)
+ 1 (initial misalignment angle) parameter model.

\subsection{\label{subsec:Resonance}Resonance and transition}

In \pA, we derived analytic expressions for the isocurvature power
spectrum in the underdamped cases. These cases occur when the mass-squared
term $c_{+}H^{2}$ of the $\phi_{+}$ field slightly exceeds $9H^{2}/4$
and the kinetic energy of the background fields $\phi_{\pm}$ is within
specific parametric bounds when at their crossing. To analyze these
scenarios, the authors in \pA ~employed a combination of perturbative
and non-perturbative methods, due to the deviation of the fields from
the flat direction described in Eq.~(\ref{eq:xidef}) which leads
to non-adiabatic effects when the background fields are approaching
the potential minimum. This deviation is caused by the $\phi_{+}$
field tending to zero while the total energy of the system is $O(F^{4}$).
Consequently, there is a significant increase in the kinetic energy
during the field crossing. In comparison to an overdamped scenario,
the authors in \pA ~discovered that this substantial kinetic energy
at the crossing leads to diverse spectral shapes with multiple bumps.
Moreover, the underdamped cases examined in \pA~ exhibited an amplification
of the isocurvature spectral amplitude by at least $O(30)$ relative
to the massless plateau.

More explicitly, consider the zeroth order perturbed solution of Eqs.~(\ref{eq:backgroundeom0})
and (\ref{eq:backgroundeom}) which is valid in the limit $\sqrt{\phi_{-}/\phi_{+}}\ll1$
where we assume that $\phi_{+}$ field rolls down along the flat direction
from an initial displacement much greater than $F$ (typically $O(M_{P}/H)$).
Hence, we can approximate $\phi_{+}$ as 
\begin{align}
\phi_{+}(T)\approx\phi_{+}^{(0)}(T) & =\phi_{+}(0)e^{-\frac{3}{2}T}\sec(\varphi)\cos(\omega T-\varphi)\label{eq:approxsol}
\end{align}
where 
\begin{equation}
\omega=\sqrt{c_{+}-9/4},\label{eq:omega}
\end{equation}
\begin{equation}
\tan\varphi\equiv\frac{3/2+\epsilon_{0}}{\omega},\label{eq:tanphi}
\end{equation}
and 
\begin{equation}
\epsilon_{0}\equiv\frac{\dot{\phi}_{+}(0)}{\phi_{+}(0)}\label{eq:eps0}
\end{equation}
describes the initial velocity. Note that underdamped cases imply
that $c_{+}>9/4$. The matching order $\phi_{-}(T)$ solution is given
as 
\begin{equation}
\phi_{-}(T)\approx\phi_{-}^{(0)}(T)=\frac{F^{2}}{\phi_{+}^{(0)}}.\label{eq:phimapprox}
\end{equation}
In \pA, the analysis was carried out by defining a new parameter
\begin{align}
\alpha & \equiv\frac{\left|\partial_{T}\phi_{+}^{(0)}(T_{z})\right|}{F^{2}}\label{eq:alphadefinition}\\
 & =\omega\frac{\phi_{+}(0)}{F^{2}}\,\sec\varphi\,e^{-3/2T_{z}}\label{eq:alpha-formula}
\end{align}
that characterizes the kinetic energy of the underdamped background
fields close to a zero-crossing. Here, $T_{z}$ is the zero of the
$\phi_{+}^{(0)}$ field defined by 
\begin{equation}
\phi_{+}^{(0)}\left(T_{z}\right)=0.\label{eq:Tz}
\end{equation}
\begin{figure}[t]
\begin{centering}
\includegraphics[scale=0.45]{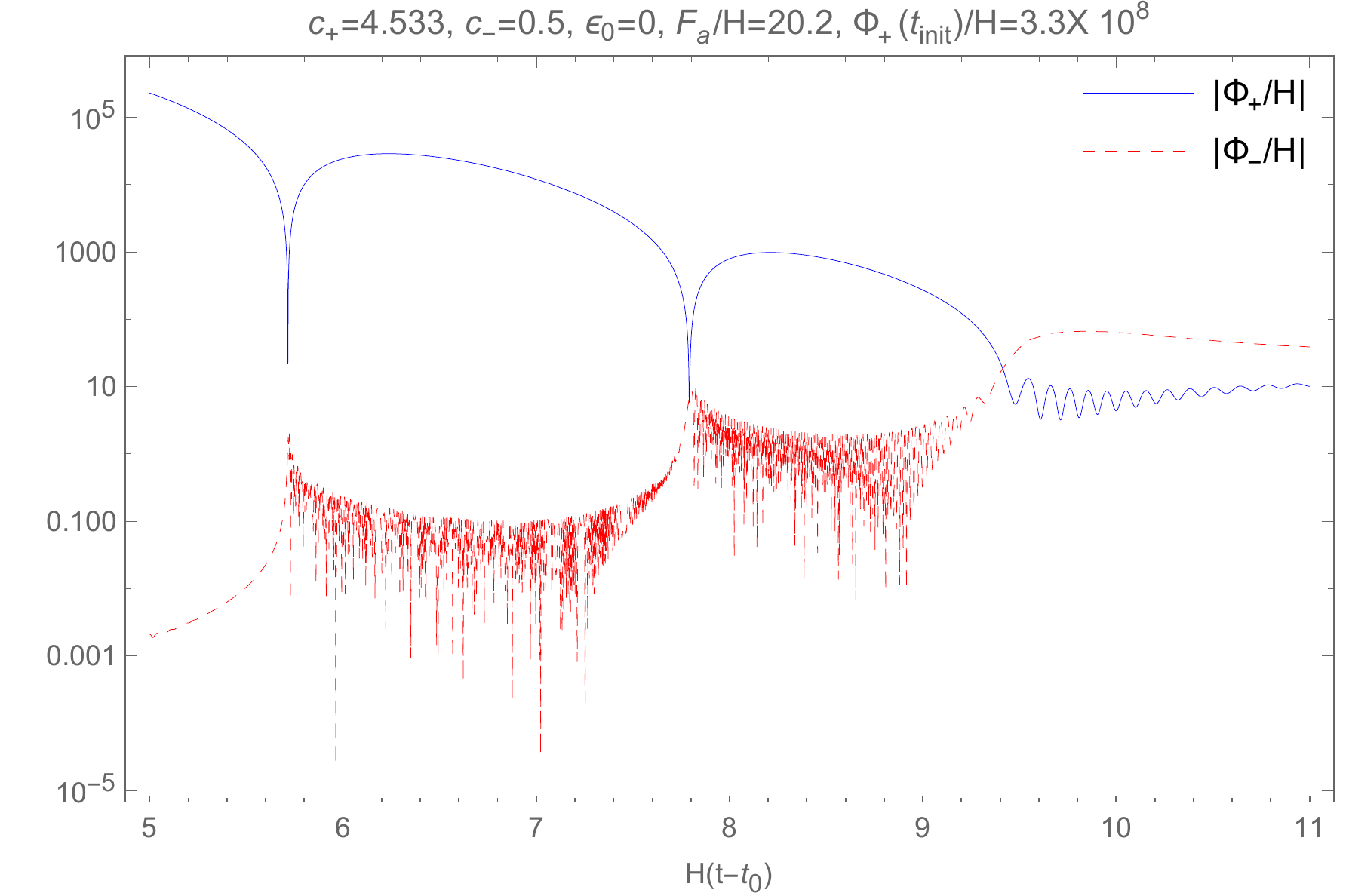} 
\par\end{centering}
\caption{\label{multiplecrossings}Plot highlighting the background field dynamics
for $c_{+}=4.533$ where the $\phi_{+}$ field undergoes multiple
zero-crossings before transitioning at approximately $T_{c}\approx9.4$.
The remaining Lagrangian parameters are set at $F_{a}/H=20.2$, $c_{-}=0.5$,
$\epsilon_{0}=0$ and $\phi_{+}(0)/H=3.32\times10^{8}$.}
\end{figure}

As the $\phi_{+}$ field approaches zero, it intersects with the $\phi_{-}$
field. The number of these crossings is uniquely determined by the
value of $c_{+}$ through $\omega$ defined in Eq.~(\ref{eq:omega}).
Fig.~\ref{multiplecrossings} shows the rolling down of the background
fields for a fiducial value of $c_{+}=4.533$ that exhibits multiple
zero-crossings of $\phi_{+}^{(0)}$ occurring at times $T_{z}\approx1.56,\;3.63,\;5.7,\;7.79,\;\mbox{and }9.87$.
The plot is obtained for an illustrative fiducial set $P_{A}$ of
parameters that we will often use throughout this paper: 
\begin{equation}
P_{A}\equiv\{F=20.2,\,c_{-}=0.5,\,\epsilon_{0}=0,\,\phi_{+}(0)=0.1M_{p}/H\}.\label{eq:Pset}
\end{equation}
At each crossing, we can assess the influence of the $\phi_{-}$ field
on $\phi_{+}$ by evaluating the force $\xi\phi_{-}$. These forces
can induce displacements of $\phi_{+}$ towards the ``steep'' direction
of the potential (perpendicular to the flat direction), where $\xi$
is significant. This in turn causes strong oscillatory behavior of
both $\phi_{+}$ and the order unity coupled $\phi_{-}$. Thus, during
each crossing at a time $T_{{\rm cross}}$ when $\phi_{+}(T_{{\rm cross}})=\phi_{-}(T_{{\rm cross}})$,
we can express the effective coupling force $f_{+}$ on $\phi_{+}$
as follows: 
\begin{equation}
f_{+}(T_{{\rm cross}})=-\xi\phi_{-}|_{T_{{\rm cross}}}\label{eq:force at crossing}
\end{equation}
whose magnitude measures the deviation of $\phi_{+}$ from the flat
direction trajectory or the zeroth order solution given in Eq.~(\ref{eq:approxsol}).
This deviation is a sufficient condition for the force in the steep
direction to be significant. Hence, we define resonant transition
time $T_{c}$ as the first $T_{{\rm cross}}$ that satisfies the following
two conditions: 
\begin{equation}
\left.f_{+}\right|_{T=T_{c}}\gtrsim O(0.1)\left|\ddot{\phi}_{+}(T_{c})\right|\label{eq:cond1}
\end{equation}
and 
\begin{equation}
|\dot{\phi}_{+}(T_{{\rm c}})|\gtrsim\alpha_{{\rm L}}F^{2}.\label{eq:cond2}
\end{equation}
The first of the conditions ensure sufficient coupling force from
$\phi_{-}$ such that $\phi_{+}$ deviates significantly from the
unperturbed solution $\phi_{+}^{(0)}$ while the second condition
here is required for $\phi_{+}$ to oscillate with an amplitude such
that the kinetic energy is sufficiently large, $O(F^{4})$. When the
above two conditions are satisfied, the two background fields oscillate
with a frequency of $O(F)$. We call this situation ``resonance''.
In \pA, we chose $\alpha_{{\rm L}}\approx0.2$ as the threshold for
resonance. For the fiducial example presented in Fig.~\ref{multiplecrossings},
the transition occurs at $T_{c}\approx9.5$ which is close to the
$5$th zero-crossing, $T_{z,5}=9.87$. Note that the force $f_{+}$
is directly proportional to the amplitude of $\phi_{-}$, and from
Eq.~(\ref{eq:phimapprox}), we see that $\phi_{-}$ becomes $O(F)$
when $\phi_{+}\sim O(F)$. Therefore, the transition typically occurs
in the vicinity of a zero-crossing of $\phi_{+}^{(0)}$. The moment
of transition serves as a pivotal time in the coupled dynamics of
the background fields. It marks the point in time when the axion begins
to make a dynamical transition to a massless final state. From an
observational perspective, it defines the wavenumber, $k_{c}$, which
corresponds to the location of the cutoff where the isocurvature spectrum
smoothly departs from a blue-tilted power law.

Unlike the case presented in Fig.~\ref{multiplecrossings}, the analysis
and findings described in \pA ~were focused on $c_{+}$ values for
which the background fields $\phi_{\pm}$ undergo a transition close
to the first zero-crossing of the $\phi_{+}^{(0)}$ field. Additionally,
the parameter $\alpha$ was constrained within approximate bounds
of $[0.2,1)$. However, for larger $c_{+}$ values, the condition
specified in Eq.~(\ref{eq:cond1}) may not be fulfilled at the first
zero-crossing. In the next section, we will estimate the $n$th zero-crossing
point which is closest to the transition $T_{c}$ and hence satisfies
the conditions outlined in Eqs.~(\ref{eq:cond1}) and (\ref{eq:cond2}).
Then, by determining the value of $\alpha$ at this zero-crossing,
we can effectively characterize the dynamics of the background fields
after the transition.

\section{\label{Sec:Estimate-Tzjc}Estimation of the $T_{z}$ closest to $T_{c}$}

As discussed above and illustrated in Fig.~\ref{multiplecrossings},
in case of underdamped scenarios, the background fields $\phi_{\pm}$
intersect each other near each zero-crossing of $\phi_{+}^{(0)}$
until the transition occurs at time $T_{c}$. The zero-crossings of
the zeroth order solution $\phi_{+}^{(0)}$ in Eq.~(\ref{eq:approxsol})
are given by the expression 
\begin{equation}
T_{z,j}=\frac{1}{\omega}\left(\left(j-\frac{1}{2}\right)\pi+\varphi\right)\label{eq:Tzj}
\end{equation}
where $j$ gives us the location of the $j$th zero-crossing. Using
Eq.~(\ref{eq:alphadefinition}), we define the quantity $\alpha_{j}$
at each $T_{z,j}$ as 
\begin{equation}
\alpha_{j}=\omega\frac{\phi_{+}(0)}{F^{2}}e^{-\frac{3}{2}T_{z,j}}\sec(\varphi).\label{eq:alpha-j}
\end{equation}
From Fig.~\ref{multiplecrossings} we observe that close to each
$T_{z,j}$, the two background fields cross each other at $T_{{\rm cross},j}$.
The amplitude of the fields at each $T_{{\rm cross},j}$ is controlled
by the parameter $\alpha_{j}$. In Appendix.~\ref{sec:-phim_Transient},
we demonstrate that a higher value of $\alpha_{j}$ results in a greater
incoming velocity of the $\phi_{+}$ field, resulting in a smaller
crossing amplitude $\phi_{\pm}\left(T_{{\rm cross},j}\right)$. Since
the force $f_{+}$ as defined in Eq.~(\ref{eq:force at crossing})
is proportional to the amplitudes of the two fields at the crossing,
there exists an upper limit on the value of $\alpha_{j}$ for the
transition to occur while satisfying the condition described in Eq.~(\ref{eq:cond1}).
Thus, for $\alpha_{j}$ higher than the upper limit, the field amplitudes
are too small at the crossings and correspondingly the force $f_{+}$
isn't significant to cause the transition.

To estimate the value of $T_{z,j}$ which is closest to the transition,
we consider the first condition stated in Eq.~(\ref{eq:cond1}) and
perform an integration in a small neighborhood around $T_{c}$. By
integrating this condition, we determine the smallest value of $j$
that satisfies the condition and corresponds to the occurrence of
the transition. Thus, we obtain 
\begin{equation}
\int_{T_{c}-\delta T_{2}}^{T_{c}+\delta T_{1}}dT\left(\phi_{+}\phi_{-}^{2}-F^{2}\phi_{-}\right)\gtrsim O(0.1)\left|\left.\dot{\phi}_{+}\right|_{T_{c}-\delta T_{2}}^{T_{c}+\delta T_{1}}\right|.\label{eq:integraleqn}
\end{equation}
Since the fields must cross before the zero-crossing of $\phi_{+}$,
we choose the lower and upper boundaries as $T_{c}-\delta T_{2}=T_{s}$
and $T_{c}+\delta T_{1}=T_{z,j}$ where 
\begin{equation}
T_{s}=T_{z,j}-2\epsilon_{j}
\end{equation}
for $\epsilon_{j}=1/\left(F\sqrt{\alpha_{j}}\right)$. In \pA, this
choice of $\epsilon_{j}$ was defined to be approximately when the
deviation of the $\phi_{-}$ from $\phi_{-}^{(0)}$ of Eq.~(\ref{eq:phimapprox})
is roughly 10$\%$. Since we have $F\gg1$, the time interval $\epsilon_{j}$
is far less than unity for values of $\alpha_{j}\gg1$, and thus we
can expand $\phi_{\pm}$ within the interval of integration as 
\begin{align}
\phi_{-}(T) & \approx\phi_{-}^{(0)}(T_{s})+\dot{\phi}_{-}^{(0)}(T_{s})(T-T_{s})\\
\phi_{+}(T) & \approx\phi_{+}^{(0)}(T)\approx\alpha_{j}F^{2}\left((T-T_{z,j})-\frac{3}{2}(T-T_{z,j})^{2}\right).\label{eq:phiexpn}
\end{align}
In Appendix A of \pA, it was shown explicitly that neglecting higher
order derivatives for the $\phi_{-}$ field near transition ($T\sim T_{c}-O(2/F)$)
leads to a much better approximation and hence we restrict ourselves
to a first-order expansion for $\phi_{-}$. Upon substitution into
Eq.~(\ref{eq:integraleqn}) we get

\begin{align}
\int_{T_{s}}^{T_{z}}\left(\phi_{+}^{(0)}\phi_{-}^{2}-F^{2}\phi_{-}\right)dt & \gtrsim O(0.1)\left|\left.\dot{\phi}_{+}^{(0)}\right|_{T_{s}}^{T_{z,j}}\right|\\
\frac{F^{3}\left(1944+1458F\sqrt{\alpha_{j}}+384F^{2}\alpha_{j}+35F^{3}\alpha_{j}^{3/2}\right)}{60\sqrt{\alpha_{j}}\left(3+F\sqrt{\alpha_{j}}\right)^{4}} & \gtrsim O(0.1)\;6F\sqrt{\alpha_{j}}\\
\frac{7F^{2}}{12\alpha_{j}}-\frac{3F}{5\alpha_{j}^{3/2}}+O\left(\frac{1}{F\alpha_{j}^{5/2}}\right) & \gtrsim O(0.1)\;6F\sqrt{\alpha_{j}}
\end{align}
which when solved in the limit $F\gg1$ yields 
\begin{equation}
\alpha_{j}\lesssim\left(\frac{7F}{72\,O(0.1)}\right)^{\frac{2}{3}}.\label{eq:alphaj}
\end{equation}
We note that the upper limit in the right hand side (RHS) of Eq.~(\ref{eq:alphaj})
is sensitive to the choice of an $O(0.1)$ number. By comparing with
the numerical results, we choose a value of $0.07$, and thus obtain
\begin{equation}
\alpha_{j}\lesssim1.244\times F^{\frac{2}{3}}
\end{equation}
as the upper limit on the value of $\alpha_{j}$ for the transition
to occur close to the zero-crossing at $T_{z,j}$. Hence, using Eqs.~(\ref{eq:alphadefinition}),
(\ref{eq:cond1}) and (\ref{eq:alphaj}) we infer that the background
fields transition when 
\begin{equation}
\left|\dot{\phi}_{+}^{(0)}\right|_{T_{z,j}}=\alpha_{j}F^{2}\lesssim F^{\frac{2}{3}+2}.\label{eq:dphipbound}
\end{equation}
Eq.~(\ref{eq:dphipbound}) is isomorphic to Eq.~(\ref{eq:cond1}).
Since $F\gg1$, our assumption that $\alpha_{j}\gg1$ above Eq.~(\ref{eq:phiexpn})
is justified. For instance, if $F=20$ we obtain the upper bound as
$\alpha\approx9$ which also satisfies the second condition given
in Eq.~(\ref{eq:cond2}). We will now estimate the smallest value
of $j$ that satisfies the condition in Eq.~(\ref{eq:dphipbound})
for a given $c_{+}$ and initial conditions. Defining the upper bound
as $\alpha_{{\rm max}}=1.244F^{\frac{2}{3}}$ from Eq.~(\ref{eq:alphaj})
and $j_{c}$ as the $j$th index corresponding to the zero-crossing
closest to transition $T_{c}$, we require that 
\begin{align}
\alpha_{j_{c}} & <\alpha_{{\rm max}}\\
\omega\frac{\phi_{+}(0)}{F^{2}}e^{-\frac{3}{2}T_{z,j_{c}}}\sec(\varphi) & <\alpha_{{\rm max}}\\
T_{z,j_{c}} & >\frac{-2}{3}\ln\left(\frac{\alpha_{{\rm max}}F^{2}}{\omega\phi_{+}(0)\sec(\varphi)}\right)\\
\frac{1}{\omega}\left(\left(j_{c}-\frac{1}{2}\right)\pi+\varphi\right) & >\frac{-2}{3}\ln\left(\frac{\alpha_{\max}F^{2}}{\omega\phi_{+}(0)\sec(\varphi)}\right)\\
j_{c} & >\frac{1}{2}-\frac{1}{\pi}\left(\frac{2\omega}{3}\ln\left(\frac{\alpha_{{\rm max}}F^{2}}{\omega\phi_{+}(0)\sec(\varphi)}\right)+\varphi\right).
\end{align}
Therefore, the zero-crossing $T_{z,j_{c}}$ closest to the transition
$T_{c}$ is given by the expression 
\begin{equation}
T_{z,j_{c}}=\frac{1}{\omega}\left(\left(\text{Ceiling}\left[\frac{1}{2}-\frac{1}{\pi}\left(\varphi+\frac{2\omega}{3}\ln\left(\frac{\alpha_{{\rm max}}F^{2}}{\omega\phi_{+}(0)\sec(\varphi)}\right)\right)\right]-\frac{1}{2}\right)\pi+\varphi\right)\label{eq:transitontimeTc}
\end{equation}
which has a limiting behavior 
\begin{align}
\lim_{\omega\gg1}T_{z,j_{c}} & \approx\frac{\pi}{\omega}\text{Ceiling}\left[-\frac{2\omega}{\pi3}\ln\left(\frac{\alpha_{{\rm max}}F^{2}}{\omega\phi_{+}(0)\sec(\varphi)}\right)-\frac{1}{2}\right].
\end{align}
\begin{figure}[t]
\begin{centering}
\includegraphics[scale=0.63]{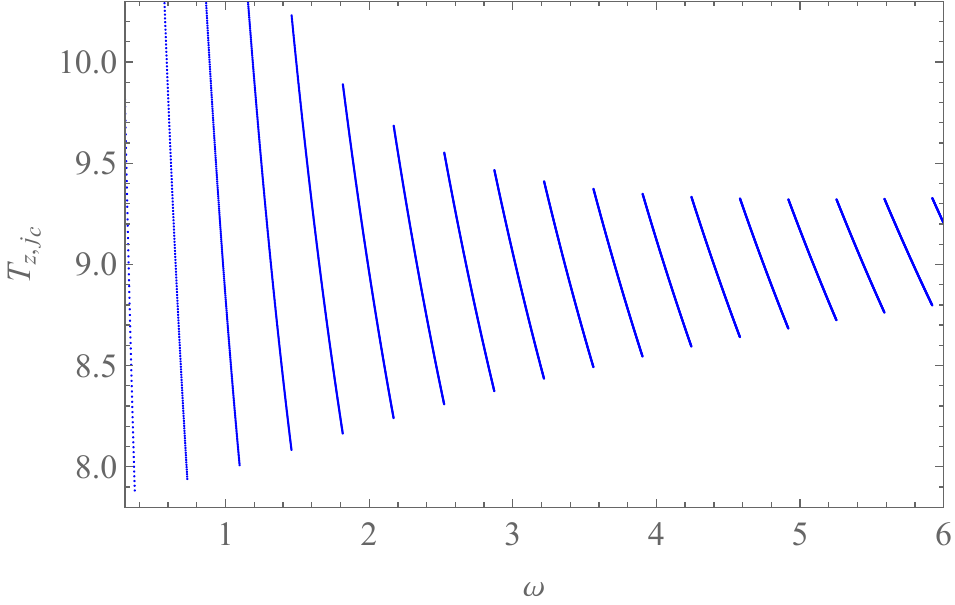} 
\par\end{centering}
\caption{\label{fig:Tzvscp}Plot of $T_{z,j_{c}}$ with respect to $\omega$
providing an approximate time at which the background fields transition
for different values of $c_{+}$. The curve is generated using Eq.~(\ref{eq:transitontimeTc})
and exhibits discontinuous curves corresponding to increasing integer
values of $j=j_{c}$ starting with $j_{c}=1$ for the first branch
corresponding to $\omega\lesssim0.4$. Once $T_{z,j_{c}}$ is known
for given Lagrangian parameters, an analytical estimate for the location
of the transition $T_{c}$ can be made using Eq.~(\ref{eq:Tc-emperical}).
The above plot is obtained using the standard fiducial set $P_{A}$
given in Eq.~(\ref{eq:Pset}).}
\end{figure}

\begin{figure}
\begin{centering}
\includegraphics[scale=0.65]{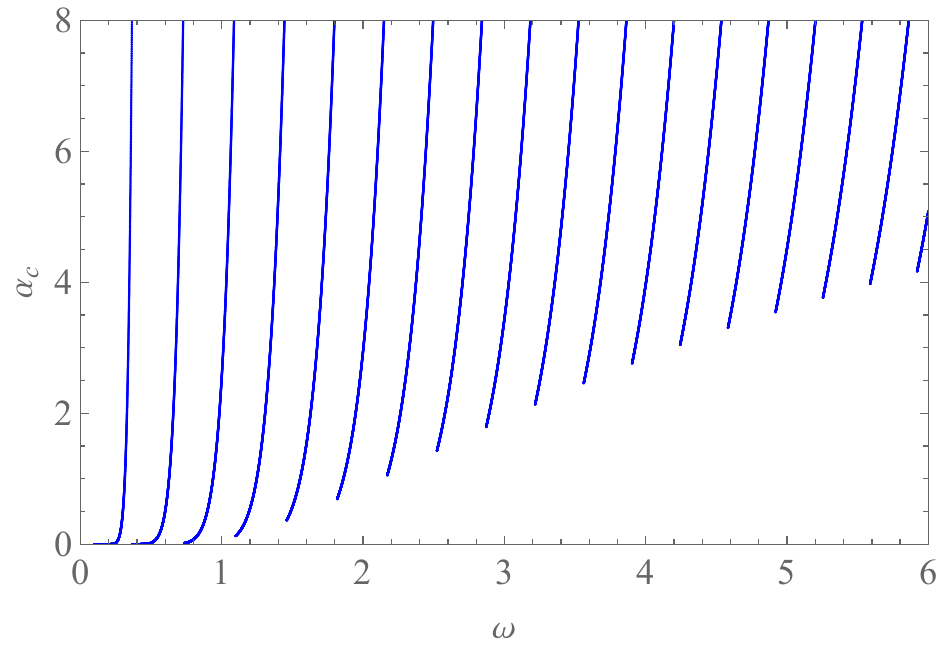} 
\par\end{centering}
\centering{}\caption{\label{fig:alphavscp}Plot highlighting different monotonic branches
in the $\omega(c_{+})-\alpha_{c}$ phase space showing the $c_{+}$
dependence of $\alpha_{c}$ for standard fiducial set $P_{A}$ defined
in Eq.~(\ref{eq:Pset}). Each of the branches in the above figures
corresponds to an increasing value of $j_{c}$ where $j_{c}=1$ for
the first branch starting from $c_{+}=9/4$ and similarly $j_{c}=4$
for the branch where $\omega\in[1.11,1.4]$ corresponding to $c_{+}\in[3.5,4.21]$.}
\end{figure}

In Fig.~\ref{fig:Tzvscp} we plot $T_{z,j_{c}}$ with respect to
$\omega$ using Eq.~(\ref{eq:transitontimeTc}). The corresponding
value of the parameter $\alpha$ at $T_{z,j_{c}}$ is 
\begin{equation}
\alpha_{c}=\omega\frac{\phi_{+}(0)}{F^{2}}e^{-\frac{3}{2}T_{z,j_{c}}}\sec(\varphi)\label{eq:alpha_c}
\end{equation}
where we introduce the index $c$ on $\alpha$ to distinguish it from
other non-transition $\alpha_{j}$ values. Note that for the rest
of our discussion, $\alpha_{c}$ refers to the value at $T_{z,j_{c}}$
close to transition $T_{c}$. In Fig.~\ref{fig:alphavscp} we show
the value of $\alpha_{c}$ as a function of $\omega$ where the individual
discontinuous curves are bounded from above by the upper bound $\alpha_{{\rm max}}$
given in Eq.~(\ref{eq:alphaj}), while the lower bound in each successive
branch increases with $\omega$. The plot highlights distinct monotonic
branches in the $\omega(c_{+})-\alpha_{c}$ phase space. Within a
branch, such as the range $c_{+}\in[3.5,4.21]$, $\alpha_{c}$ exhibits
a monotonically increasing behavior, reaching the maxima defined in
Eq.~(\ref{eq:alphaj}), before abruptly transitioning to the next
branch. 

This can be understood by recalling that, for the background fields
to transition, the condition described in Eq.~(\ref{eq:cond1}) must
be fulfilled. As $\alpha$ increases, the LHS in Eq.~(\ref{eq:cond1})
decreases while the RHS increases. Consequently, for sufficiently
large values of $\alpha$, the left-hand side can drop below the threshold
required to meet the condition, prompting the fields to transition
at the next zero-crossing (succeeding branch). This occurs as they
dissipate enough total kinetic energy through Hubble friction, leading
to a reduction in the value of $\alpha$.

Furthermore, we observe that for cases with larger values of $c_{+}$,
the value of $\alpha$ at transition is larger than $O(1)$. This
will have intriguing implications when we study the isocurvature power
spectrum in Sec.~\ref{sec:Isocurvature-power-spectrum} for the cases
where the background fields tend to exhibit chaotic behavior.

To summarize the above analysis, as the parameter $c_{+}$ increases,
the background fields undergo more frequent oscillations with shorter
time periods. Since the system of background fields must lose sufficient
energy before transition, the number of zero-crossings before transition
increases with $c_{+}$. Close to each zero-crossing, the fields momentarily
cross each other. If the field velocity, characterized by the parameter
$\alpha$, is sufficiently small at a given crossing $T_{{\rm cross},j_{c}}$
while still satisfying Eq.~(\ref{eq:cond1}), then the two fields
are said to transition. This imposes an upper bound on the value of
$\alpha$ for the transition to occur.

In the above analysis, we have focused on the adiabatic approximation,
neglecting the nonadiabatic effects originating from all zero-crossings
($j<j_{c}$) prior to transition. The analysis and examples considered
in \pA ~had $T_{c}$ close to the first zero-crossing ($j_{c}=1$)
and therefore nonadiabatic effects due to previous zero-crossings
$T_{z,j}$ were absent. These nonadiabatic oscillations can be understood
as rapid transient (homogeneous) oscillations of the $\phi_{-}$ field,
generated at $T_{z,j}$, due to the quartic interaction term, $(\phi_{+}\phi_{-}-F^{2})^{2}$,
in the reduced Lagrangian. Consequently, the homogeneous component,
$\phi_{-}\approx\phi_{-,{\rm Tr}}$, oscillates rapidly with an effective
frequency controlled by $\phi_{+}$ and an amplitude approximately
proportional to $\alpha_{j}^{-3/4}$. In Appendix \ref{sec:-phim_Transient}
we provide a detailed derivation of $\phi_{-,{\rm Tr}}$.

However, if $T_{z,j}$ is at least two e-folds time interval prior
to $T_{c}$, with the corresponding value of $\alpha_{j}$ much larger
than unity, the nonadiabatic effects from previous zero-crossings
corresponding to values of $\alpha_{j}\gg\alpha_{\max}$ do not induce
significant changes in the mode functions. This is because the dynamical
effects differing from that of a constant mass decays as 
\begin{equation}
\frac{\Delta I_{\pm}}{I_{\pm}}\sim O\left(\frac{1}{\alpha_{j}}\right)\sim O\left(e^{-\frac{3}{2}(T_{c}-T_{z,j})}\right).\label{eq:deviation}
\end{equation}
These $\alpha_{j}^{-1}$ effects are the non perturbative resonant
oscillations that occur at each crossing of the background fields.

For cases where a previous zero-crossing has occurred within about
two e-folds time interval of $T_{c}$ (applicable to $c_{+}\gtrsim5$),
the nonadiabatic effects from the zero-crossing at $T_{z,j_{c}-1}$
can have a significant impact. The rapid oscillations of $\phi_{-,{\rm Tr}}$
lead to an increase in the effective mass of the $\phi_{+}$ field,
resulting in a mass-squared function $m_{+}^{2}\approx c_{+}+\phi_{-}^{2}$
of the $\phi_{+}$ field that can exceed $c_{+}$ for $T_{z,j_{c}-1}<T<T_{c}$.
As a consequence, the zero-crossing $T_{z,j_{c}}$ corresponding to
$T_{c}$ is modified and occurs slightly earlier than predicted by
Eq.~(\ref{eq:transitontimeTc}) due to an increased frequency of
$\phi_{+}$ during the time-interval $T_{z,j_{c}-1}<T<T_{c}$.

In Appendix \ref{sec:-phim_Transient} we analytically solve the equation
of motion (EoM) for the $\phi_{+}$ field in the time-region $T_{z,j_{c}-1}<T<T_{c}$,
taking into account the finite nonadiabatic effects arising due to
the UV oscillations of $\phi_{-}$ field. We derive a general expression
for the deviation of the $\phi_{+}$ field from the zeroth order solution
$\phi_{+}^{(0)}$. This modified analytic solution for the $\phi_{+}$
given in Eq.~(\ref{eq:phip-general}) is useful in predicting the
location of the next zero-crossing at $T_{z,j_{c}}$. Consequently
we can state that the next zero-crossing at $T_{z,j_{c}}$ occurs
at 
\begin{equation}
T_{z,j_{c}}-T_{z,j_{c}-1}=\frac{\pi}{\omega}-\Delta T\left(A_{j_{c}-1}\right)\label{eq:Tzjc with delta_T}
\end{equation}
where $\Delta T\left(A_{j_{c}-1}\right)$ is a function of the amplitude
of $\phi_{-,{\rm Tr}}$ at $T_{z,j_{c}-1}$.\footnote{See Eq.~(\ref{eq:Aj}) for an expression of $A_{j_{c}-1}$.}
We can obtain an estimate for $\Delta T$ by solving the transcendental
equation corresponding to $\phi_{+}(T_{z,j_{c}})=0$ using the analytic
solution of $\phi_{+}$ given in Eq.~(\ref{eq:phip-general}).

\begin{figure}
\begin{centering}
\includegraphics[scale=0.75]{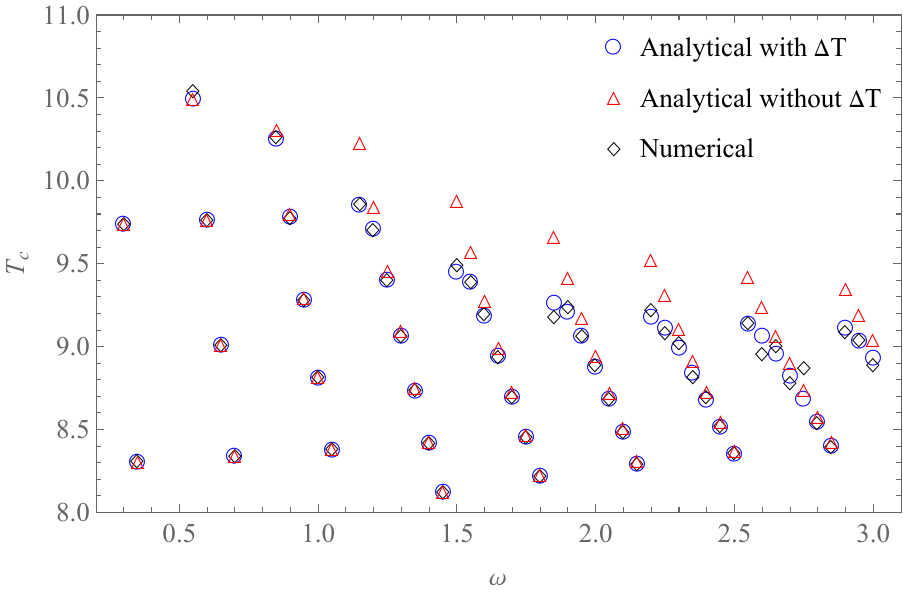} 
\par\end{centering}
\caption{\label{fig:Tc-plot}Plot showing the transition time $T_{c}$ as a
function of $\omega$ for the standard fiducial set $P_{A}\equiv\{F=20.2,\,c_{-}=0.5,\,\epsilon_{0}=0,\,\phi_{+}(0)=0.1M_{p}/H\}$.
For comparison, we plot the values using our analytical estimate from
Eq.~(\ref{eq:Tc-emperical}) with and without the $\Delta T$ correction
for $T_{z,j_{c}}$ as described in Eq.~(\ref{eq:Tzjc with delta_T}).
We observe that our analytical prediction including the $\Delta T$
correction matches with the numerical values with an accuracy $\gtrsim90\%$.
As noted in the main text, the $\Delta T$ correction becomes significant
for $\omega\gtrsim1.5$ which corresponds to a time-interval $\pi/\omega\lesssim2$
e-folds between the adjacent zero-crossings and hence the non-adiabatic
corrections described in App.~\ref{sec:-phim_Transient} cannot be
neglected. We find that these corrections are important when $\alpha_{c}\lesssim1$
since the corresponding non-adiabatic effects from a previous zero-crossing
scale as $\approx e^{\frac{-\pi}{\omega}}\alpha_{c}^{-1/2}$.}
\end{figure}

For resonant underdamped cases where $\alpha_{c}\gtrsim\alpha_{{\rm L}}\approx0.2$,
the transition time $T_{c}$ can be estimated using the expression
\begin{equation}
T_{c}\approx T_{z,j_{c}}-\frac{0.7}{F\alpha_{c}}\qquad\alpha_{c}\gtrsim\alpha_{{\rm L}}
\end{equation}
given in Sec.~4 of \pA. For a broad range of $\alpha_{c}$ values
including non-resonant cases, we propose the following fitting formula:
\begin{equation}
T_{c}\approx T_{z,j_{c}}-\left(-2.396+\frac{9.8047}{1+1.112\left(F\alpha_{c}/20.2\right)-6.013\left(F\alpha_{c}/20.2\right)^{0.5}+7.935\left(F\alpha_{c}/20.2\right)^{0.25}}\right)\qquad\alpha_{c}\gtrsim10^{-4}\label{eq:Tc-emperical}
\end{equation}
where $T_{z,j_{c}}$ is obtained from Eq.~(\ref{eq:Tzjc with delta_T}).
In Fig.~\ref{fig:Tc-plot} we plot the transition time, $T_{c}$,
as a function of $\omega$ for fiducial set $P_{A}\equiv\{F=20.2,\,c_{-}=0.5,\,\epsilon_{0}=0,\,\phi_{+}(0)=0.1M_{p}/H\}$.
We compare our analytic estimates obtained from the expressions given
in Eq.~(\ref{eq:Tc-emperical}) with the numerical values and find
an accuracy $\gtrsim90\%$. We observe that the non-adiabatic effects
generated from the previous zero-crossing must be taken into account
using the $\Delta T$ correction for $\omega\gtrsim1.5$ corresponding
to $c_{+}\gtrsim4.5$. Note that these corrections are important when
$\alpha_{c}\lesssim1$ since the corresponding non-adiabatic effects
from a previous zero-crossing scale as $A_{j_{c}-1}^{2}\sim O\left(\alpha_{j_{c}-1}^{-1/2}F^{3}\right)$
where 
\begin{equation}
\alpha_{j_{c}-1}^{-1/2}\approx\left(\alpha_{c}e^{\frac{3\pi}{2\omega}}\right)^{-1/2}\propto\alpha_{c}^{-1/2}
\end{equation}
where we remind the reader that $\alpha_{c}\equiv\alpha_{j_{c}}$
is the value of the $\alpha$ parameter at the zero-crossing $T_{z,j_{c}}$.

In \pA, the authors observed that the location of the first bump
in the isocurvature power spectrum is approximately $k_{{\rm first-bump}}\approx2Ha(0)\exp\left(T_{c}\right)$.
Since $T_{c}\approx T_{z,j_{c}}+O(1/F)$ for $\alpha_{c}>\alpha_{{\rm L}}$,
we find that for $c_{+}\gg9/4$, 
\begin{align}
T_{c}\approx T_{z,j_{c}} & \sim O\left(\text{\ensuremath{\frac{2}{3}\ln\left(\frac{\phi_{+}(0)}{F^{8/3}}\right)}}\right).
\end{align}
Therefore, the location of the first bump or the cutoff scale from
a blue-tilted part of the spectrum to a flat plateau is related to
the hierarchy between the initial displacement of the $\phi_{+}$
field and the $f_{PQ}\equiv F_{a}$ scale. In order to hide the isocurvature
spectrum at large CMB scales, we require a large hierarchy or displacement
along the flat direction. This is a generic requirement for the blue
axion models where the blue index is generated during the slow roll
along a flat direction. Conversely, it is possible to move the cutoff
scale towards lower $k$ values by choosing a larger $F$, while keeping
all other cosmological parameters such as the number of inflationary
efolds $N_{{\rm inf}}$, $\phi_{+}(0)$, $H$, and reheating temperature
$T_{{\rm RH}}$ fixed. The estimation of $T_{c}$, given in this section,
allows us to predict the cutoff scale $k_{c}$ for the high-blue isocurvature
spectrum.

In the next section, we will discuss the isocurvature power spectrum
in the blue-tilted region for for the massive $c_{+}$ cases and present
plots highlighting the $\alpha_{c}$ dependence of the final isocurvature
amplitudes.

\section{\label{sec:Isocurvature-power-spectrum} Isocurvature power spectrum
in the blue region}

For the axion model considered in this work, the isocurvature power
spectrum generated during inflation is evaluated using the expression
given in Eq.~(\ref{eqspec}) by solving the coupled differential
system in Eq.~(\ref{eq:modeeq}) for the associated mode fluctuations
$\delta a_{\pm}/2\equiv I_{\text{\ensuremath{\pm}}}$. Since we are
considering massive underdamped cases where $c_{+}>9/4$, the isocurvature
power spectrum generically has a high-blue spectral index $\text{Re}\left[n_{{\rm I}}\right]\approx4$.
This blue-tilted region of the power spectrum extends to all modes
$k<k_{c}$ that exit the horizon well before the transition of the
background fields $\phi_{\pm}$ at time $T_{c}$, where the cutoff
scale $k_{c}$, is associated with transition $T_{c}$, and marks
the region in the spectrum where the power spectrum begins to settle
into a scale-invariant massless plateau.

In \pA, the authors uncovered that for specific cases where the value
of the parameter $\alpha_{c}$ at transition is $\gtrsim O(0.1)$,
there is a significant increase in the kinetic energy during the field
crossing. This substantial kinetic energy at the transition leads
to nonadiabatic effects during the period when the background fields
are approaching the potential minimum that results in diverse spectral
shapes with multiple bumps/oscillations. Moreover, for the cases covered
in \pA, the authors found that these resonant non-adiabatic effects
for the underdamped cases lead to an amplification of at least $O(30)$
relative to the massless plateau. In summary, the isocurvature power
spectrum for the massive underdamped fields consists of two regions:
a blue-titled spectrum for $k\lesssim k_{c}$ and a region with multiple
bumps for $k>k_{c}$ that eventually settles to a scale-invariant
massless plateau. The cutoff scale $k_{c}$ that separates these two
regions of the spectrum is a function of $T_{c}$.

\subsection{\label{subsec:Zero-mode}Zero-mode $I_{0}$}

Because the superhorizon inhomogeneous modes $I_{\pm}(k)$ behave
similarly as the background fields (as explained below), the superhorizon
modes in principle only need to be solved until the horizon crossing
and matched to the background fields, similar to what happens to curvature
perturbation variables during the quasi-dS era. This simplifies the
computation if one has access to an accurate computation of background
field solutions. The inhomogeneous modes $I_{\pm}(k)$ can be solved
trivially for $T<T_{c}$ for the small $k$ region analytically, which
allows this matching program to be efficient for the rising blue part
of the power spectrum characterized by a simple power law. The main
nontriviality is to argue that despite the nonadiabaticities of the
mass matrix that affects the mode equations for $T>T_{c}$, we can
compute the functional behavior of the power spectrum (to a matching
condition-dependent $r_{a}$ accuracy). This will allow us to obtain
an expression for the approximate $k$ dependence of the power spectrum
for long wavelength modes. Furthermore, because of the accidental
duality that exists between the superhorizon modes and the background
fields, we will be able to semi-quantatively explain the chaotic map
between the Lagrangian parameters such as $c_{+}$ and the isocurvature
amplitude.

Let us begin with the mode function governed by Eq.~(\ref{eq:modeeq})
for the axion model. After normalizing with the BD adiabatic vacuum,
the mode function for $T<T_{c}$ when $\phi_{+}^{2}\gg F^{2}$ is
given as 
\begin{equation}
\lim_{K\exp(-T)\gg1}I(K,T\ll T_{c})\approx I^{(\mathrm{early})}(K,T)\equiv\left(\frac{e^{-T}}{2a(0)}\right)^{3/2}\sqrt{\frac{\pi}{H}}e^{-\omega\frac{\pi}{2}+i\frac{\pi}{4}}H_{i\omega}^{(1)}\left(Ke^{-T}\right)\left[\begin{array}{c}
1\\
0
\end{array}\right]+O\left(\frac{F^{2}}{\phi_{+}^{2}(T)}\right)\left[\begin{array}{c}
0\\
1
\end{array}\right]\label{eq:Imode-earlytime}
\end{equation}
where $\omega=\sqrt{c_{+}-9/4}$, $a(0)$ is the scale factor at a
chosen initial time $T_{0}=0$, and the normalized wave vector $K$
is defined in Eq.~(\ref{eq:KDEF}). Note that the normalization here
is different from $1/\sqrt{2ka^{2}}$ by another factor of $1/\sqrt{2}$.
Here we emphasize that the BD boundary condition is aligned along
the lightest normalized real eigenvector, $e_{1}$, of the mode mass-matrix
$\tilde{M}^{2}$ which was defined in Eq.~(\ref{eq:massmat}). In
terms of a parameter, $\lambda=F^{2}/\phi_{+}^{2}$, the lightest
eigenvector of $\tilde{M}^{2}$ in the limit $\lambda\ll1$ can be
given as 
\begin{equation}
e_{1}(T)=\left[\begin{array}{c}
1\\
0
\end{array}\right]-\lambda\left[\begin{array}{c}
0\\
1
\end{array}\right]+O(\lambda^{2}).\label{eq:e1}
\end{equation}
while the heavier eigenvector, $e_{2}$, is 
\begin{equation}
e_{2}(T)=\left[\begin{array}{c}
0\\
1
\end{array}\right]+\lambda\left[\begin{array}{c}
1\\
0
\end{array}\right]+O(\lambda^{2}).\label{eq:e2}
\end{equation}
Hence, Eq.~(\ref{eq:Imode-earlytime}) can be equivalently written
as 
\begin{equation}
I^{(\mathrm{early})}(K,T\ll T_{c})\approx\left(\frac{e^{-T}}{2a(0)}\right)^{3/2}\sqrt{\frac{\pi}{H}}e^{-\omega\frac{\pi}{2}+i\frac{\pi}{4}}H_{i\omega}^{1}\left(Ke^{-T}\right)e_{1}(T)\label{eq:I_along_e1}
\end{equation}
which has a peculiar normalization in that in the $K\rightarrow\infty$
limit, it has an extra power of $1/\sqrt{2}$ coming from a choice
of axion normalization and a physically irrelevant extra overall minus
sign coming from a phase choice. We note that Eq.~(\ref{eq:I_along_e1})
is an approximation that holds only when the contribution from the
heavier mode can be neglected. The most general expression for $I(K,T)$
valid at all times can be written as 
\begin{equation}
I(K,T)=y_{1}(K,T)e_{1}(T)+y_{2}(K,T)e_{2}(T)\label{eq:instanteigen}
\end{equation}
where $y_{1,2}$ are the corresponding mode functions in the instantaneous
eigenstate basis. In \pA ~the authors demonstrated that as the background
fields approach transition at $T_{c}$, substantial mode mixing occurs.
Through their analysis, they revealed that the heavy-mode mixing is
most significant when the hierarchy between the lightest and the heaviest
mass eigenvalues is minimal. As a result, the expression in Eq.~(\ref{eq:I_along_e1})
is generically invalid as $T\rightarrow T_{c}$. However, when the
background fields have settled to the minima, the mode function asymptotes
to 
\begin{equation}
I(K,T_{\infty})\approx y_{1}(K,T_{\infty})e_{1}(T_{\infty})\label{eq:futureasymptote}
\end{equation}
 where $e_{1}(T_{\infty})$ corresponds to the Goldstone mode of the
axionic system.

Let us now consider modes that exit the horizon at $\mathcal{T}_{K}<T_{c}$
which is defined to be when 
\begin{equation}
K^{2}\exp\left(-2\mathcal{T}_{K}\right)=r_{a}(c_{+}-2)
\end{equation}
where $r_{a}\sim0.1$ represents the accuracy with which one wants
to estimate the amplitude. Because of Eq.~(\ref{eq:I_along_e1}),
we know
\begin{equation}
y_{1}(K,\mathcal{T}_{K})\approx\left(\frac{e^{-\mathcal{T}_{K}}}{2a(0)}\right)^{3/2}\sqrt{\frac{\pi}{H}}e^{-\omega\frac{\pi}{2}+i\frac{\pi}{4}}H_{i\omega}^{(1)}\left(Ke^{-\mathcal{T}_{K}}\right).\label{eq:y1early}
\end{equation}
One can also check that $y_{2}(K,\mathcal{T}_{K})$ term is comparatively
negligible for $\mathcal{T}_{K}<T_{c}$.Now let us consider Eq.~(\ref{eq:modeeq})
with $k=0$ which describes a mode a priori distinct from any physical
modes because $k=0$ is always outside of the horizon:
\begin{equation}
\left(\partial_{T}^{2}+3\partial_{T}\right)I_{0}+\tilde{M}^{2}I_{0}=0.\label{eq:zero-mode-eqn}
\end{equation}
 Note that the variable change 
\begin{equation}
I_{0}\rightarrow I_{0}^{(\mathrm{dual})}\equiv(f_{0}\phi_{+},-f_{0}\phi_{-})\label{eq:dualitymap}
\end{equation}
maps the zero-mode system in Eq.~(\ref{eq:zero-mode-eqn}) to the
background field EoMs in Eqs.~(\ref{eq:backgroundeom0}) and (\ref{eq:backgroundeom})
for a nonvanishing constant $f_{0}$. This is a type of an accidental
duality in which the background equations become identical to the
perturbation equations even though the background equations are nonlinear.
Although $I$ that we seek appearing in Eq.~(\ref{eq:instanteigen})
is fundamentally different from $I_{0}^{\mathrm{(dual)}}$ since $I_{0}^{\mathrm{(dual)}}$
is real up to a time-independent phase, we know that one linear combination
of $I_{0}$ and $I_{0}^{*}$ can be made to equal $I_{0}^{\mathrm{(dual)}}$.
This, in particular, means that if $\phi_{\pm}$ solutions exhibit
exponential sensitivity to parameters, then $I_{0}$ will as well.
Such exponentially sensitive parametric dependence will be presented
later in this section.

We impose boundary conditions for Eq.~(\ref{eq:zero-mode-eqn}) for
the zero-mode $I_{0}$ at a time $T_{0}\ll T_{c}$ along the direction
of the lightest eigenvector $e_{1}$, following a similar phase expression
as shown in Eq.~(\ref{eq:I_along_e1}):
\begin{align}
I_{0}(T_{0}) & =e^{-\left(3/2+i\omega\right)T_{0}}e_{1}\label{eq:initi0}\\
\left.\partial_{T}I_{0}\right|_{T=T_{0}} & =-\left(3/2+i\omega\right)e^{-\left(3/2+i\omega\right)T_{0}}e_{1}\label{eq:zeromodeBC}
\end{align}
which is not the same as the BD condition since these modes are already
outside of the horizon. On the other hand, unlike the dual $I_{0}^{\mathrm{(dual)}}$
which can be made real by dividing by $f_{0}$, the zero mode $I_{0}(T)$
here is complex with a time-dependent phase just like $I(K,T)$ which
means that $I_{0}^{*}$ will be an independent solution once $I_{0}$
is known owing to the real valued nature of the differential equation
system. At the horizon exit time $\mathcal{T}_{K}<T_{c}$, the zero-mode
near time $\mathcal{T}_{K}$ can be given by the expression 
\begin{equation}
I_{0}(T)\approx e^{-\left(3/2+i\omega\right)T}e_{1}(T)\,\,\,\,\,\,\,\,\,\,\,\,\,\mbox{ near }T\sim\mathcal{T}_{K}\label{eq:Itk}
\end{equation}
because the heavier mode contribution can still be neglected at that
time. The $k$-dependent mode function $I(K,\mathcal{T}_{K}$) can
be written as 
\begin{equation}
y_{1}(K,T)e_{1}(T)\approx c_{1}(K)I_{0}(T)+c_{2}(K)I_{0}^{*}(T)\label{eq:matching}
\end{equation}
for $T\sim\mathcal{T}_{K}$. Note that we have conveniently avoided
any contribution from the heavier eigenmode $e_{2}$ at $\mathcal{T}_{K}$
by ensuring that $\lambda(\mathcal{T}_{K})\ll1$. The coefficients
$c_{1,2}(K)$ are obtained by using Eqs.~(\ref{eq:e1}), (\ref{eq:y1early}),
(\ref{eq:Itk}), and (\ref{eq:matching}). For $T>T_{c}>\mathcal{T}_{K}$,
the complete $I_{0}(T)$ solution incorporates all significant interactions
arising from the mixing of nontrivial heavier mode such that $I(T\geq\mathcal{T}_{K})$
can be evaluated as 
\begin{equation}
I(K,T\geq\mathcal{T}_{K})=c_{1}(K)I_{0}(T)+c_{2}(K)I_{0}^{*}(T)\label{eq:Ifactorizaton}
\end{equation}
where we obtain the matched coefficients $c_{1,2}(k)$ to be 
\begin{equation}
c_{1}(k)=\left(1+i\right)2^{-2-i\omega}\left(\frac{1}{a(0)}\right)^{3/2}\left(\frac{k}{a(0)H}\right)^{i\omega}\exp\left(-\pi\omega/2\right)\sqrt{\frac{\pi}{H}}\frac{\left(1+\coth\left[\omega\pi\right]\right)}{\Gamma\left(1+i\omega\right)}+O(r_{a})
\end{equation}
\begin{eqnarray}
c_{2}(k) & = & \left(1-i\right)2^{-2+i\omega}\left(\frac{1}{a(0)}\right)^{3/2}\left(\frac{k}{a(0)H}\right)^{-i\omega}\exp\left(-\pi\omega/2\right)\frac{1}{\sqrt{\pi H}}\Gamma(i\omega)+O(r_{a})\label{eq:c1c2}
\end{eqnarray}
where $k=Ka(0)H$. Since we require that $\mathcal{T}_{K}<T_{c}$,
the above method is valid for $k$ modes that satisfy 
\begin{equation}
\ln\left[\frac{k}{a(0)H\sqrt{r_{a}c_{+}}}\right]<T_{c}.\label{eq:k-restriction}
\end{equation}
Numerically we found an accuracy up to $90\%$ for modes $K\lesssim0.35\exp\left(T_{c}\right)$.
Note that $|c_{1,2}|^{2}$ is independent of $k$ because $i\omega$
is imaginary.

The axion isocurvature power spectrum in Eq.~(\ref{eqspec}) can
be expressed in terms of the zero-mode by expanding 
\[
I^{\dagger}\left(\begin{array}{cc}
r_{+}^{2} & 0\\
0 & r_{-}^{2}
\end{array}\right)I
\]
in terms of $I_{0}$ evaluated at a time $T=T_{\infty}$. At $T_{\infty}$,
the Goldstone theorem is satisfied and the $\tilde{M}^{2}$ mass-matrix
yields one massless and one massive eigenvalue where the massless
mode corresponds to the axion. Therefore, for the normalized massless
eigenvector $\psi_{0}$ at the end ($T\rightarrow T_{\infty}$): 
\begin{equation}
\tilde{M}^{2}\psi_{0}=0
\end{equation}
or more explicitly 
\begin{equation}
\psi_{0}=\frac{1}{\sqrt{c_{+}+c_{-}}}\left(\begin{array}{c}
-\sqrt{c_{-}}\\
\sqrt{c_{+}}
\end{array}\right).
\end{equation}
Since as $T\rightarrow T_{\infty},~\tilde{M}^{2}\psi_{0}\rightarrow0$,
we find 
\begin{equation}
I_{0}(T)\approx\left(\mathcal{N}e^{i\theta}+\mathcal{A}e^{-3T}\right)\psi_{0}\label{eq:z-factorization}
\end{equation}
where $\mathcal{A}$ is a complex number, $\mathcal{N}$ and $\theta$
are real numbers independent of $k$. In terms of $\psi_{0}$, 
\begin{equation}
\left(\begin{array}{cc}
r_{+}^{2} & 0\\
0 & r_{-}^{2}
\end{array}\right)_{T=T_{\infty}}=C\left[\frac{c_{-}+c_{+}}{c_{-}}\psi_{0}\psi_{0}^{\dagger}+\sqrt{\frac{c_{+}}{c_{-}}}\sigma_{x}\right]
\end{equation}
for a constant $C$ that depends upon $F$, $\theta_{+}$ and $c_{\pm}$
and where $\sigma_{x}$ is a Pauli matrix. Using the above definitions,
the isocurvature power spectrum for $k$ modes that satisfy Eq.~(\ref{eq:k-restriction})
is given as 
\begin{equation}
\Delta_{s}^{2}(k)\approx\mathcal{N}^{2}\left(4\frac{\omega_{a}^{2}}{\theta_{+}^{2}}\right)\left(\frac{k^{3}}{2\pi^{2}}\right)\frac{\sqrt{c_{+}c_{-}}(c_{+}^{2}+c_{-}^{2})|c_{1}(k)e^{i\theta}+c_{2}(k)e^{-i\theta}|^{2}}{(c_{+}+c_{-})^{3}(F_{a}^{2}-\sqrt{c_{-}c_{+}}H^{2})}\label{spectrum-1}
\end{equation}
where the $k$-independent real coefficients $\mathcal{N}$ and $\theta$
cannot be analytically computed in the present approach.

\begin{figure}[H]
\begin{centering}
\includegraphics[scale=0.55]{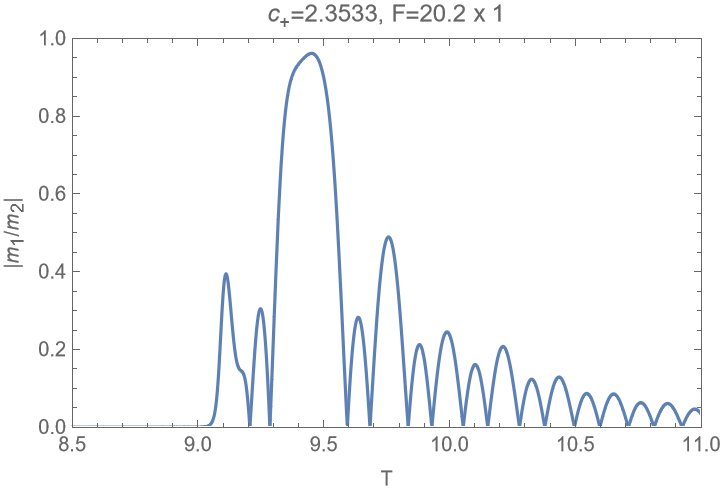}$\quad$\includegraphics[scale=0.55]{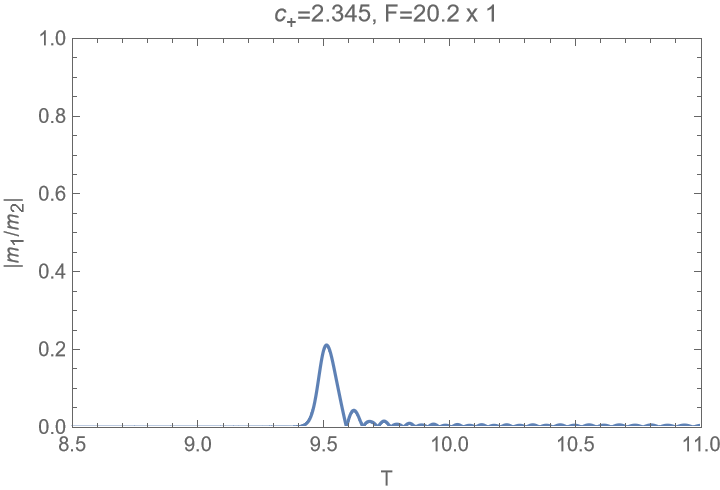} 
\par\end{centering}
\begin{centering}
\includegraphics[scale=0.38]{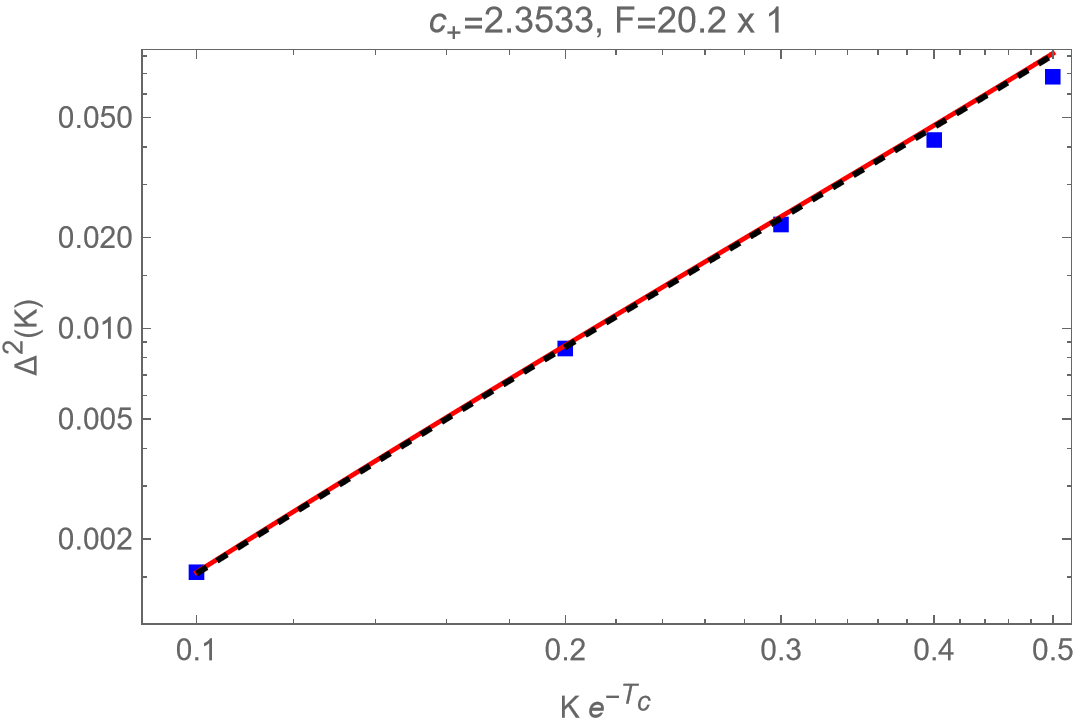}$\quad$\includegraphics[scale=0.38]{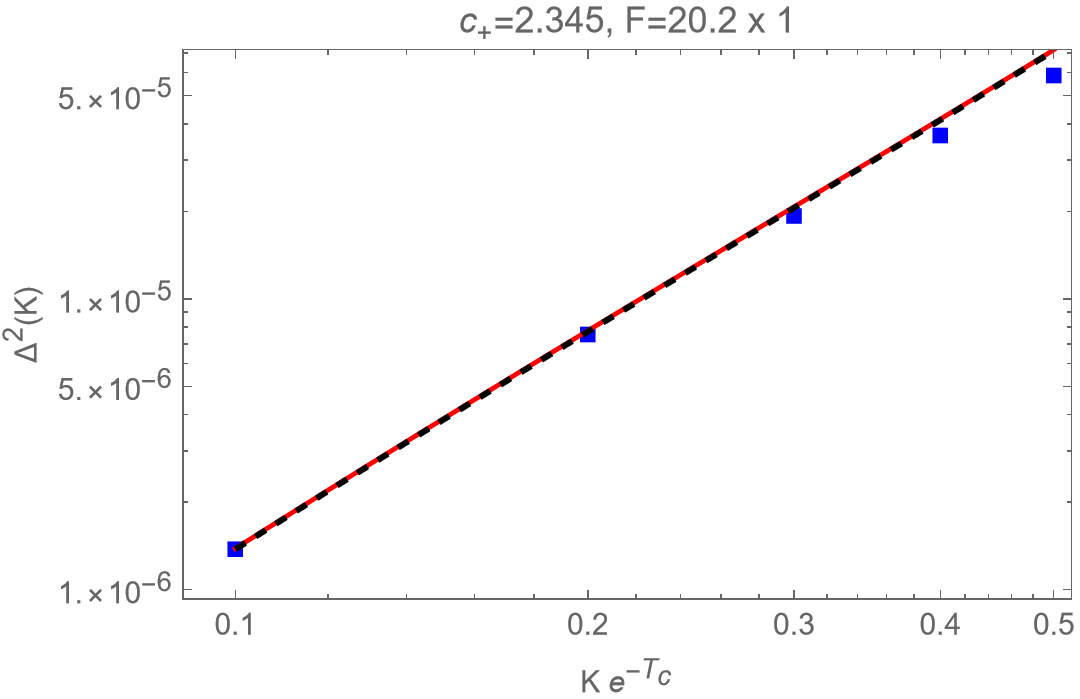} 
\par\end{centering}
\caption{\label{fig:comparison}In the top row, we plot the ratio $|m_{1}/m_{2}|$
of the lightest over heaviest mass eigenvalues of the mass-matrix
$\tilde{M}^{2}$ for two fiducial cases with $c_{+}$ values $2.3533$
and $2.345$, both at $F_{a}/H=20.2$ and $c_{-}=0.5$. A substantial
hierarchy between the lighter and heavier eigenvalues renders the
impact of the mode mixing and the contribution from the heavier mode
$e_{2}$ negligible. Examining the two plots in the top row, we observe
that the heavy mode mixing is most pronounced for $c_{+}=2.3533$
(left) compared to $c_{+}=2.345$ (right). In the bottom row, we compare
the isocurvature power spectrum obtained from solving the $k$-dependent
mode equations as given in Eq.~(\ref{eq:modeeq}) (solid markers)
with our approximation (solid and dashed curves) given in Eq.~(\ref{spectrum-1})
constructed from the numerical solution to the zero-mode equation.
The solid red (dashed black) curves are plotted using the numerical
value of $\theta$ (approximation $\theta=-\omega T_{c}$). By construction,
the zero-mode solution incorporates all significant nontrivial interactions
arising from the heavy mode mixing. Note that the slope of the power
spectrum in the $k$-range shown above is $\approx2.6$, signaling
the nontrivial correction from the sinusoid in log $k$.}
\end{figure}

The power spectrum is directly proportional to the square of the amplitude
$\mathcal{N}$ of the scalar mode $z$. The $c_{1}^{*}(k)c_{2}(k)$
have only logarithmic $k$ dependence, which makes this solvable situation
having an approximately $k^{3}$ dependence in the long wavelength
limit as long as $c_{1}(k)e^{i\theta}+c_{2}(k)e^{-i\theta}$ does
not vanish. In that sense, the prediction in this parametric region
($k^{3}$ part of the spectrum with small sinusoidal oscillations
in log $k$) is not particularly interesting in $k$ dependence. However,
the amplitude computation is nontrivial, and that is what is captured
by the intricate numerical computation of $|I_{0}|$. Through the
factorization of the zero-mode in Eq.~(\ref{eq:z-factorization}),
the normalization of the isocurvature power spectrum at late times
as given by the expression in Eq.~(\ref{spectrum-1}) now depends
upon the mode amplitude $\mathcal{N}$, while the dependence on the
wavenumber $k$ is determined by the $\theta$ parameter. In Appendix
\ref{sec:Approx_theta} we show that $\theta$ can be approximated
as $-\omega T_{c}.$ Consequently, we can conveniently eliminate $\theta$
when evaluating the power spectrum using Eq.~(\ref{spectrum-1}).

In Fig.~\ref{fig:comparison}, we compare the isocurvature power
spectrum obtained from solving the $k$-dependent mode equations as
outlined in Eq.~(\ref{eq:modeeq}) (blue markers) with our approximation
given in Eq.~(\ref{spectrum-1}) constructed from the zero-mode solution.
Through the plots, we highlight that the zero-mode solution $I_{0}$,
along with the suitable matching conditions given in Eqs.~(\ref{eq:c1c2}),
can be used to construct the power spectrum for long-wavelength modes
$K\exp\left(-T_{c}\right)\lesssim0.4$ that exit the horizon well
before transition. By construction, the zero-mode solution $I_{0}$
incorporates all significant interactions arising from the mixing
of nontrivial heavier modes. To evaluate the isocurvature power spectrum
for long wavelength $K$ modes using Eq.~(\ref{spectrum-1}), we
solved the zero-mode system in Eq.~(\ref{eq:zero-mode-eqn}) and
obtained the values of parameters $\mathcal{N}$ and $\theta$ numerically.
For the fiducial cases characterized by $c_{+}=2.3533$ and $2.345$
with $F_{a}/H=20.2$ and $c_{-}=0.5$, we find
\begin{enumerate}
\item $c_{+}=2.3533$, $\alpha_{c}\approx1.434$ 
\begin{equation}
\left(\mathcal{N},\theta\right)\equiv\left(1.254\times10^{-4},-2.93035\right),\label{eq:num-1}
\end{equation}
\item $c_{+}=2.345$, $\alpha_{c}\approx0.765$: 
\begin{equation}
\left(\mathcal{N},\theta\right)\equiv\left(2.0549\times10^{-6},-2.92929\right).\label{eq:num-2}
\end{equation}
\end{enumerate}
Numerically, the shape of the power spectrum is not very sensitive
to the exact value of the parameter $\theta$ since it enters the
expression in Eq.~(\ref{spectrum-1}) as a phase shift of the sinusoid
function whose argument is logarithmic in $k$. To elucidate this
further, we compare the solid-red and dashed-black curves as shown
in Fig.~\ref{fig:comparison} which are plotted respectively using
the numerical value of $\theta$, and our approximation $\theta=-\omega T_{c}$.

For the massive fields where $\omega>0.75$ or $c_{+}>2.8$, one can
show that 
\begin{equation}
|c_{1}(k)|\gg|c_{2}(k)|
\end{equation}
and hence the $\theta$ parameter is insignificant for these cases
and the power spectrum $\propto k^{3}$ with negligible sinusoidal
oscillations. Thus, for the massive underdamped fields, the $c_{1}(k)$
contribution can be approximated as 
\begin{equation}
|c_{1}(k)|^{2}\approx\left(\frac{1}{a_{0}}\right)^{3}\frac{\pi}{H}\frac{2^{-1}}{|\Gamma(1+i\omega)|^{2}}e^{-\pi\omega}
\end{equation}
for $\omega\gtrsim0.75$. Using the approximation
\begin{equation}
\left|\frac{e^{-\pi\omega/2}}{\Gamma(1+i\omega)}\right|^{2}\approx\frac{1}{2\pi}\frac{e^{2+2\omega\tan^{-1}(\omega)-\pi\omega}}{\sqrt{1+\omega^{2}}},
\end{equation}
valid for $\omega>1$, we obtain the following expression for the
dimensionless isocurvature power spectrum 
\begin{equation}
\Delta_{s}^{2}(k)\approx\mathcal{N}^{2}\frac{\omega_{a}^{2}}{\theta_{+}^{2}}\frac{1}{2\pi^{2}}\left(\frac{k}{a_{0}H}\right)^{3}\frac{\sqrt{c_{+}c_{-}}(c_{+}^{2}+c_{-}^{2})}{(c_{+}+c_{-})^{3}F^{2}}\frac{e^{2+2\omega\tan^{-1}(\omega)-\pi\omega}}{\sqrt{1+\omega^{2}}}
\end{equation}
valid for $\omega\gtrsim1$. Using the approximation 
\begin{equation}
\frac{e^{2+2\omega\tan^{-1}(\omega)-\pi\omega}}{\sqrt{1+\omega^{2}}}\approx\frac{1}{\omega}
\end{equation}
for $\omega\gg1$ that is applicable for very massive underdamped
fields having the mass parameter $c_{+}\gg9/4$, the dimensionless
isocurvature power spectrum can be given as 
\begin{equation}
\Delta_{s}^{2}(k)\approx\mathcal{N}^{2}\frac{\omega_{a}^{2}}{\theta_{+}^{2}}\frac{1}{2\pi^{2}}\left(\frac{k}{a_{0}H}\right)^{3}\frac{\sqrt{c_{+}c_{-}}(c_{+}^{2}+c_{-}^{2})}{(c_{+}+c_{-})^{3}F^{2}}\frac{1}{\sqrt{c_{+}}}
\end{equation}
for $\omega\gg1$. Furthermore, if $c_{-}\ll c_{+}$, the above expression
reduces to a compact expression: 
\begin{equation}
\lim_{c_{+}\gg9/4,\,\,c_{-}\ll c_{+}}\frac{\Delta_{s}^{2}(k)\theta_{+}^{2}}{\omega_{a}^{2}}\approx\mathcal{N}^{2}\frac{1}{2\pi^{2}}\left(\frac{k}{a_{0}H}\right)^{3}\frac{\sqrt{c_{-}}}{c_{+}F^{2}}.
\end{equation}
Therefore, we conclude that the axion isocurvature power spectrum
is power-law suppressed ( $\Delta_{s}^{2}\propto\sqrt{c_{-}}c_{+}^{-1}$)
for massive background fields.

This power law suppression with $\sqrt{c_{-}}/c_{+}$ is interesting
for a couple of reasons. First, one might naively expect the correlation
function to exhibit exponential decay for large mass values (a form
of decoupling). However, in this case, the exponential dependence
cancels out in both the numerator and denominator due to the definition
of the isocurvature perturbations. Secondly, the power law is multiplying
a coefficient $\mathcal{N}$ which we will explain below is almost
stochastic whose distribution amplitude depends on $c_{+}$. Hence,
even though this wave function squared leading to a $\sqrt{c_{-}}/c_{+}$
suppression seems intuitive, it's important to note that the factor
$\sqrt{c_{-}}/c_{+}$ will be multiplying an effectively stochastic
variable, and thus the amplitude dependence on the mass parameters
in this ``decoupling'' parametric region is nontrivial.

\begin{figure}
\begin{centering}
\includegraphics[scale=0.6]{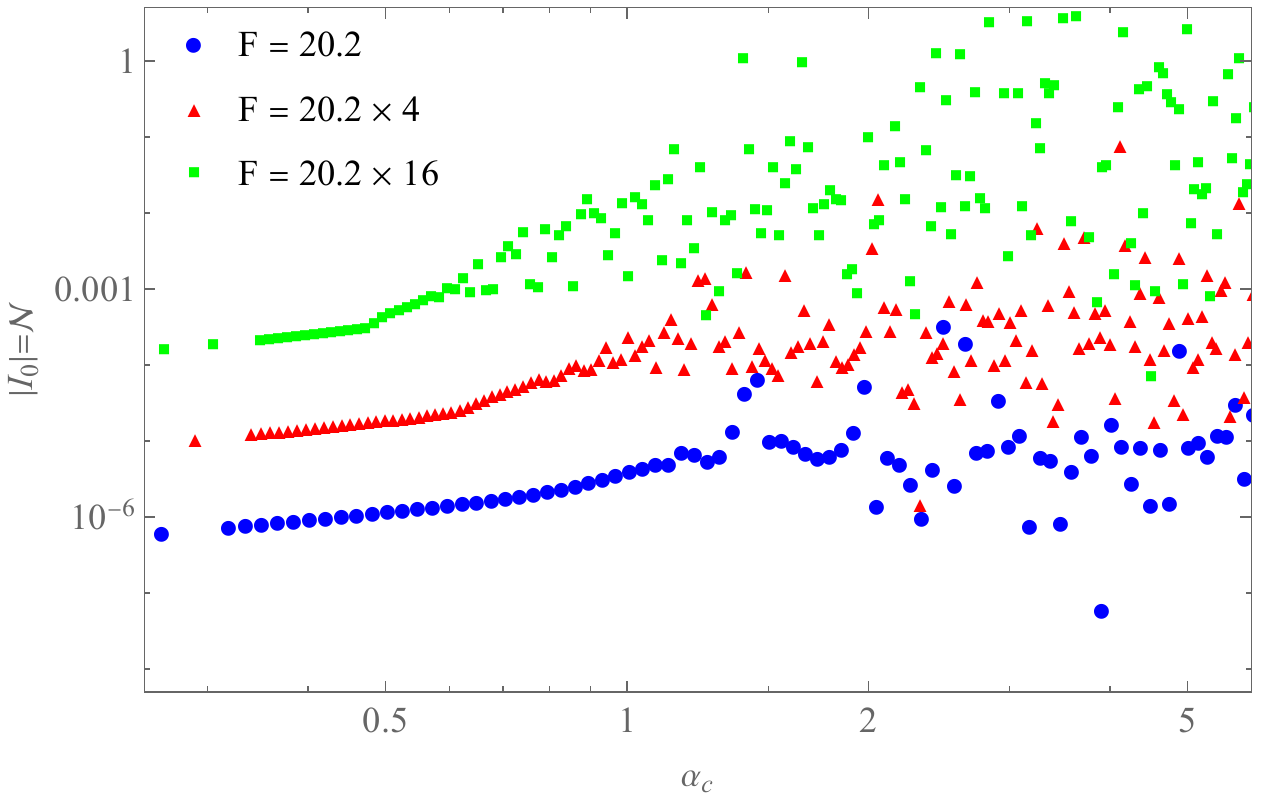} 
\par\end{centering}
\caption{\label{fig:zcpalpha}Plot showing the $\alpha_{c}$ dependence of
the zero-mode amplitude, $|I_{0}|=\mathcal{N}$, for three different
values of $F=F_{a}/H$ with the remaining Lagrangian parameters given
by the $P_{A}$ set. The data points are obtained by numerically solving
Eqs.~(\ref{eq:backgroundeom0}), (\ref{eq:backgroundeom}) and (\ref{eq:zero-mode-eqn}).}
\end{figure}

In Fig.~\ref{fig:zcpalpha}, we plot the final zero-mode amplitude,
$\lim_{T\rightarrow\infty}|I_{0}|\equiv\mathcal{N}$, with respect
to $\alpha_{c}$ for different values of $F=F_{a}/H$. The data is
obtained by numerically solving Eqs.~(\ref{eq:backgroundeom0}),
(\ref{eq:backgroundeom}) and (\ref{eq:zero-mode-eqn}) for a large
set of $c_{+}$ values using an Runge-Kutta solver (RK-solver) to
a high numerical precision. From the plots, we observe that the amplitude
of the zero-mode initially exhibits a continuous increase with respect
to $\alpha_{c}$. However, as we explore in Appendix \ref{sec:Chaotic-structure},
for $\alpha_{c}$ greater than a cutoff $\alpha_{{\rm Ch}}$ (where
${\rm Ch}$ stands for chaotic), the trajectory of the background
fields $\phi_{\pm}$ becomes chaotic. Since the effective mass of
the zero-mode is controlled by the dynamics of the background fields
through the mass-squared matrix $\tilde{M}^{2}(\phi_{\pm})$, the
zero-mode amplitude starts fluctuating chaotically.\footnote{We have verified numerically that the chaotic data points in Fig.~\ref{fig:zcpalpha}
exhibit a self-similar fractal structure.} Consequently, in this region, $\mathcal{N}$ can be seen as a stochastic
variable whose distribution depends on $c_{+}$ or $\alpha_{c}$.
In such cases, the axionic fluctuation amplitudes have only a distributional
prediction from the underlying Lagrangian parameters.

To understand the chaotic behavior, we note that the EoMs for the
background fields given in Eqs.~(\ref{eq:backgroundeom0}) and (\ref{eq:backgroundeom})
represent a set of two quartically coupled oscillators. In the absence
of dissipative and linear-force terms, the effective EoMs are described
by a classical Yang-Mills-like potential ($V=x^{2}y^{2}/2$). It is
well known in the literature \citep{Carnegie_1984,Dahlqvist:1990zz,Marcinek:1994}
that the classical Yang-Mills-like potential leads to chaotic motion,
except for a very small set of initial conditions, due to a nonlinear
mapping of the initial conditions through the nonlinear interactions.
In the presence of dissipation term such as Hubble friction, the background
fields must eventually settle to one of the local energy minima. Hence,
the presence of dissipative Hubble term tends to make chaotic motion
more orderly causing the system to converge to one of the equilibrium
states. If the interaction and kinetic energy are considerable during
the transition, we expect a transient chaotic phase until dissipation
brings the system back to an ordered state.

Consequently, there is a critical threshold for the value of kinetic
energy controlling parameter $\alpha_{c}$ below which chaos does
not set in. In Appendix.~\ref{sec:Chaotic-structure}, we show examples
of $\phi_{\pm}$ field trajectories for a chaotic case in Fig.~\ref{fig:field-trajectory}
and further derive the condition for the onset of the transient chaotic
motion. We show that to minimize transient chaos, the fast UV mode
of the background fields, which is induced by $\xi$, should be negligible
at the moment when the two fields cross each other.\footnote{During each crossing, the nonlinear force $f_{\pm}=\phi_{\pm}^{2}\phi_{\mp}$
acting on the two fields becomes comparable. Consequently, the presence
of substantial UV components during these crossings can trigger significant
trajectory shifts orthogonal to the flat direction ($\xi=0$), which
can lead to instabilities. When considering scenarios where the fields
cross for the first time at the transition, the UV modes generated
can induce transient chaos at the next crossing. However, if there
are multiple crossings prior to the transition, those occurring within
$O(1)$ e-folds before $T_{c}$ can also induce chaotic motion at
$T_{c}$.} Assuming there are no crossings before transition, we obtain the
condition 
\begin{equation}
\left\langle \xi^{2}\right\rangle _{T_{{\rm 1}}}<2r^{2}F^{4}
\end{equation}
to avoid transient chaos at the first crossing $T_{1}$ after transition
time $T_{c}$, where $r\approx0.2$ is an $O(0.1)$ number. In case
of multiple crossings prior to the transition, the above condition
must be applied to each crossing (including transition) where the
UV modes are significant. Hence, in cases where the total energy during
the crossings is large $\gtrsim O(F^{4})$, we expect the fields to
behave chaotically. 
\begin{figure}
\begin{centering}
\includegraphics{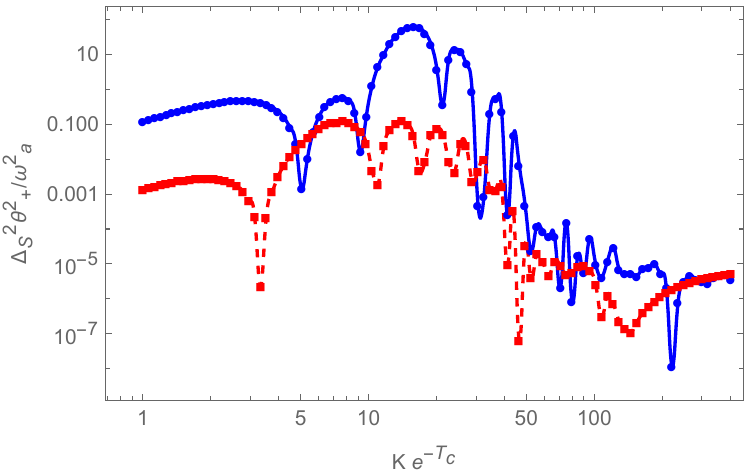} 
\par\end{centering}
\caption{\label{fig:chaos_isospec}Plot showing axionic isocurvature power
spectra for two fiducial chaotic cases with a $0.5\%$ deviation in
the values of $\Phi_{+}(0)$. Due to a large sensitivity of the field
trajectories to the initial conditions, the resulting power spectrum
amplitudes in the rising part of the spectrum and the initial few
bumps can differ by orders of magnitude as seen in this example. For
very short wavelength modes which exit the horizon when the field
trajectories have either settled or track similar points in phase
space, the power spectrum has similar amplitudes.}
\end{figure}

Due to the dependence of axion isocurvature fluctuation mode functions
$I_{\pm}$ on the background fields via the mass-squared matrix $\tilde{M}^{2}$,
significant variations in the trajectories of the background fields
can lead to large changes in the amplitude of the final isocurvature
modes. In Fig.~\ref{fig:chaos_isospec}, we present an example of
the isocurvature spectrum obtained for a fiducial chaotic case, where
we see a large deviation in the rising part of the spectrum and the
first few bumps after the cutoff due to a small sub-percent deviation
in the initial condition value of $\Phi_{+}(0)$. As highlighted in
Fig.~\ref{fig:chaos_isospec}, the exponential sensitivity of the
field trajectory on initial conditions can result in either a strong
amplification or significant attenuation of the mode amplitude. Large
amplification may occur when the background fields follow a trajectory
that results in additional dips in the effective potential (leading
to tachyonic masses),\footnote{We note from Fig.~\ref{fig:zcpalpha} that very large amplifications
can break perturbativity requirement of $\delta_{S}\ll1$ which is
required for neglecting backreaction on the homogeneous components.
Such cases that violate linearization assumptions are not covered
in this work.} while attenuation can be caused by mode mixing corrections (large
heavy mixing) or a slow roll of the background fields along a flat
direction, leading to an exponential decay of the mode amplitude until
the fields stabilize. We refer the interested reader to \pA ~for
further discussion on mode attenuation through an $m_{{\rm B}}^{2}$
parameter.

Another semi-quantitative way to see that the transient chaos of the
background fields can lead to a chaotic isocurvature amplitude comes
from the duality discussed earlier near Eq.~(\ref{eq:dualitymap}).
We know that there exists a time independent $\Xi_{1,2}=\Xi_{1,2}(c_{+},c_{-},F,\phi_{+}(0),\epsilon_{0})$
such that
\begin{equation}
\Xi_{1}I_{0}+\Xi_{1}^{*}I_{0}^{*}+\Xi_{2}I_{02}+\Xi_{2}^{*}I_{02}^{*}=\frac{I_{0}^{(\mathrm{dual})}}{f_{0}}=(\phi_{+},-\phi_{-})\label{eq:identified}
\end{equation}
where the right hand side is real and $I_{02}$ are zero modes independent
of $I_{0}$ defined earlier that spans the solution space. When the
right hand is a solution with different boundary conditions (e.g.
for example change $\phi_{+}(0)$ without changing $c_{+}$), then
the right hand side picks out a very different phase space trajectory
because of the chaotic nature (as evidenced by Fig.~\ref{fig:chaos_isospec}).
Let's call this solution $\phi_{\pm}^{(2)}$:
\begin{equation}
\Xi_{1}^{(2)}I_{0}^{(2)}+\Xi_{1}^{(2)*}I_{0}^{(2)*}+\Xi_{2}^{(2)}I_{02}^{(2)}+\Xi_{2}^{(2)*}I_{02}^{(2)*}=(\phi_{+}^{(2)},-\phi_{-}^{(2)})
\end{equation}
where $\{\Xi_{n}^{(2)},I_{0}^{(2)},I_{02}^{(2)}\}$ constitute a new
set analogous to Eq.~(\ref{eq:identified}). Because $I_{0}\not\propto I_{0}^{(\mathrm{dual})}$
satisfies a different differential equation than $\phi_{\pm}$ and
different boundary conditions, the new $\phi_{\pm}^{(2)}$ solution
requires $\Xi_{n}^{(2)}$ that is drastically different than the original
$\Xi_{n}$ (exponentially sensitive to the change in the initial conditions
$\phi_{\pm}(0)$) to \emph{functionally} match the different right
hand side phase space trajectory.\footnote{Note that $I_{0}^{(2)}$ and $I_{02}^{(2)}$ can be very different
from $I_{0}$ and $I_{02}$ because the mass matrix in the differential
equation that governs these basis modes have $\phi_{\pm}$ dependences
that have chaotically different phase space trajectories.}

Now, let's see how this drastic change in $\Xi_{n}$ changes the coefficient
$\mathcal{N}$ in Eq.~(\ref{eq:z-factorization}) which characterize
the amplitude at $T_{\infty}$ as 
\begin{equation}
I_{0}(T_{\infty})\approx\mathcal{N}e^{i\theta}\psi_{0}\label{eq:match}
\end{equation}
after the background fields have settled to their minima. Note that
Eq.~(\ref{eq:match}) only applies to situations in which the zero-mode
matching of the mode function at the horizon exit can be identified
with a single mode $I_{0}$ without mixing $I_{02}$. In such situations,
substituting Eq.~(\ref{eq:match}) into Eq.~(\ref{eq:identified})
evaluated at $T_{\infty}$, we see
\begin{equation}
\mathcal{N}\left(\Xi_{1}e^{i\theta}+\Xi_{1}^{*}e^{-i\theta}\right)\psi_{0}+\Xi_{2}I_{02}+\Xi_{2}^{*}I_{02}^{*}=(\phi_{+}(T_{\infty}),-\phi_{-}(T_{\infty}))\label{eq:const}
\end{equation}
where at $T_{\infty}$, the $\phi_{\pm}$ goes to an attractor (i.e.~the
global minimum) despite the different initial conditions. However,
as we have said above, since with different $\phi_{\pm}$ initial
conditions $\Xi_{n}$ now has changed drastically say to $\Xi_{n}^{(2)}$,
the other terms change ($\mathcal{N}\rightarrow\mathcal{N}^{(2)}$,
$\theta\rightarrow\theta^{(2)}$) drastically to match the attractor
of the right hand side: i.e.
\begin{align}
 & \mathcal{N}^{(2)}\left(\Xi_{1}^{(2)}e^{i\theta^{(2)}}+\Xi_{1}^{(2)*}e^{-i\theta^{(2)}}\right)+\left(\Xi_{2}^{(2)}I_{02}^{(2)}+\Xi_{2}^{(2)*}I_{02}^{(2)*}\right)\cdot\psi_{0}\nonumber \\
 & =\mathcal{N}\left(\Xi_{1}e^{i\theta}+\Xi_{1}^{*}e^{-i\theta}\right)+\left(\Xi_{2}I_{02}+\Xi_{2}^{*}I_{02}^{*}\right)\cdot\psi_{0}\label{eq:variation}
\end{align}
where $\mathcal{N}\rightarrow\mathcal{N}^{(2)}$ change is chaotic
with different initial conditions for $\phi_{\pm}$. Although one
may naively not expect the parameter changes (such as $c_{+}$ changes)
have much to do with the initial condition changes that translate
to chaotically different phase space trajectories of $(\phi_{+},\phi_{-})$,
small $c_{+}$ changes do map to different initial phase space points
for $(\phi_{+},\phi_{-})$ through Eq.~(\ref{eq:approxsol}) when
the nonlinear forces (governed by $\xi\phi_{-}$ of Eq.~(\ref{eq:force at crossing}))
start to dominate say at time $T_{1}$ (with the linear force region
initial conditions fixed).\footnote{Small changes in $c_{+}$, unlike the direct changes in the initial
conditions of $\phi_{+}$, do change the boundary conditions of $I_{0}$
through Eqs.~(\ref{eq:initi0}) and (\ref{eq:zeromodeBC}), but the
way the effective boundary conditions change in entering the chaotic
dynamical time period is not matched, and therefore, the large change
in $\Xi_{n}$ is still expected.}

On the other hand, different $K$ values do not control $(\phi_{+},\phi_{-})$
and will therefore not change their phase space trajectory. That is
why the $\Delta_{s}^{2}$ has a $K$ spectral dependence that is still
smooth (unlike the chaotic jumps in the overall normalization $\mathcal{N}$
as a function of $c_{+}$) in the $K$ region for which $\mathcal{T}_{K}<T_{c}$.
When $\mathcal{T}_{K}$ is in the intermediate time region defined
to be when the $\phi_{\pm}$ undergoes oscillations, the isocurvature
amplitude oscillations as a function of $K$ can be complicated since
the matching condition analogous to Eq.~(\ref{eq:matching}) but
with a strongly oscillating function of time on the left hand side
produces an oscillatory $c_{n}(K)$. That translates to different
matching time $\mathcal{T}_{K}$ sampling the oscillations as a function
of time and not strongly divergent phase space trajectories (chaos)
due to effectively different initial conditions. In Secs.~\ref{sec:Generic-mass-model}
and \ref{sec:sine-model-ftting} we will present general fitting formulas
for $\Delta_{s}^{2}(K)$ that can be used for practical fitting situations
which would be sensitive to the first three bumps. In other words,
the chaotic behavior of the map between the Lagrangian parameters
and the amplitude does not present an obstacle for fitting $\Delta_{s}^{2}$
characteristic of this class of models as a function of $K$.

Of course, if the $K$ value is sufficiently large that $\mathcal{T}_{K}\gg T_{c}$,
Eq.~(\ref{eq:matching}) is irrelevant and the amplitude is fixed
by the usual massless mode nearly-exact solution in quasi-dS space:
\begin{align}
I & \approx-\frac{iH}{2k^{3/2}}\psi_{0}=\mathcal{N}\left(c_{1}(K)e^{i\theta}+c_{2}(K)e^{-i\theta}\right)\psi_{0}\label{eq:latematch}
\end{align}
which enters directly into Eq.~(\ref{spectrum-1}). Even if $(\mathcal{N},\theta)$
changes drastically say with $c_{+}$ variation (discussed in Eq.~(\ref{eq:variation})),
$c_{1,2}(K)$ also changes drastically to compensate for the $(\mathcal{N},\theta)$
variation. There is no such compensating matching condition for $c_{n}(K)$
for smaller $K$s, as can be seen explicitly in Eq.~(\ref{eq:c1c2})
as they are fixed by a different function than the final nearly-exact
solution.

In the next section, we provide analytic fitting functions for the
zero-mode amplitude $\mathcal{N}$ as a function of $\alpha_{c},$
$F$, and $c_{-}$ for the non-chaotic cases where $\alpha_{c}<\alpha_{{\rm Ch}}$.
For $c_{+}$ values where the background fields are chaotic, a closed-form
analytic expression for the zero-mode function may not be feasible.
For these cases, we propose a numerically motivated stochastic amplitude
model with a log-normal distribution. Using the estimated value of
$\mathcal{N}$ from these fits, one can approximate the amplitude
of the axion isocurvature power spectrum from Eq.~(\ref{spectrum-1})
for long-wavelength modes that exit the horizon prior to the transition
of the background fields.

\section{\label{sec:Fitting-functions}Fitting functions for zero-mode amplitude}

In Eq.~(\ref{eq:zeromodeBC}), we provided the initial conditions
for the zero-mode, $I_{0}$, at time $T_{0}\ll T_{c}$. Furthermore,
in Sec.~(\ref{subsec:Zero-mode}), we noted that when $\phi_{+}\gg F$,
the lightest mass eigenvalue is $m_{1}\approx c_{+}$, and during
the time $T<T_{c}$ when the expansion parameter $\lambda=F^{2}/\phi_{+}^{2}<1$,
the zero-mode solution can be approximated as 
\begin{equation}
I_{0}(T)\approx e^{-\left(3/2+i\omega\right)T}\left[\begin{array}{c}
1\\
-\lambda
\end{array}\right].
\end{equation}
The amplitude of $I_{0}$ during this time can be expressed as 
\begin{equation}
\left|I_{0}(T)\right|\approx e^{-\frac{3}{2}T}.
\end{equation}
As the background fields approach transition at $T_{c}$, we expect
non-adiabatic behavior and heavy mode mixing due to a large interaction
energy $\sim O(\xi^{2})\sim O(F^{4})\gg O(H^{2}F^{2})$. Hence, we
define an amplification factor $\mathcal{Z}$ as 
\begin{equation}
\mathcal{N}=e^{-\frac{3}{2}T_{c}}\mathcal{Z}\label{eq:amplificationfactorZ}
\end{equation}
that captures the nontrivial amplification or attenuation of the zero-mode
after the transition. In the succeeding subsections, we will provide
an analytic fitting function and a distribution function for $\mathcal{Z}$
in the non-chaotic and chaotic cases respectively.

\subsection{\label{subsec:Non-chaotic}Non-chaotic: $\alpha_{c}<\alpha_{{\rm Ch}}$}

For the non-chaotic cases where $\alpha_{c}<\alpha_{{\rm Ch}}$, the
zero-mode amplitude increases monotonically with $\alpha_{c}$. From
Fig.~\ref{fig:zcpalpha} we notice that the slope of the zero-mode
amplitude in this parametric region changes at an intermediate value
of $\alpha_{c}\approx\alpha_{2}$. As we have verified numerically
and briefly explained in \pA, the change of slope corresponds to
situations where large kinetic energy causes the background fields
to cross at least once after the transition. Hence, for $c_{+}$ values
where $\alpha_{c}<\alpha_{2}$, the background fields do not cross
each other again after transition in the limit $c_{-}<O(1)$.\footnote{A value of $c_{-}>O(9/4)$ can lead to the crossing of the background
fields after transition. However, we limit the definition of $\alpha_{2}$
solely based on $\alpha_{c}$ which is fairly independent of $c_{-}$
for values of $c_{-}\ll F^{2}$.} At each crossing of the background fields, the system has a brief
period of tachyonic mass dip in the effective mass of the lightest
eigenmode. As a result, the mode function undergoes a brief period
of amplification whenever the two background fields cross each other
with a large relative kinetic energy. See Sec.~6 in \pA ~for more
details. Hence, for all $c_{+}$ values where the background fields
cross again after transition, we expect additional amplification of
the mode function. This explains the change in the slope of the zero-mode
at $\alpha_{2}$.

\begin{figure}
\begin{centering}
\includegraphics[scale=0.6]{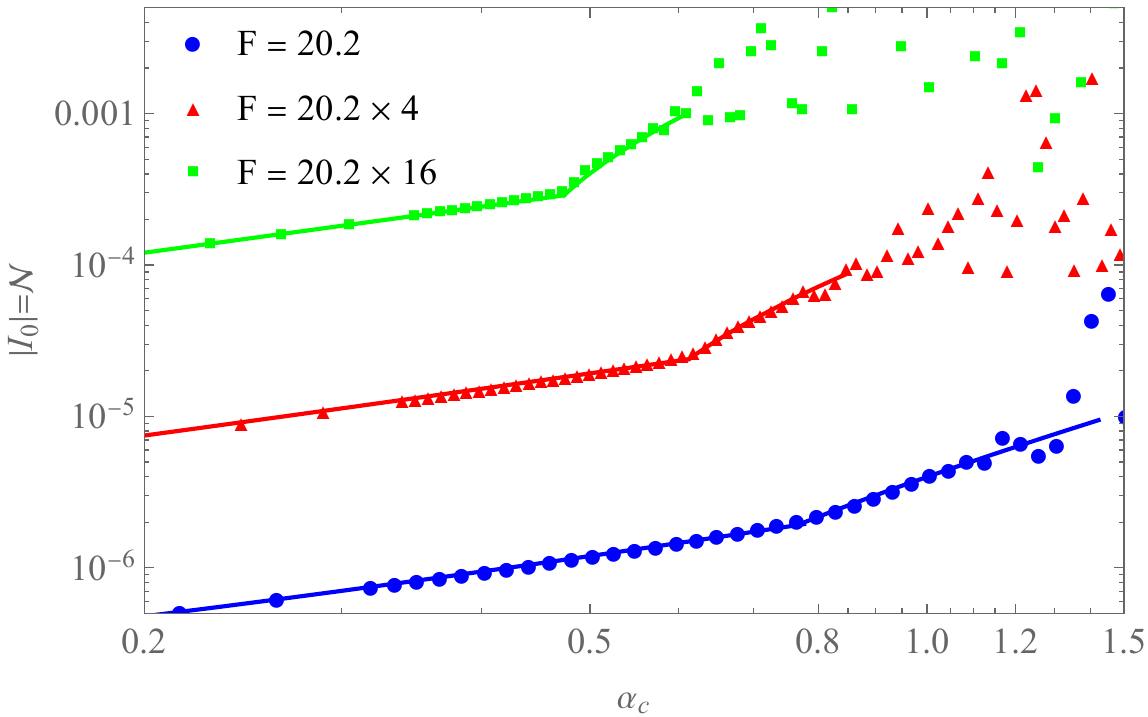} 
\par\end{centering}
\caption{\label{fig:fitting_fn_comparison}In this figure, we compare our fitting
results with the numerical data for the zero-mode amplitude $\mathcal{N}$
as a function of $\alpha_{c}$, for three different values of $F=F_{a}/H$.
The numerical data, plotted in solid markers, is the same as in Fig.~\ref{fig:zcpalpha},
and we construct our fitting curve using Eq.~(\ref{eq:amplificationfactorZ})
where we take the fitting function for $\mathcal{Z}$ from Eq.~(\ref{eq:fit-Z-non-chaotic})
and the analytical estimation of $T_{c}$ from Eq.~(\ref{eq:Tc-emperical}).}
\end{figure}

Using a large set of numerical evaluations, we provide the following
fitting functions for the zero-mode amplitude for non-chaotic cases
for Lagrangian parameters $F$ and $c_{-}$ in the range, $F\equiv F_{a}/H\in[10,400]$
and $c_{-}\in[0.1,9/4]$: 
\begin{align}
\mathcal{Z} & \approx f_{-}(c_{-})\times\begin{cases}
c_{0}+c_{1}\sqrt{\alpha_{c}} & \alpha_{{\rm 1}}<\alpha_{c}<\alpha_{2}\\
8.0+c_{2}\left(\alpha_{c}-\alpha_{{\rm Ch}}\right) & \alpha_{2}<\alpha_{c}<\alpha_{{\rm Ch}}
\end{cases}\label{eq:fit-Z-non-chaotic}
\end{align}
where the prefactor $f_{-}$ gives an approximate dependence of the
zero-mode amplitude on $c_{-}$, 
\begin{equation}
f_{-}(c_{-})\approx0.26c_{-}^{0.5}+0.66c_{-}^{-0.29}\qquad0.1<c_{-}<9/4.
\end{equation}
For values of $c_{-}$ that lie outside the range $[0.1,9/4]$, we
observe a multi-functional dependence of the mode amplitude on $F$
and $c_{\pm}$ as summarized in Tab.~\ref{tab:Table-c+-}. We briefly
discuss these cases in Appendix \ref{sec:cminus}. The fitting parameters
are approximated using the following expressions: 
\begin{align}
c_{0} & =2.99703-2.419\left(\frac{20.2}{F}\right)+2.5927\left(\frac{20.2}{F}\right)^{2}-1.2713\left(\frac{20.2}{F}\right)^{3},\\
c_{1} & =0.07975+3.193\left(\frac{20.2}{F}\right)-4.2829\left(\frac{20.2}{F}\right)^{2}+2.2462\left(\frac{20.2}{F}\right)^{3},\\
c_{2} & =\frac{9.127}{0.181+\left(20.2/F\right)},
\end{align}
and 
\begin{equation}
\alpha_{{\rm Ch}}=1.0185\left(\frac{20.2}{F}\right)^{0.194}+0.4065\left(\frac{20.2}{F}\right)^{1.26}.\label{eq:alpha-Ch}
\end{equation}
We set $\alpha_{2}$ as approximately,
\begin{equation}
\alpha_{2}=\alpha_{{\rm Ch}}-\frac{5.0}{c_{2}}\label{eq:alpha-1}
\end{equation}
and
\begin{equation}
\alpha_{1}\approx10^{-3}.
\end{equation}
In Eq.~(\ref{eq:fit-Z-non-chaotic}), the specific form of $\mathcal{Z}$
for $\alpha_{c}<\alpha_{2}$ is motivated from Eq\@.~(239) of \pA
~where the authors derived analytic expressions for the mode amplitude
in this parametric region. The lower cutoff is set at $\alpha_{1}\ll\alpha_{{\rm L}}\approx0.2$
where $\alpha_{c}$ greater (lesser) than $\alpha_{{\rm L}}$ corresponds
to resonant (non-resonant) underdamped fields. In Fig.~\ref{fig:fitting_fn_comparison},
we compare our fitting function with the numerical results where we
construct our fitting curves (solid lines) using Eq.~(\ref{eq:amplificationfactorZ})
by taking the fitting function for $\mathcal{Z}$ from Eq.~(\ref{eq:fit-Z-non-chaotic})
and the analytical estimation of $T_{c}$ as given in Eq.~(\ref{eq:Tc-emperical}).
Over the range $\alpha_{{\rm L}}<\alpha_{c}<\alpha_{{\rm Ch}}$, the
amplitude of zero-mode varies by $\sim O(10)$ which implies an $O(100)$
variation in the amplitude of the isocurvature power spectrum.

\begin{figure}
\centering{}\includegraphics[scale=0.45]{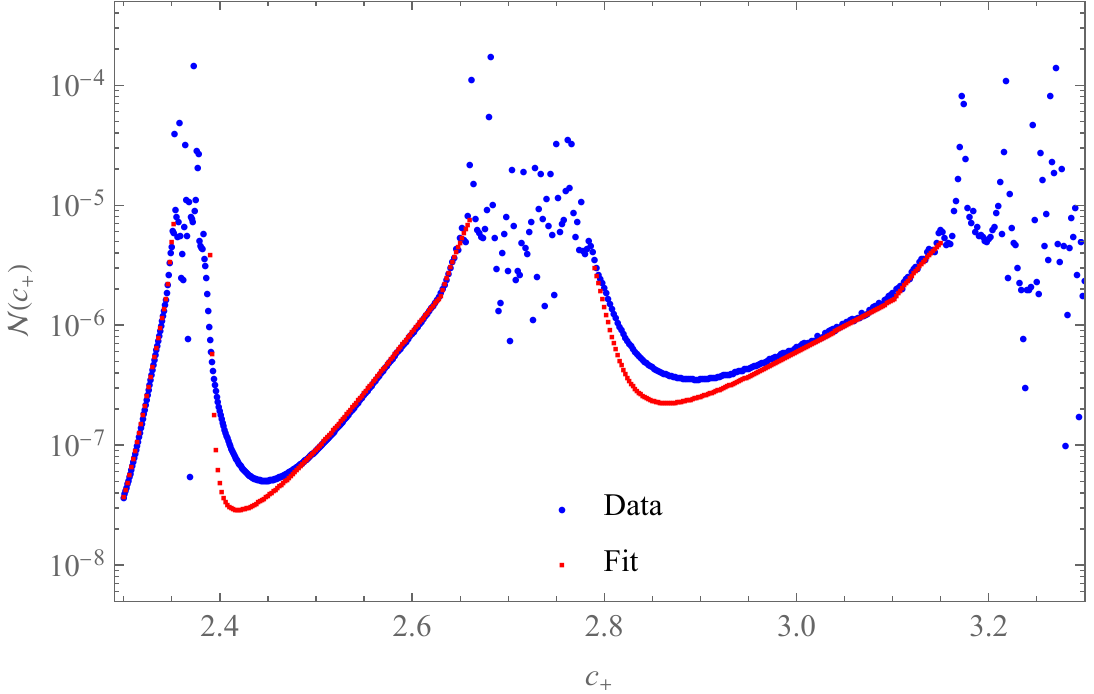}\includegraphics[scale=0.52]{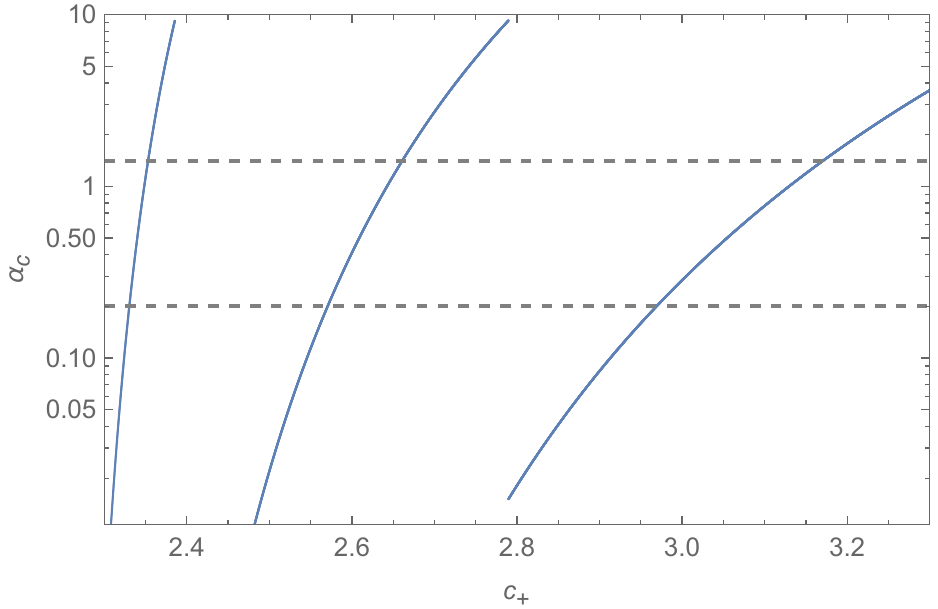}\caption{\label{fig:z_with3branch}In the plot on the left, we compare the
zero-mode amplitude $\mathcal{N}$ obtained from the fitting formula
in Eq.~(\ref{eq:fit-Z-non-chaotic}) (red square marker) with the
numerical data (blue circular marker) for the first three branches
highlighting the distinction between the non-chaotic and chaotic classes
of fields based on the value of $\alpha_{c}$ as $c_{+}$ is varied
from one branch to another. In each branch, we first estimate $\alpha_{c}$
and $T_{c}$ for the corresponding values of $c_{+}$ from the expressions
given in Sec.~\ref{Sec:Estimate-Tzjc}, and use the fitting function
in Eq.~(\ref{eq:fit-Z-non-chaotic}) to obtain an approximation for
the zero-mode amplitude in the non-chaotic region. For values of $c_{+}$
where $\alpha_{c}$ lies outside the fitting range of Eq.~(\ref{eq:fit-Z-non-chaotic}),
the prediction is very sensitive to the values of $c_{+}$ such that
only an approximate band of prediction can be realistically given.
The mismatch between the fitting formula and the numerical results
near the bottom of the troughs in the left plot occurs for non-resonant
underdamped cases due to the smallness of the $\alpha_{c}$ value
such that the separation between $T_{z,j_{c}}$ and $T_{c}$ is large
($\sim O(1)$). On the right, we plot $\alpha_{c}$ as a function
of $c_{+}$ for the first 3 branches shown on the left. The dashed
lines in this plot represents the lower and upper cutoffs at $\alpha_{{\rm L}}$
and $\alpha_{{\rm Ch}}$ respectively.}
\end{figure}

In Fig.~\ref{fig:zcpalpha}, the $c_{+}$ values are restricted within
the first branch ($j_{c}=1$). As we increase $c_{+}$, the parameter
$\alpha_{c}$ makes a jump to the next branch $(j_{c}=2)$ and successive
branches (See Fig.~\ref{fig:alphavscp}). In each branch, $c_{+}$
values that correspond to $\alpha\lesssim\alpha_{{\rm Ch}}$ belong
to non-chaotic class of fields. Fig.~\ref{fig:z_with3branch} gives
a plot of $\mathcal{N}$ for the first three branches ($j_{c}\leq3$)
for $F=20.2$ and shows the transition of mode amplitude within each
branch from a smooth predictive behavior to random fluctuations due
to chaotic background field dynamics. 

\subsection{\label{subsec:Noise-model}Noise model: $\alpha_{c}>\alpha_{{\rm Ch}}$}

From Fig.~\ref{fig:zcpalpha} we infer that a closed form prediction
of the final mode amplitude as a function of $\alpha_{c}$ is not
feasible, unlike the fitting functions presented in Sec.~\ref{subsec:Non-chaotic}.
For the fiducial case with $F=20.2$, the estimated cutoff $\alpha_{{\rm Ch}}\approx1.4$
as given by Eq.~(\ref{eq:alpha-Ch}). In Fig.~\ref{fig:alphavscp},
we have plotted the expected value of $\alpha_{c}$ against $\omega$
for $F=20.2$. Based on that figure, we can conclude that for $F=20.2$,
values of $c_{+}\gtrsim8.5$ will transition with $\alpha_{c}\gtrsim\alpha_{{\rm Ch}}$.
In the limit, $F\gtrsim O(100)$, we estimate that massive underdamped
fields with $c_{+}>O(5)$ do not have a stable predictive solution
due to the chaotic behavior of the background fields. This represents
a loss of predictability of the axionic model for massive fields.
On the other hand, for any $\alpha_{c}>\alpha_{{\rm Ch}}$, there
generically exists the phenomenological possibility of large amplitude
enhancement without fine tuning of $\alpha_{c}$.

\begin{figure}
\begin{centering}
\includegraphics[scale=0.5]{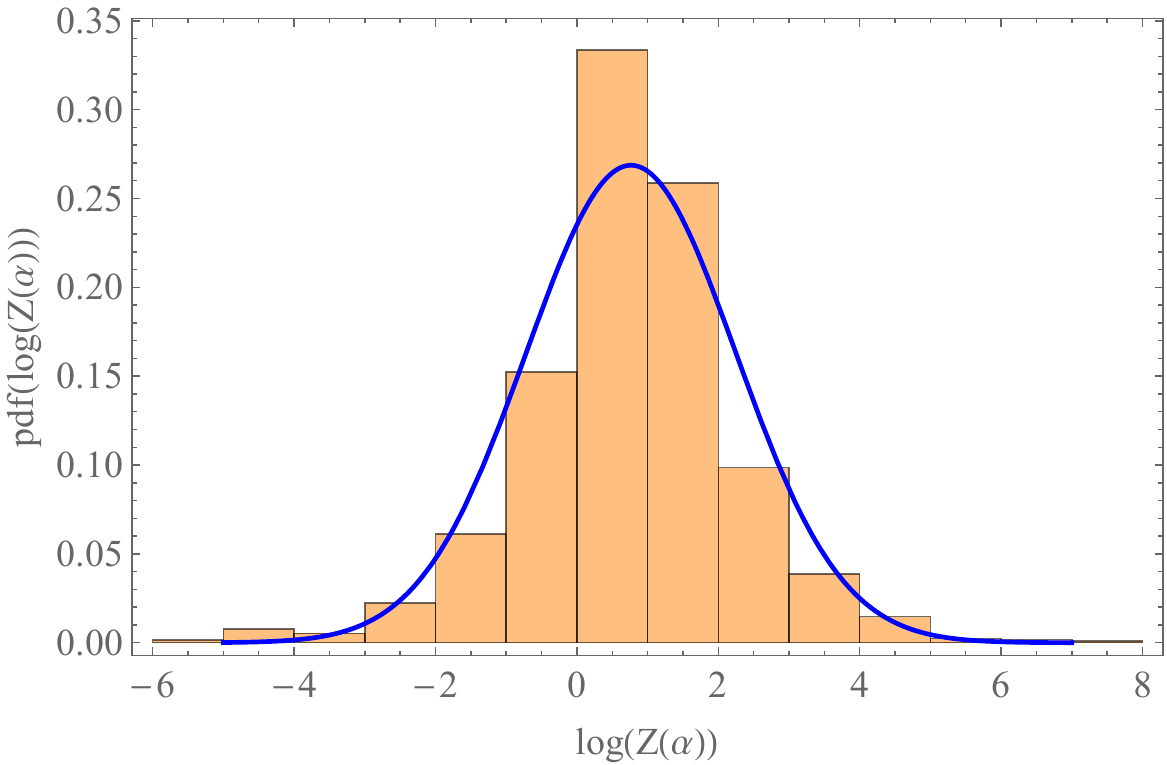} 
\par\end{centering}
\caption{\label{fig:histograms}Histogram of the ${\rm \log}\left(\mathcal{Z}\right)$
over the noisy region for the fiducial choice of $F=20.2$ compared
with an approximate normal distribution (solid blue curve).}
\end{figure}

To complete our analysis of the massive axionic underdamped fields,
we show in Fig.~\ref{fig:histograms} that the histogram of ${\rm \log}\left(\mathcal{Z}\right)$,
for cases with $\alpha_{c}>\alpha_{{\rm Ch}}$, resembles a normal
distribution. Therefore, we propose that for $\alpha_{c}>\alpha_{\mathrm{Ch}}$,
the numerical data for $\mathcal{Z}$ can be approximated by a log-normal
distribution given by the expression: 
\begin{equation}
{\rm pdf}\left({\rm log}\left(\mathcal{Z}\right)\right)=N\left(\mu,\sigma\right)\label{eq:noisemodel}
\end{equation}
where the normal distribution $N\left(\mu,\sigma\right)$ has mean
$\mu$ and variance $\sigma^{2}$. As seen in Fig.~\ref{fig:zcpalpha},
the numerical data points in the chaotic region appear to be centered
around a mean value. Using the log-normal distribution function, we
estimate this mean amplification as 
\begin{equation}
\left\langle \mathcal{Z}\right\rangle \approx e^{\mu+\frac{\sigma^{2}}{2}}\label{eq:avgZ}
\end{equation}
which would approximately correspond to the mode amplification from
an average trajectory following the transition for the chaotic background
fields. By fitting the numerical data, we find that the mean $\mu$
and variance $\sigma^{2}$ have the following approximate $F$ dependencies:
\begin{align}
\mu(F) & \approx1.01+0.25\left(\frac{F}{20}\right)\\
\sigma(F) & \approx1.25+0.12\left(\frac{F}{20}\right).\label{eq:noisemodelFdep}
\end{align}

Through the distribution function presented in this subsection and
using the expression for the isocurvature power spectra for modes
$k<k_{c}$ as outlined in Eq.~(\ref{spectrum-1}), one can, at best,
offer an approximate estimation for the average power within the blue
region of the axion isocurvature spectrum. This applies specifically
to massive underdamped fields that exhibit chaotic behavior owing
to significant nonadiabaticity during their transition. This analysis
is complementary to the fitting functions provided in subSec.~\ref{subsec:Non-chaotic}
for the non-chaotic dynamical system of background fields. However,
these fits are limited to the blue-tilted region of the isocurvature
power spectrum, and restricted to modes satisfying Eq.~(\ref{eq:k-restriction}).

We do not have a simple, smooth prediction for the small-scale modes
that lie within the oscillating part of the spectrum. This region
of the power spectrum has a complex dependence on the underlying dynamical
eigen-masses of the system of fields and as we shall see in the next
section consists of a series of bumps (oscillations) with varying
amplitudes. It was shown in \pA ~that even the simplest cases require
various analytical tools and the situation is compounded by the chaotic
nature of the background fields implying that any fairly approximate
estimation of the height of these bumps would require a complete numerical
analysis of the mode equations as in Eq.~(\ref{eq:modeeq}).

In \pA, the authors presented an empirical piecewise mass-model motivated
from a large set of analytical calculations including UV integration,
nonlinear field redefinitions, and other techniques to approximate
the shape and amplitude of the isocurvature power spectrum in the
oscillating region. They found that for most non-chaotic cases, a
generic non-minimal model with two negative square wells can be used
to approximately map the power spectrum with variable bump heights
in the small-scale region. In the next section, we will review the
mass-model of \pA ~and apply it to a few sample cases.

\section{\label{sec:Generic-mass-model}Empirical mass model}

The mass-model presented in \pA ~can be generalized into a Lagrangian
independent numerical model. A Lagrangian-free model benefits from
the variability and freedom of choice for the values of different
parameters that can give rise to unique spectral shapes and amplitudes
regardless of an underlying known or unknown action. The model will
also allow us to fit isocurvature power spectra for cases that were
beyond the scope of analytical methods presented in \pA. This is
likely to be useful for fitting data and discovering isocurvature
signatures. We begin by first introducing the mass-model and then
use it to fit numerically obtained isocurvature power spectrum for
the QCD axion toy-model presented in Sec.~\ref{subsec:Review}.

Consider the following second-order linear differential equation for
a scalar perturbation $y\left(K,T\right)$ in an expanding FRW spacetime:
\begin{equation}
\ddot{y}\left(K,T\right)+3\dot{y}\left(K,T\right)+\left(K^{2}e^{-2T}+m^{2}(T)\right)y\left(K,T\right)=0
\end{equation}
where $m^{2}(T)$ encapsulates information regarding the form of the
potential (mass and interactions) of the effective Lagrangian. Next,
we define an effective mode-dependent mass-squared term 
\begin{equation}
m_{{\rm eff}}^{2}\left(K,T\right)\equiv K^{2}e^{-2T}+m^{2}(T),
\end{equation}
such that the differential equation takes the form 
\begin{equation}
\ddot{y}\left(K,T\right)+3\dot{y}\left(K,T\right)+m_{{\rm eff}}^{2}\left(K,T\right)y\left(K,T\right)=0\label{eq:yeffeqn}
\end{equation}
and has the general solution 
\begin{equation}
y\left(K,T\right)=c_{1}\psi_{1}\left(K,T\right)+c_{2}\psi_{2}\left(K,T\right).\label{ysol}
\end{equation}
To determine $\psi_{1,2}$ and solve the system of differential equations,
we model $m_{{\rm eff}}^{2}$ through a piecewise discontinuous mass-model
using the set of parameters 
\begin{equation}
P_{{\rm set}}=\left\{ V_{0},V_{1},V_{2},T_{1},T_{2},C_{m}\right\} +\left\{ V_{i},T_{i},\Delta_{i}\right\} _{3\leq i\leq N}.\label{Vset}
\end{equation}
Thus, in each $K$-dependent time region $R_{(j)}$, the mass-model
$m_{{\rm eff}(j)}^{2}$ takes the form 
\begin{equation}
m_{{\rm eff}(j)}^{2}\left(K,T\right)\equiv A_{(j)}\left(K\right)e^{-n_{(j)}T}+c_{(j)}\quad\quad T\in R_{(j)}(K)\label{meff}
\end{equation}
where $A_{(j)}(K)$ and $c_{(j)}$ are functions of the parameters
in Eq.~(\ref{Vset}) and $n_{(j)}$ will take on a value from the
set $\{3,5/2,2\}$ depending on the region $R_{(j)}$.The mass-model
is based on the assumption that it is sufficient to follow the smooth
(IR) behavior of the effective mass. Consequently, it incorporates
a positive exponentially decaying term (derived from integrating out
fast $O(F)$ UV oscillations of the lightest eigen-mass with a Hubble-driven
decaying envelope) and a negative tachyonic mass-dip (associated with
transient non-adiabatic effects) within each sub-region. In terms
of the effective mass-squared $m_{{\rm eff}(j)}^{2}\left(K,T\right)$
in Eq.~(\ref{meff}), the linearly independent solutions to the equation
of motion presented in Eq.~(\ref{eq:yeffeqn}) in each region take
the form 
\begin{equation}
\psi_{1,2(j)}\left(K,T\right)=e^{-\frac{3}{2}T}J_{\pm\frac{\sqrt{9-4c_{(j)}}}{n_{(j)}}}\left(\frac{2}{n_{(j)}}A_{(j)}\left(K\right)e^{-\frac{n_{(j)}}{2}T}\right)\label{Psi12}
\end{equation}
where $J_{\nu}$ is the cylindrical Bessel function. Thus, starting
from $y\left(K,T_{0}\right)$, the final mode amplitude $y\left(K,T_{\infty}\right)$
is obtained by evaluating appropriate scattering matrices $S\left(K,R_{(j)}\right)$
in each piecewise region $R_{(j)}$ using the linearly independent
functions given in Eq.~(\ref{Psi12}) and the derived parameters
$A_{(j)}(K)$, $n_{(j)}$, and $c_{(j)}$ that specify the $m_{{\rm eff}(j)}^{2}$
in each region.

\subsection{Mass-Model Implementation}

Using the parameter set $P_{{\rm set}}$ presented in Eq.~(\ref{Vset}),
we will now define the mass-model to fit the blue axion isocurvature
power spectrum for the axion toy model presented in \pA. The motivation
for the mass-model and its parameterization in terms of the Lagrangian
variables were covered previously in \pA. We define the model $m_{{\rm eff}(j)}^{2}$
as
\begin{equation}
m_{{\rm eff}(j)}^{2}-K^{2}e^{-2T}=\begin{cases}
V_{0} & 0\leq T\leq T_{1}\\
-V_{1} & T_{1}\leq T\leq T_{2}\\
V_{2}e^{-3\left(T-T_{2}\right)}+\sum_{i=3}^{N}\left(-V_{i}\right){\rm sqw}\left(T,T_{i},\Delta_{i}\right) & T_{2}\leq T<T_{\infty}
\end{cases}
\end{equation}
where
\begin{equation}
{\rm sqw}\left(T,T_{i},\Delta_{i}\right)=\begin{cases}
1 & T_{i}\leq T\leq T_{i}+\Delta_{i}\\
0 & {\rm otherwise}
\end{cases}
\end{equation}
and we set $T_{0}=0$.\footnote{For a more generic model that can be fitted to a larger class of perturbative
systems, we can replace the exponentially decaying potential with
$V_{2}e^{-d\left(T-T_{2}\right)}$ where $d\in R$.} Similarly we choose $T_{\infty}$ long after the background fields
have settled to their minima. The initial parameter $V_{0}$ models
the mass-squared term induced for the lighter mass eigenmode (that
tracks axion) as the $\phi_{+}$ field is rolling down along the flat
direction and $\phi_{-}\ll\phi_{+}$. Negative mass-squared terms
(dips) like $V_{1}$ and $V_{i}$ are due to the nonadiabatic effects
induced during the crossing of the two fields, where the nonadiabaticity
is controlled by the relative velocity of the fields as they cross.
The exponential term $V_{2}$ signifies a positive (stabilizing) mass-squared
term induced due to the high frequency $O(F)$ resonant oscillations
of the two fields along the steeper direction in the overall potential.

Except for the exponentially decaying $V_{2}$ term, all of the remaining
$V_{i}$ parameters act as $c_{(j)}$ within our model and they define
a constant mass-squared term where it is either stabilizing (positive)
or tachyonic (negative). For the region $\left[T_{2},T_{\infty}\right]$,
we can write the effective mass-squared term as 
\begin{equation}
m_{{\rm eff}(j)}^{2}(K,T)-c_{(j)}=K^{2}e^{-2T}+V_{2}e^{-3\left(T-T_{2}\right)}\quad T\in\left[T_{2},T_{\infty}\right]\label{eq:massinV2region}
\end{equation}
Thus, the effective mass-squared is a sum of exponentials with different
decay constants and amplitudes. However, by construction, we require
that the value of decay constant $n_{(j)}$ within each piecewise
region $R_{(j)}$ be constant. Therefore, to specify a single decay
constant $n_{(j)}$ and an amplitude $A_{(j)}(K)$, we perform our
evaluations by dividing the region $\left[T_{2},T_{\infty}\right]$
into two more regions $\left[T_{2},T_{K}\right]$ and $\left[T_{K},T_{\infty}\right]$
such that $m_{{\rm eff}(j)}^{2}\left(K,T\right)$ in Eq.~(\ref{eq:massinV2region})
is now given by 
\begin{equation}
m_{{\rm eff}(j)}^{2}\left(K,T\right)-c_{(j)}\equiv\begin{cases}
B_{1}\left(K\right)e^{-3T} & T_{2}\leq T\leq T_{K}\\
B_{2}\left(K\right)e^{-2T} & T_{K}\leq T<T_{\infty}
\end{cases}
\end{equation}
where $T_{K}$ and $B_{1,2}$ are $k$-dependent boundary and amplitudes,
respectively. These are not model parameters and are only required
for internal calculations. We define $T_{K}$ as the time when 
\begin{equation}
\left(K^{2}e^{-2T}-V_{2}e^{-3\left(T-T_{2}\right)}\right)_{T=T_{K}}=0,
\end{equation}
therefore 
\begin{equation}
T_{K}=T_{2}+\ln\left(\frac{V_{2}}{K^{2}e^{-2T_{2}}}\right).\label{eq:TK}
\end{equation}
With the above definitions, we now give the amplitudes $B_{1,2}(K)$
in each region as 
\begin{align}
B_{1}\left(K\right)=\frac{\int_{T_{2}}^{T_{K}}dT\,\left(K^{2}e^{-2T}+V_{2}e^{-3\left(T-T_{2}\right)}\right)}{\int_{T_{2}}^{T_{K}}dT\,e^{-3T}}=\left(V_{2}e^{3T_{2}}+\frac{3}{2}K^{2}\left(\frac{e^{T_{2}}+e^{T_{K}}}{1+2\cosh\left(T_{K}-T_{2}\right)}\right)\right)\label{eq:B1}
\end{align}
and 
\begin{align}
B_{2}\left(K\right)=\frac{\int_{T_{K}}^{T_{\infty}}dT\,\left(K^{2}e^{-2T}+V_{2}e^{-3\left(T-T_{2}\right)}\right)}{\int_{T_{K}}^{T_{\infty}}dT\,e^{-2T}}=\left(K^{2}+\frac{2}{3}Ve^{3T_{2}}\left(\frac{1+2\cosh\left(T_{K}-T_{\infty}\right)}{e^{T_{\infty}}+e^{T_{K}}}\right)\right).\label{eq:B2}
\end{align}

For long wavelengths, $K$ can be small enough such that $T_{K}$
is very large. Let us then consider a long wavelength mode, such that
an additional $-V_{3}$ dip lies within the first section $\left[T_{2},T_{K}\right]$.
In this case, we can divide the region $\left[T_{2},T_{K}\right]$
into 3 sub-regions as follows: 
\begin{equation}
\left[T_{2},T_{K}\right]=\left[T_{2},T_{3}\right]\cup\left[T_{3},T_{3}+\Delta_{3}\right]\cup\left[T_{3}+\Delta_{3},T_{K}\right],
\end{equation}
while the last region $\left[T_{K},T_{\infty}\right]$ remains unchanged.
Similarly, for a sufficiently large $K$ mode (short wavelength) if
the dip $-V_{3}$ lies within the region $\left[T_{K},T_{\infty}\right]$
we obtain 3 new sub-regions 
\begin{equation}
\left[T_{K},T_{\infty}\right]=\left[T_{K},T_{3}\right]\cup\left[T_{3},T_{3}+\Delta_{3}\right]\cup\left[T_{3}+\Delta_{3},T_{\infty}\right].
\end{equation}

\subsection{Scattering matrices}

After all the regions/sub-regions have been determined, we evaluate
the scattering matrices in each region $R_{(j)}$ using the linearly
independent functions $\psi_{1,2(j)}$ from Eq.~(\ref{Psi12}) and
the derived parameters $A_{(j)}(K)$, $n_{(j)}$ and $c_{(j)}$. Since
$m_{{\rm eff}(j)}^{2}$ has the same form in each region given by
Eq.~(\ref{meff}), the scattering matrix that we will provide below
is generic and applicable in all regions.

For a set of $N$ piecewise regions $R_{(\{1,...,N\})}$, the final
mode amplitude is given by the expression 
\begin{equation}
Y(K,T_{N})=\prod_{j=1}^{N}S(K,R_{(j)})Y(K,T_{0})\label{eq:final-S}
\end{equation}
where 
\begin{equation}
Y(K,T)=\left[\begin{array}{c}
y(K,T)\\
\partial_{T}y(K,T)
\end{array}\right]
\end{equation}
for mode function $y(K,T)$.

The scattering-propagator matrix $S(K,R_{(j)})\equiv S(K,T_{U(j)},T_{L(j)})$
for a region $R_{(j)}=\left[T_{L(j)},T_{U(j)}\right]$ where the indices
$U,L$ indicate upper and lower bounds for the region, is 
\begin{align}
S(K,T_{U(j)},T_{L(j)}) & =\left[\begin{array}{cc}
\psi_{1(j)} & \psi_{2(j)}\\
\dot{\psi}_{1(j)} & \dot{\psi}_{2(j)}
\end{array}\right]_{T=T_{U(j)}}\left[\begin{array}{cc}
\psi_{1(j)} & \psi_{2(j)}\\
\dot{\psi}_{1(j)} & \dot{\psi}_{2(j)}
\end{array}\right]_{T=T_{L(j)}}^{-1}\label{eq:scatteringsol}\\
 & =\Psi_{(j)}\left(T_{U(j)}\right)\Psi_{(j)}^{-1}\left(T_{L(j)}\right)
\end{align}
where $\Psi_{(j)}^{-1}$ represents an inverse operation on matrix
$\Psi_{(j)}$, and the two square matrices on the RHS are evaluated
at $T_{L(j)}$ and $T_{U(j)}$ respectively. Using Eq.~(\ref{Psi12}),
the matrix $\Psi_{(j)}$ in the RHS of Eq.~(\ref{eq:scatteringsol})
is explicitly given as 
\begin{equation}
\Psi_{(j)}\left(T\right)=\left[\begin{array}{cc}
\psi_{1(j)} & \psi_{2(j)}\\
\dot{\psi}_{1(j)} & \dot{\psi}_{2(j)}
\end{array}\right]_{T}\equiv e^{-\frac{3}{2}T}\left[\begin{array}{cc}
J_{r}\left(z\right) & J_{-r}\left(z\right)\\
\left(-3/2J_{r}\left(z\right)+\partial_{T}J_{r}\left(z\right)\right) & \left(-3/2J_{-r}\left(z\right)+\partial_{T}J_{-r}\left(z\right)\right)
\end{array}\right]\label{eq:Psimatrix}
\end{equation}
where $z=2A_{(j)}\left(K\right)\exp\left(-n_{(j)}T/2\right)/n_{(j)}$
and order $r=\sqrt{9-4c_{(j)}}/n_{(j)}$ with the parameters $A_{(j)}(K)$,
$n_{(j)}$ and $c_{(j)}$ for a region $R_{(j)}$ as defined previously.

\subsection{Numerical Fitting}

We will now present a few examples where we utilize the mass-model
to fit axion isocurvature power spectra obtained by numerically solving
the axion toy model in Eqs.~(\ref{eq:backgroundeom0}), (\ref{eq:backgroundeom})
and (\ref{eq:modeeq}) for different Lagrangian parameters using an
RK-solver. The fitted isocurvature power spectrum can be expressed
as 
\begin{equation}
\Delta_{S,{\rm fit}}^{2}(K,T_{\infty})=C_{m}K^{3}\left|Y(K,T_{\infty})\right|^{2}
\end{equation}
where $C_{m}$ is a normalization parameter from the parameter set
$P_{{\rm set}}$ of Eq.~(\ref{Vset}) and $Y(K,T_{\infty})$ is obtained
from Eq.~(\ref{eq:final-S}) using the scattering matrices by taking
$T_{N}\rightarrow T_{\infty}$. In Eq.~(\ref{eq:final-S}), we set
the initial mode amplitude $Y(K,0)$ using the adiabatic boundary
conditions of the BD vacuum.

\begin{figure}
\begin{centering}
\includegraphics[scale=0.31]{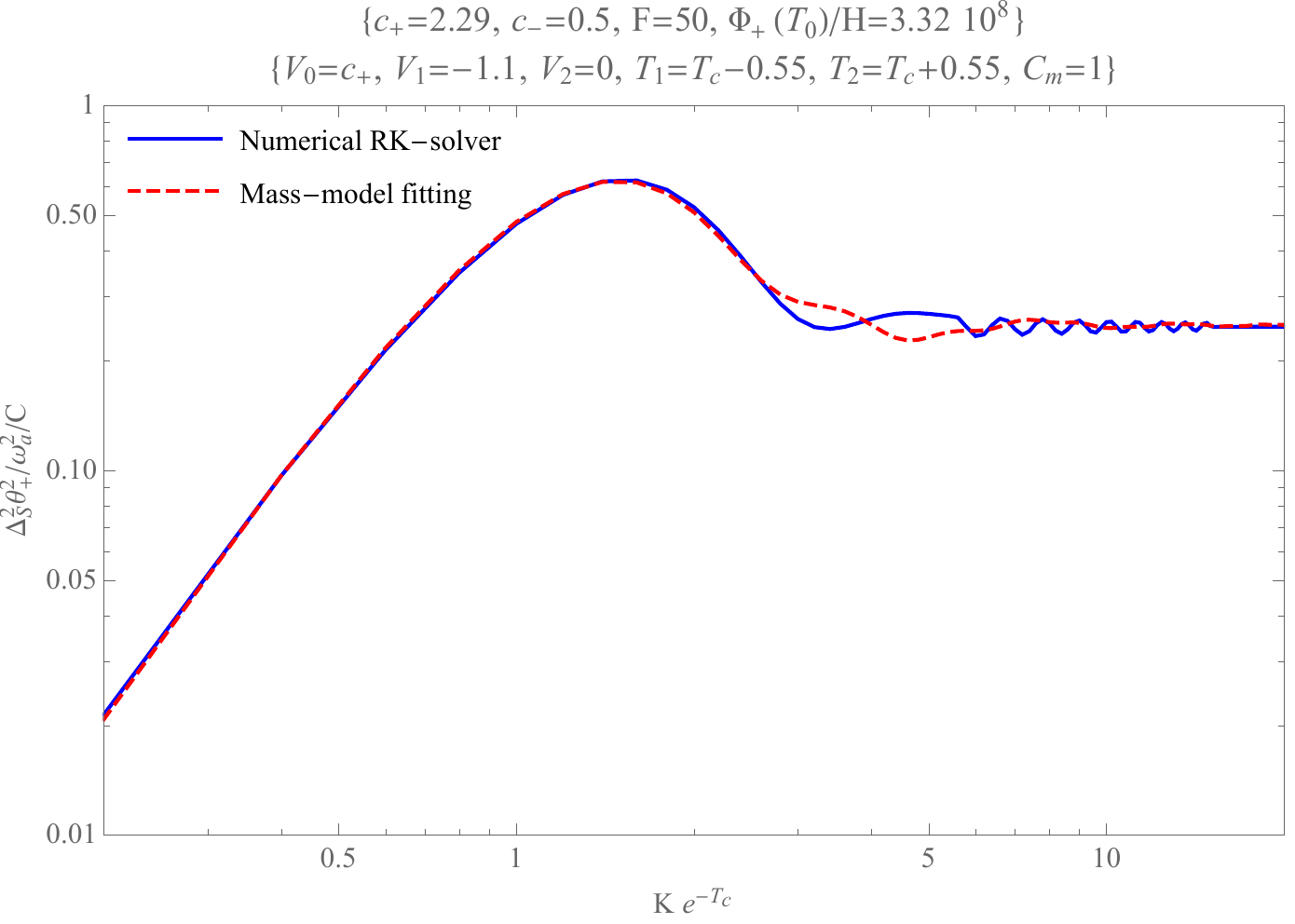}\qquad{}\includegraphics[scale=0.29]{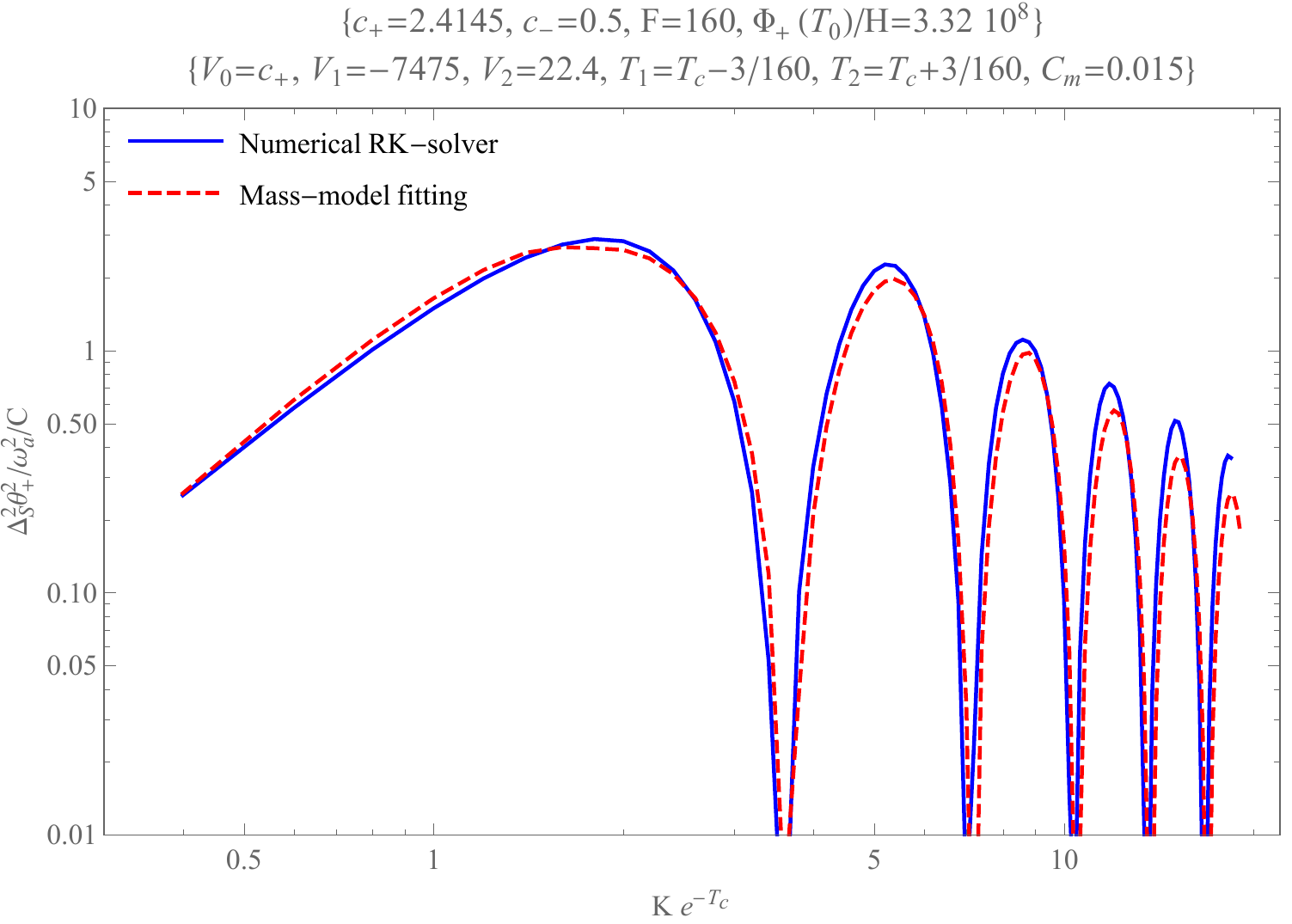}
\par\end{centering}
\begin{centering}
\includegraphics[scale=0.26]{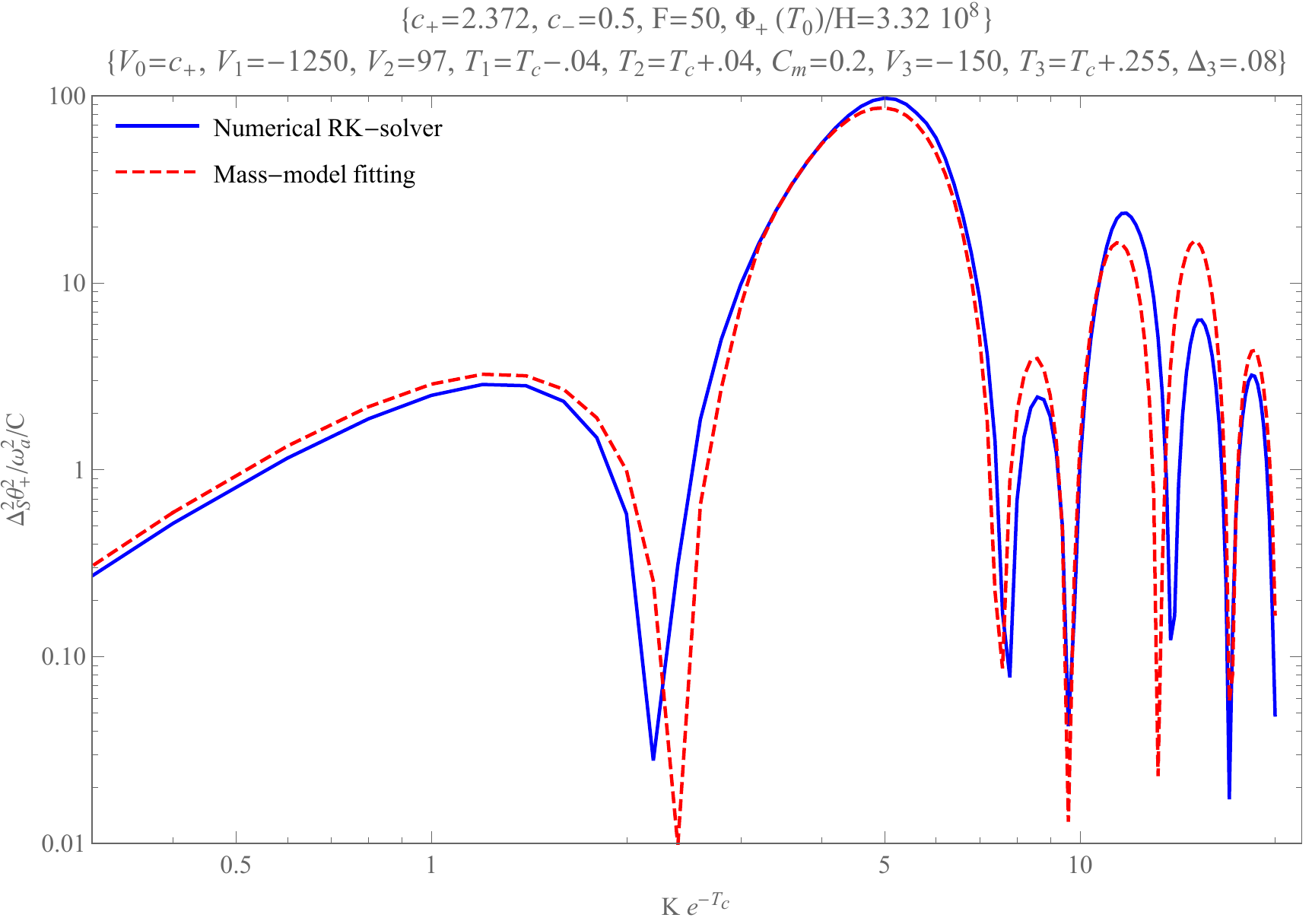}\qquad{}\includegraphics[scale=0.26]{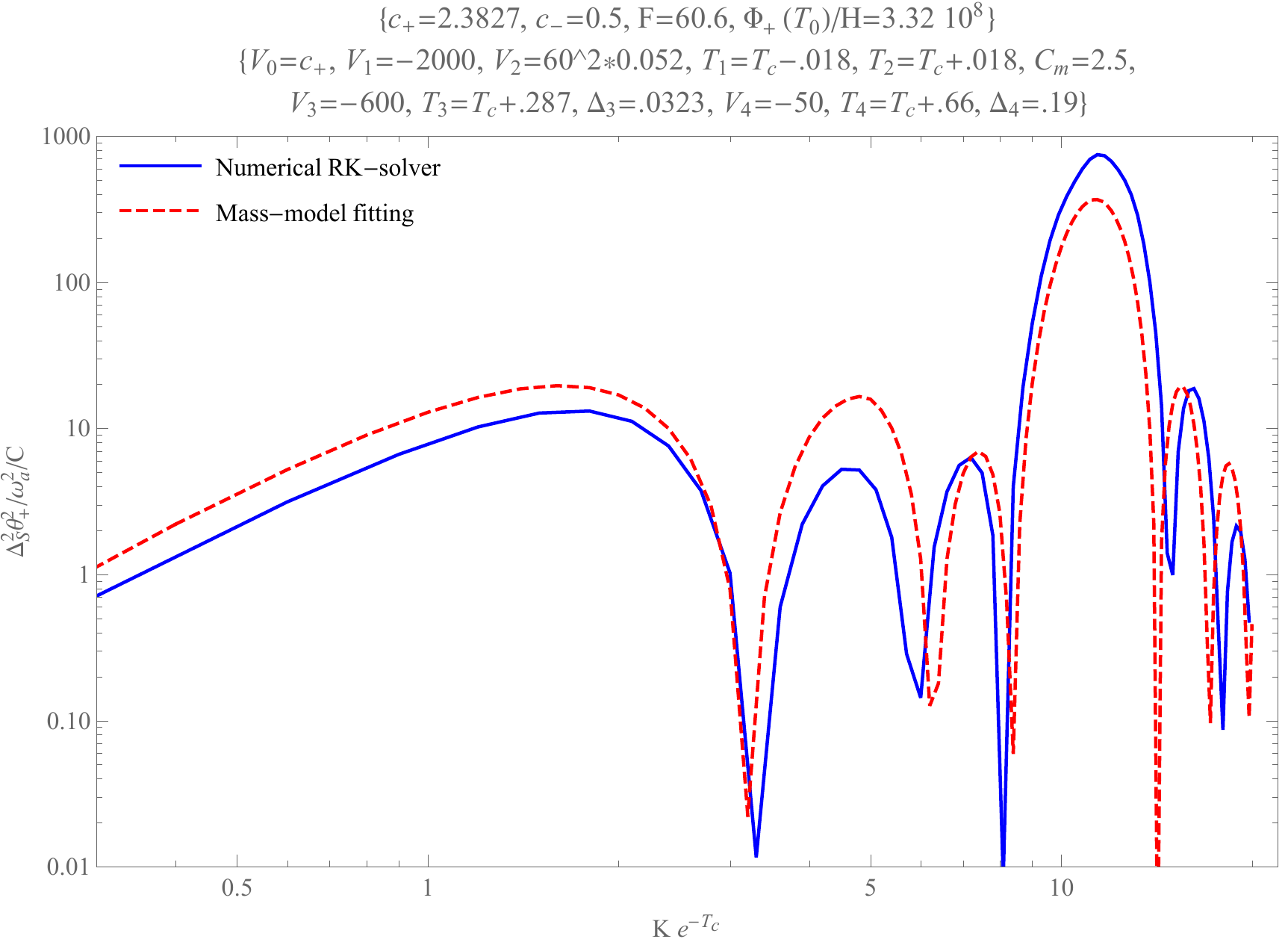}
\par\end{centering}
\caption{\label{fig:Fitting}Fitting axionic blue isocurvature power spectra
for examples where $\alpha<\alpha_{{\rm Ch}}$ using mass-model presented
in Sec.~\ref{sec:Generic-mass-model}. In each plot, solid (blue)
curve represents the power spectrum obtained by solving Eqs.~(\ref{eq:backgroundeom0}),
(\ref{eq:backgroundeom}) and (\ref{eq:modeeq}) using an RK-solver
while the dashed (red) curve is the fit from the mass-model. The plots
in the top row are fitted using the minimal mass-model with $P_{{\rm set}}$
in Eq.~(\ref{Vset}) restricted to $N=2$, while the plots in the
bottom row are examples of large resonant underdamped cases that are
fitted using non-minimal mass-model with $N=3$ $(4)$ such that the
model consists of additional $-V_{3}$ $(-V_{4})$ dips respectively.}
\end{figure}

\begin{figure}
\begin{centering}
\includegraphics[scale=0.3]{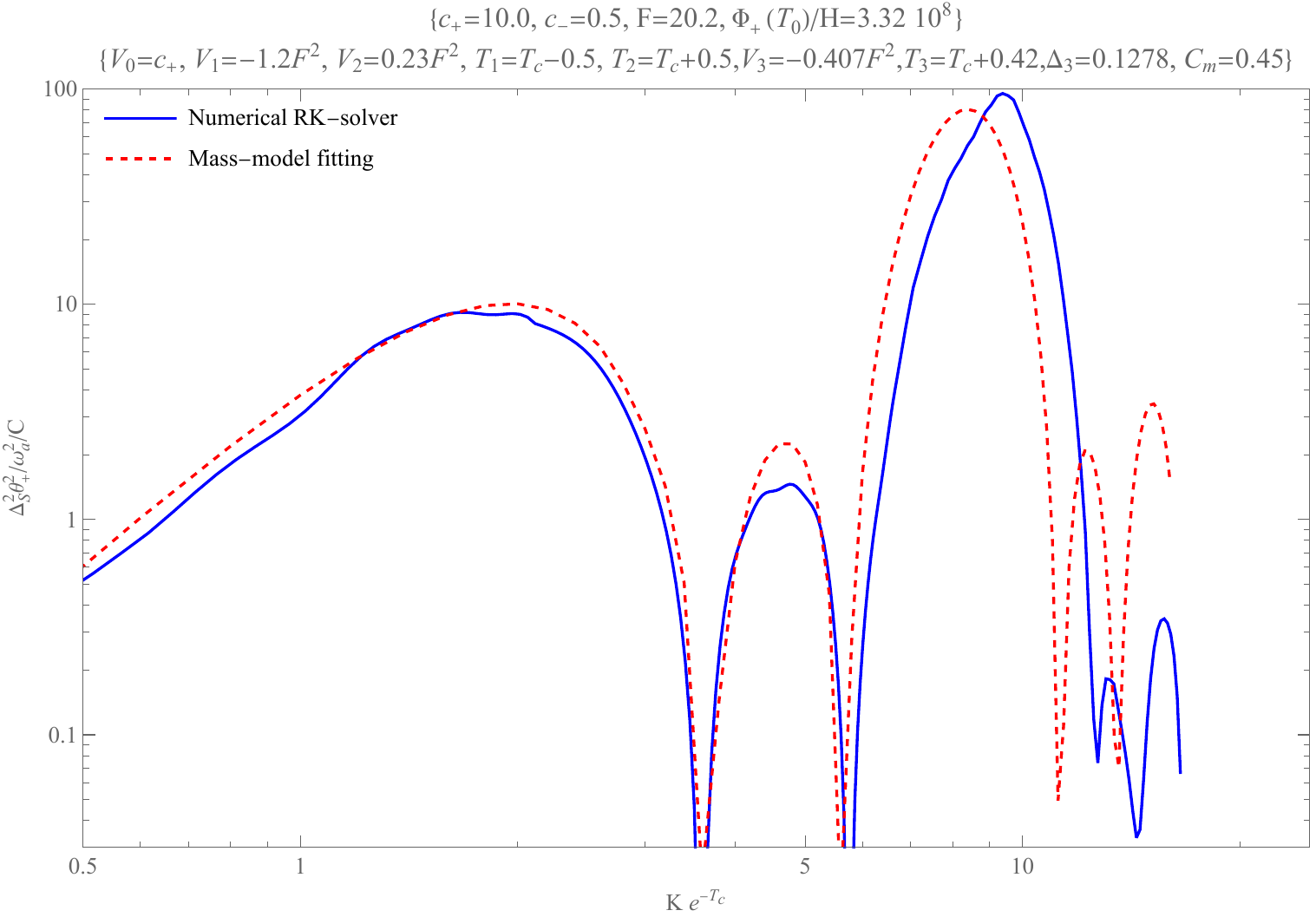} 
\par\end{centering}
\caption{\label{fig:Fitting-largecp-chaotic}Fitting axionic blue isocurvature
power spectrum for a large $c_{+}$ case where $\alpha_{c}\gg\alpha_{{\rm Ch}}$
using mass-model presented in Sec.~\ref{sec:Generic-mass-model}.}
\end{figure}

Let us consider a minimal case where we restrict ourselves to $i=2$
in Eq.~(\ref{Vset}). This minimal model is sufficient to fit isocurvature
power spectra for both over as well as underdamped cases where the
background fields do not cross each other again after transition at
$T_{c}$. This generically refers to all situations with $\alpha<\alpha_{2}$
where $\alpha_{2}$ is an $F$ dependent cutoff and given in Eq.~(\ref{eq:alpha-1}).
Hence, the minimal model consists of $V_{0}$, a single $-V_{1}$
dip at transition followed by an exponentially decaying $V_{2}$ term.
The minimal mass-model consists of four piecewise regions $R_{\{1,..,4\}}$
as shown below 
\begin{align}
\left[T_{0},T_{\infty}\right] & =R_{(1)}\cup R_{(2)}\cup R_{(3)}\cup R_{(4)}\\
 & =\left[T_{0},T_{1}\right]\cup\left[T_{1},T_{2}\right]\cup\left[T_{2},T_{K}\right]\cup\left[T_{K},T_{\infty}\right].\label{eq:minimalregions}
\end{align}
The final mode amplitude when evaluated using Eq.~(\ref{eq:final-S})
can be expressed as 
\begin{equation}
Y(K,T_{\infty})=S(K,R_{(4)})S(K,R_{(3)})S(K,R_{(2)})S(K,R_{(1)})Y(K,T_{0}).
\end{equation}
where we set $Y(K,T_{0})$ to the BD initial conditions
\begin{equation}
Y(K,T_{0})=\left[\begin{array}{c}
y\\
\partial_{T}y
\end{array}\right]_{T=T_{0}=0}=\frac{e^{iK}}{\sqrt{K}}\left[\begin{array}{c}
1\\
-iK-1
\end{array}\right]\quad\forall K^{2}\gg V_{0}-2
\end{equation}
where $T_{0}=0$ and we have dropped factors such as $2\left[a(0)\right]^{3/2}\sqrt{H}$
since we are presenting a fitting function. Additionally, the parameters
$A_{(j)}(K)$, $n_{(j)}$ and $c_{(j)}$ for each of the four regions
$R_{\{1,..,4\}}$ are set from 
\begin{align}
A_{(j)}(K) & =\left\{ K^{2},K^{2},B_{1}\left(K\right),B_{2}\left(K\right)\right\} \\
c_{(j)} & =\left\{ V_{0},-V_{1},0,0\right\} \\
n_{(j)} & =\left\{ 2,2,3,2\right\} 
\end{align}
where $B_{1,2}$ are given by Eqs.~(\ref{eq:B1}) and (\ref{eq:B2}).
As an example, consider region $R_{(3)}=\left[T_{2},T_{K}\right]$
with $A_{(3)}(K)=B_{1}(K)$, $n_{(3)}=3$ and $c_{(3)}=0$. We can
then evaluate the $\Psi_{(3)}$ matrix at upper boundary $T_{K}$
as 
\begin{equation}
\Psi_{(3)}\left(T_{K}\right)=e^{-\frac{3}{2}T_{K}}\left[\begin{array}{cc}
J_{1}\left(z\right) & J_{-1}\left(z\right)\\
\left(-\frac{3}{2}J_{1}\left(z\right)+\partial_{T}J_{1}\left(z\right)\right) & \left(-\frac{3}{2}J_{-1}\left(z\right)+\partial_{T}J_{-1}\left(z\right)\right)
\end{array}\right]
\end{equation}
with 
\begin{equation}
z=\frac{2}{3}B_{1}\left(K\right)e^{-\frac{3}{2}T_{K}}.
\end{equation}
In Figs.~\ref{fig:Fitting} and \ref{fig:Fitting-largecp-chaotic},
we present a few examples of axionic blue isocurvature power spectra
fitted using the mass-model described above. In these examples, we
have normalized the isocurvature spectra with respect to a constant
\begin{equation}
C=\frac{2r\left(1+r^{4}\right)}{\left(1+r^{2}\right)^{3}\pi^{2}F^{2}}
\end{equation}
where $r=\sqrt{c_{+}/c_{-}}$ and set $T_{0}$ and $T_{\infty}$ to
$0$ and $\sim30$ respectively. In each case, the value of the axion
Lagrangian variables and the fitted model parameters are shown in
the title of the plots. 

In Fig.~\ref{fig:Fitting} we restrict to examples where $\alpha_{c}<\alpha_{{\rm Ch}}$.
The plots in the top row belong to the class of axion models where
the background fields can be classified as non-resonant (resonant)
oscillations of the fields post-transition where $\alpha_{c}<\alpha_{2}$.
The non-resonant cases are defined by having $\alpha_{c}<\alpha_{{\rm L}}\sim0.2$,
where the shape of the isocurvature power spectrum for these cases
bears resemblance to the overdamped case studied in \cite{Chung:2016wvv,Chung:2017uzc}.
These are fitted using the minimal mass-model with $P_{{\rm set}}$
in Eq.~(\ref{Vset}) restricted to $N=2$ such that the model consists
of a single $-V_{1}$ dip followed by an exponential $V_{2}$ potential.
In the bottom row of Fig.~\ref{fig:Fitting} we fit the isocurvature
spectra for cases with $\alpha_{c}>\alpha_{2}$. We observe that fitting
these highly resonant underdamped instances, require non-minimal mass-model
with $P_{{\rm set}}$ in Eq.~(\ref{Vset}) set to $N=3$ and $4$
respectively such that the model consists of additional $-V_{3}$
and $-V_{4}$ dips respectively. Finally, in Fig.~\ref{fig:Fitting-largecp-chaotic}
we present an example where we apply the mass-model to a chaotic case
where we consider a large $c_{+}$ value such that $\alpha_{c}\gg\alpha_{{\rm Ch}}$.
Through these examples, we conclude that the mass-model is a semi-quantitatively
accurate representation of the axionic isocurvature power spectra
with rich and complex spectral shapes and bumps.\footnote{The exponential $V_{2}$ potential in our mass-model arises from the
UV integration of the high frequency resonant oscillations of the
lighter mass eigenmode. As shown in \pA, for $\alpha\gtrsim1$ the
UV integration procedure tends to breakdown such that the exponential
IR term may not be accurate enough to sufficiently model the power
spectrum within acceptable error margins. Since large $c_{+}$ cases
generally transition with $\alpha>1$ and belong to chaotic regime,
these may require additional parameters or variations within the model
to sufficiently map the spectral bumps and amplitudes. For the examples
that we have tested, we find that the generic mass-model provides
a good fitting model for the chaotic scenarios up to a factor of few
from the cutoff scale, $k_{{\rm cut}}$.} 

\section{\label{sec:sine-model-ftting}A sine function based fitting model}

\begin{figure}
\begin{centering}
\includegraphics[scale=0.5]{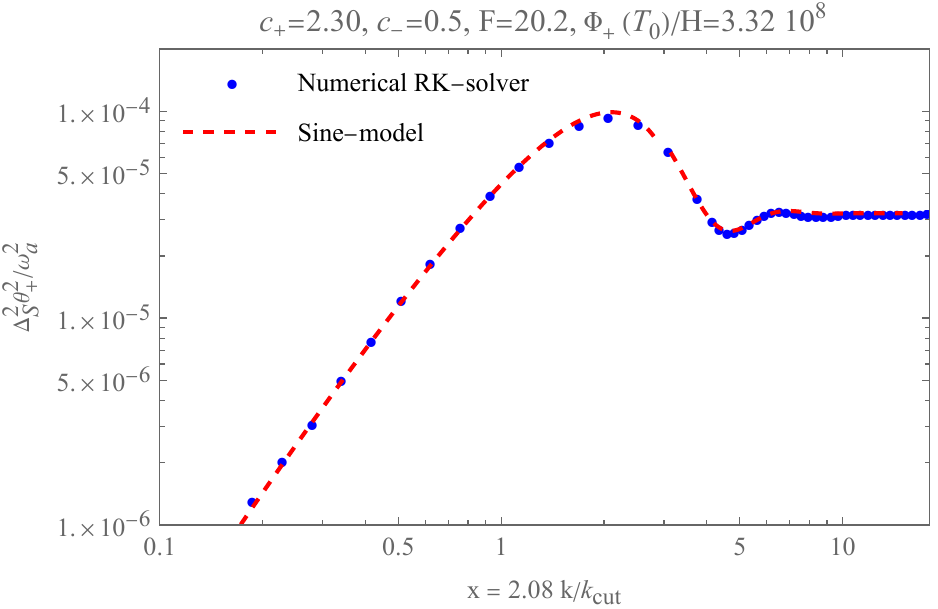}\includegraphics[scale=0.5]{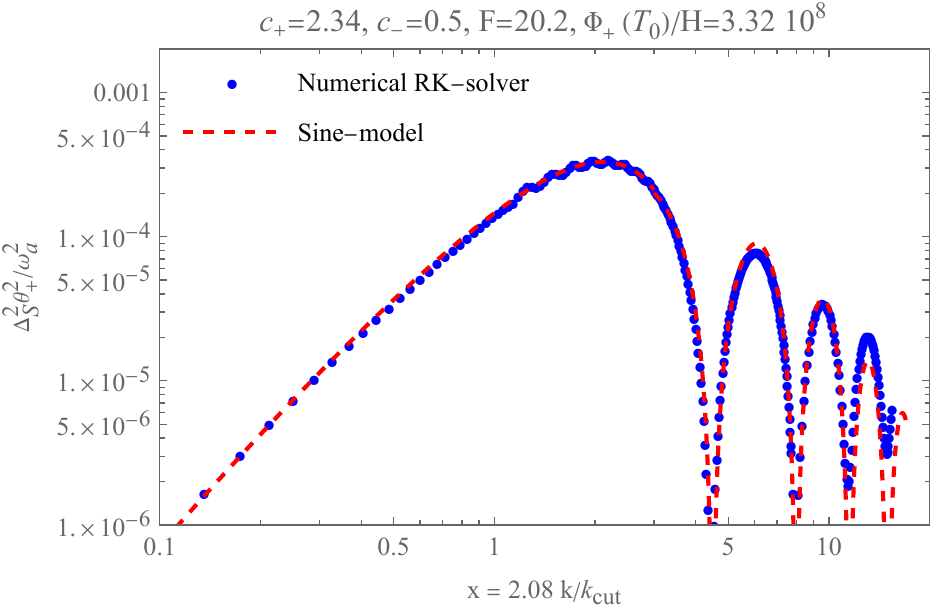}\medskip{}
 \includegraphics[scale=0.5]{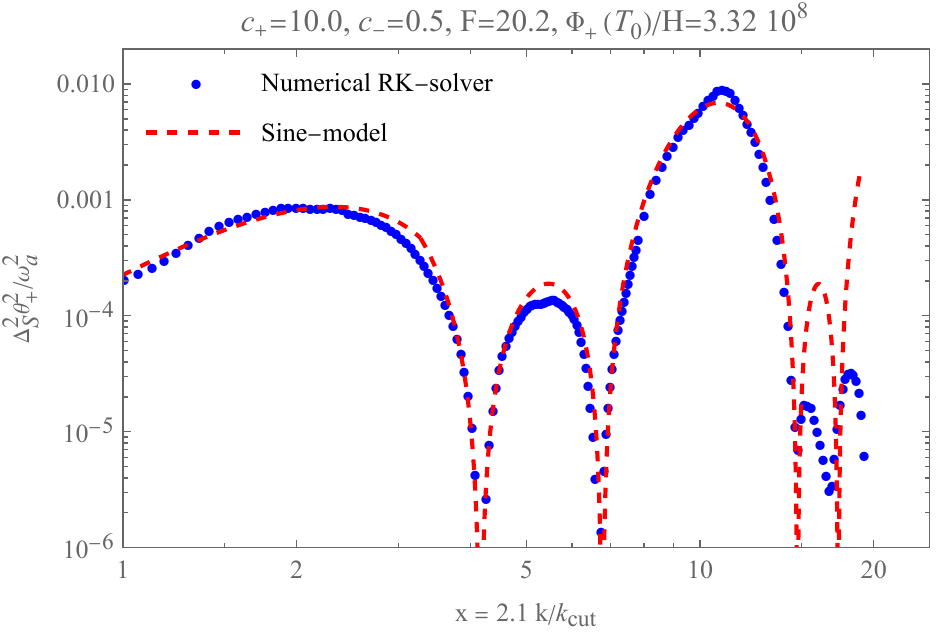}\includegraphics[scale=0.5]{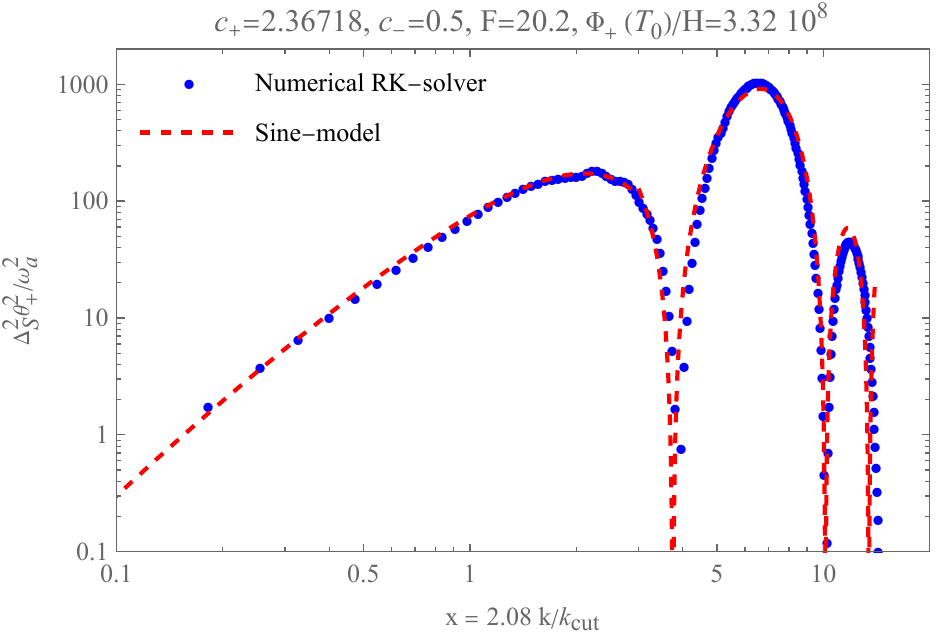} 
\par\end{centering}
\caption{\label{fig:sine-model}In this figure we show plots where we fit the
piecewise-model presented in Eq.~(\ref{eq:final sine-model}) to
the numerical data for four underdamped isocurvature power spectra
examples. Starting from the top-left and moving clockwise, the examples
presented in this figure can be further classified as non-resonant
non-chaotic, resonant non-chaotic, resonant chaotic small $c_{+}$,
and resonant chaotic large $c_{+}$ respectively.}
\end{figure}

The mass model presented earlier provides a reasonable fit to the
isocurvature spectrum over a broad range of scales. However, it involves
intricate steps, a large number of fitting parameters, and can be
time-consuming when searching for the best-fit values of the model
parameters from numerical/observational data.\footnote{For each point in the parametric space that is explored during the
fitting procedure, a naive procedure requires evaluating the entire
power spectrum for several $k$ modes using the mass-model.} From an observational perspective, the isocurvature spectrum can
be divided into two regions: pre-cutoff ($k<k_{{\rm cut}}$) and post-cutoff
scale ($k>k_{{\rm cut}}$) where we define $k_{{\rm cut}}$ as the
location of the first bump in the power spectrum, which can be identified
most simply as the first time when the slope of the power spectrum
goes to zero followed by a decrease in the power. The pre-cutoff region
is well described by a blue-tilt, while the post-cutoff region can
undergo oscillations before settling to a massless plateau. In the
context of detecting isocurvature modes and accurately fitting the
signal, the initial bumps in the post-cutoff region are of utmost
importance. To address this, we propose a simplified piecewise function
that effectively captures and fits the isocurvature spectrum up to
the first few bumps, or $k\lesssim O(5)k_{{\rm cut}}$: a $7$-parameter\footnote{$7$ real parameters: 6 $c_{i}$ and 1 cutoff scale $k_{{\rm cut}}$}
piecewise-model in terms of $x=2.08\,k/k_{{\rm cut}}$ written as
\begin{equation}
\Delta_{{\rm 2-fit}}^{2}\left(\left[c_{1-6},k_{{\rm cut}}\right],k\right)=\begin{cases}
c_{1}\left|H_{i\sqrt{c_{2}-9/4}}^{1}(x)\right|^{2}x\left(j_{1}(x)\right)^{2}, & x\lesssim x_{0}\\
\left(\sqrt{c_{1}\left|H_{i\sqrt{c_{2}-9/4}}^{1}(x_{0})\right|^{2}x_{0}\left(j_{1}(x_{0})\right)^{2}}\right.\\
\left.\phantom{\sqrt{\left|\right|^{2}}}+c_{3}\left(e^{-c_{4}x}\sin\left(c_{5}\left(x-c_{6}\right)\right)-e^{-c_{4}x_{0}}\sin\left(c_{5}\left(x_{0}-c_{6}\right)\right)\right)\right)^{2} & x_{0}\lesssim x\lesssim10
\end{cases}\label{eq:final sine-model}
\end{equation}
where $j_{1}(x)$ is the spherical Bessel function of order $1$,
and $H_{i\sqrt{c_{2}-9/4}}^{1}(x)$ is the Hankel function of order
$i\sqrt{c_{2}-9/4}$. For underdamped cases, the first bump occurs
at $x\approx2.08$, and we find that choosing $x_{0}\approx3$ yields
a better matching and a lower $\chi^{2}$ when fitting the piecewise
model to the numerical data for several examples. For the overdamped
cases, which are characteristically defined by a blue-tilt of $1<n_{{\rm I}}<4$,\footnote{This corresponds to a measurement of $c_{2}<9/4$ from blue-titled
large-scale modes.} and other scenarios with a smooth transition (without a bump) to
the plateau, we find $x_{0}=0.4$ as a suitable choice. In Appendix
\ref{sec:Fitted-sine-model-parameters} we briefly elaborate upon
the motivation for this model construction, provide fitting parameters
for the examples presented in Fig.~\ref{fig:sine-model} and give
a fitting plot for a sample overdamped case.

In Fig.~\ref{fig:sine-model}, we fit the new piecewise sine-model
to a few examples. In each case, the fiducial choice of Lagrangian
parameters is given in the title of the plots. The fit model parameters
are given in Appendix \ref{sec:Fitted-sine-model-parameters}. One
can see from the $c_{+}=10.0$ case of Fig.~\ref{fig:sine-model}
that the model does not fit the features of the spectrum for $k\gtrsim O(5)k_{{\rm cut}}$.
The reason why the $k_{\mathrm{cut}}$ fixes the $k$ scale range
over which the model is effective is because the number of oscillations
in these isocurvature models has a fundamental oscillatory $k$-space
period fixed by $k_{\mathrm{cut}}$, and one expects with a 7-parameter
model to be able to fit at most 3 bump like features (counting around
2 parameters per bump). Of course, in principle, one may be able to
add more parameters to fit more features for the higher $k$ values,
but because near future surveys may be limited in the range of $k$-scale
sensitivity, the 7-parameter model seems to strike a reasonable balance
between economy and phenomenological detection coverage of the isocurvature
perturbations in the near future. It is important to emphasize that
the form of this fitting function was inspired by the generic solutions
to the mass model that we discussed earlier.

\section{\label{sec:Conclusions}Conclusions}

In this paper, we computed the axionic blue isocurvature perturbation
power spectrum in the large radial field mass/kinetic energy limit
for which there are multiple crossings of the radial field across
the global minimum of the effective potential. This paper serves as
a companion to the paper \pA. We have derived a mass model that can
be used to compute the isocurvature spectrum based on the idea that
the fast oscillating background fields have the net effect of a square
well type of potential. Using this type of model, we have demonstrated
that one can for example fit 7 bumps using 12 parameters. This type
of model can be used for phenomenological fitting purposes if desired.
To reduce the complexity of the possible future fitting efforts, we
have also constructed a simpler 7-parameter sinusoid function based
fitting function which fits at least 3 bumps, inspired by the results
of the mass model. The new sine-model can be used to fit both underdamped
as well as overdamped scenarios of the axionic model considered in
this work, and may be applied model-independently to detect CDM isocurvature
perturbations in scenarios where the axion mass makes a dynamical
transition.

One interesting feature of the large kinetic energy cases considered
in this paper was the appearance of exponential sensitivity of the
isocurvature spectrum to Lagrangian parameters. This sensitivity arises
because there is a large kinetic energy driven resonant phenomena
that can exponentially boost or diminish amplitudes. Furthermore,
the Lagrangian parametric variations which translate to changes in
the initial conditions when the nonlinear forces temporarily dominate
generically give rise to a background field phase space mixing that
is characteristic of chaos. This means that although the spectrum
generically has an oscillatory shape whose oscillatory $k$-period
is fixed by the first collision time scale, the amplitude of the rising
part of the spectrum and the first few oscillatory bumps determined
by the background field phase space is not simply predictable as a
function of the Lagrangian parameters controlling the kinetic energy
of the radial field. That in turn implies that if we phenomenologically
detect an oscillatory spectrum of the type considered in this paper,
there will be a large theoretical uncertainty in mapping back to the
underlying Lagrangian parameters. We quantify this uncertainty using
a distribution function presented in Fig. \ref{fig:histograms}.

In the construction of the rising part of the isocurvature spectrum,
we have also noted that one can reduce the computation of the spectrum
to solving the zero-mode amplitudes. In other words, we do not need
to solve for the mode functions separately in the rising part of the
spectrum as long as we have the background field solutions. This is
owing to an accidental duality between the long wavelength mode equations
and the background field equations present in the class of axion models
considered in this paper. Using a set of numerical computations of
zero-mode amplitudes, we have constructed a formula for the isocurvature
spectrum in its rising part as a function of the underlying Lagrangian
parameters in the non-chaotic region. We have also used the duality
to explain the exponential parametric sensitivity of the spectrum
in the chaotic region.

There are many interesting future directions related to this work.
It would be interesting to carry out fits to data (or give forecasts
for future experiments) using the fitting formulae presented in this
work to look for signals of isocurvature perturbations. Most of the
previous works on blue isocurvature spectrum have not dealt with the
strongly oscillatory nature of the spectrum generic to the underdamped
isocurvature scenarios. Moreover, even without the oscillations, most
of the previous works have focused on simplified scenarios without
a plateau cutting off the rising spectrum.

Given the large magnitude of the amplification of the high $k$ bumps
(see e.g.~Fig.~\ref{fig:Fitting}), the non-Gaussianities generated
by these isocurvature amplitudes may be significant and may have a
spectral shape that is correlated with the power spectrum shape. This
would be an interesting correlated non-Gaussianity study to pursue.
One may also be able to use the large bump to generate seeds for unusually
large clumped objects in the early universe such as primordial black
holes \cite{Passaglia:2021jla}. This enhancement of the power can
also result in an overabundance of halos at high redshifts while converging
to the halo mass function shape similar to that of $\Lambda$CDM at
low redshifts \cite{Blinov:2021axd,Tkachev:2023acf}.

\appendix

\section{\label{sec:-phim_Transient}$\phi_{\pm}$ Transient solutions}

In Sec.~\ref{Sec:Estimate-Tzjc} we demonstrated that when $\omega\gtrsim O(1)$,
the background fields undergo a transition near a zero-crossing point
at $T_{z,j}$ for $j\geq2$. There we derived expressions to estimate
the zero-crossing $T_{z,j_{c}}$ closest to the transition using a
zeroth order approximate solution for the background fields. However,
as the $\phi_{+}$ field moves closer to a zero-crossing $T_{z,j}$,
significant deviations from the zeroth order solutions can occur due
to a large kinetic energy $\sim O(\alpha^{2}F^{4})$ leading to transient
effects.

In this Appendix, we will show that the non-adiabatic effects resulting
from a zero-crossing at $T_{z,j}$ results in an $O\left(F/\alpha_{j}^{3/2}\right)$
increase in the effective mass of the $\phi_{+}$ field such that
the location of the next zero-crossing at $T_{z,j+1}$ deviates from
the expression provided in Eq.~(\ref{eq:Tzj}).

\subsection{$\phi_{-,{\rm Tr}}$ solution}

\begin{figure}
\centering{}\includegraphics[scale=0.7]{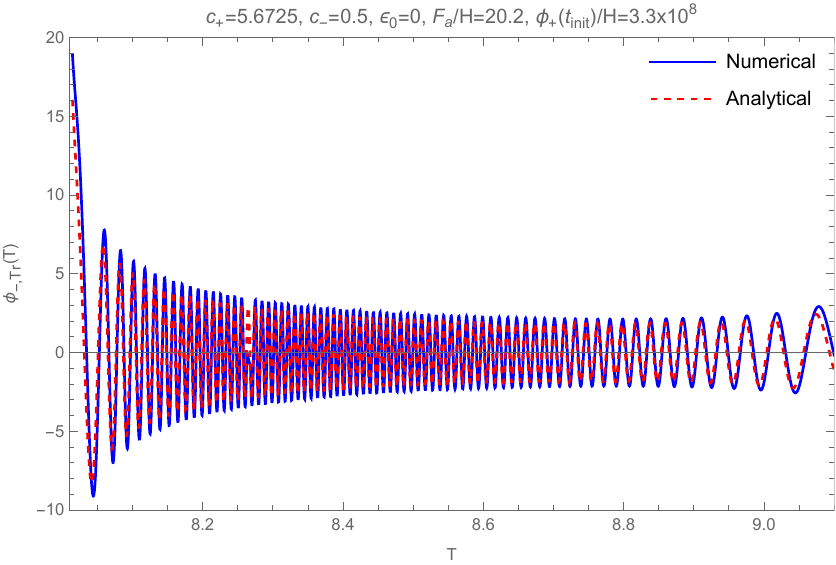}\caption{\label{fig:phim-Tr}The plot illustrates a comparison between the
analytic approximation (red, dashed line) of $\phi_{-,{\rm Tr}}$
as given by Eq.~(\ref{eq:phim-Tr}) and the corresponding numerical
result (blue, solid line) for a specific example case where $T_{z,j=4}\approx8$.}
\end{figure}

Since we assume that the background fields remain along the flat direction
before they roll down to the minimum with $\phi_{+}(T_{0})\sim O(M_{P})$,
the homogeneous solution of the $\phi_{-}$ field is negligible at
$T_{0}$. However, as the $\phi_{+}$ field moves closer to its first
zero-crossing $T_{z,j=1}$, a large kinetic energy $\sim O(\alpha^{2}F^{4})$
generates deviations in $\phi_{-}$ from the zeroth order solution.
The homogeneous (transient) solution of the $\phi_{-}$ field can
be obtained by solving the differential equation: 
\begin{equation}
\ddot{\phi}_{-,{\rm Tr}}+3\dot{\phi}_{-,{\rm Tr}}+\left(c_{-}+\phi_{+}^{2}\right)\phi_{-,{\rm Tr}}\approx0.\label{eq:homogeneous_phim}
\end{equation}
Thus, the transient component of $\phi_{-}$ oscillates rapidly with
a high frequency $\sim O(\phi_{+})$.\footnote{Here we assume $c_{-}$ to be $O(1)$.}
An approximate WKB solution to Eq.~(\ref{eq:homogeneous_phim}) can
be written as 
\begin{align}
\phi_{-,{\rm Tr}}(T) & \approx e^{-\frac{3}{2}\left(T-T_{*}\right)}\frac{A_{j}}{\sqrt{\left|\phi_{+}\right|}}\cos\left(\int_{T_{*}}^{T}dT\left|\phi_{+}\right|\right)\qquad T_{z,j}+\epsilon_{j}\lesssim T\lesssim T_{z,j+1}-\epsilon_{j+1}\label{eq:phim-Tr}
\end{align}
where $T_{*}\sim T_{z,j}+\epsilon_{j}$ for $\epsilon_{j}\approx1/\left(F\sqrt{\alpha_{j}}\right)$.
To obtain an estimate for the amplitude $A_{j}$, we match the solution
in Eq.~(\ref{eq:phim-Tr}) with the approximate solution given in
Eq.~(\ref{eq:phiexpn}) at $T_{z,j}+\epsilon_{j}$. This results
in the following expression: 
\begin{equation}
A_{j}\approx\frac{7}{4}\times\frac{F^{\frac{3}{2}}}{\alpha_{j}^{1/4}}.\label{eq:Aj}
\end{equation}
After comparing with the numerical results, we modify the factor 7/4
in Eq.~(\ref{eq:Aj}) to 2.3. This shift can be attributed to the
inaccuracy of Eq.~(\ref{eq:phiexpn}). For $T>T_{z,j}+O(1/F)$, the
complete $\phi_{-}$ solution is given by the superposition of the
forced and homogeneous solutions. In Fig.~(\ref{fig:phim-Tr}) we
plot the background field for a fiducial case showing a comparison
between the numerical solution and our approximate result in Eq.~(\ref{eq:phim-Tr}).

\subsection{Transient mass-squared term}

Due to the high frequency transient oscillations of $\phi_{-}$, the
$\phi_{+}$ field inherits a positive mass contribution which leads
to a decrease in the time period of oscillation. To see this explicitly
we consider the EoM for $\phi_{+}$ and substitute $\phi_{-}\approx\phi_{-}^{(0)}+\phi_{-,{\rm Tr}}$
to analyze the effect of $\phi_{-,{\rm Tr}}$: 
\begin{equation}
\ddot{\phi}_{+}+3\dot{\phi}_{+}+c_{+}\phi_{+}+\left(\left(\phi_{-}^{(0)}+\phi_{-,{\rm Tr}}\right)\phi_{+}-F^{2}\right)\left(\phi_{-}^{(0)}+\phi_{-,{\rm Tr}}\right)=0
\end{equation}
which is equivalent to
\begin{align}
\ddot{\phi}_{+}+3\dot{\phi}_{+}+\left(c_{+}+\phi_{-,{\rm Tr}}^{2}+2\phi_{-}^{(0)}\phi_{-,{\rm Tr}}\right)\phi_{+}-F^{2}\phi_{-,{\rm Tr}} & =0.
\end{align}
By examining the aforementioned expression, it is evident that the
$\phi_{+}$ field acquires a mass-squared term that oscillates rapidly
with a high frequency. This oscillation is a result of the homogeneous
oscillations of $\phi_{-}$. Through the UV integration procedure
described in Appendix C of \pA, these high-frequency oscillations
can be integrated out, leading to the determination of an effective
mass-squared quantity composed of a residual IR term. More explicitly,
we see that the dominant IR contribution is obtained through the UV
integration of $\phi_{-,{\rm Tr}}^{2}$: 
\begin{equation}
\phi_{-,{\rm Tr}}^{2}\underrightarrow{~~{\rm UV-Integration}~~}\frac{1}{2}\left(\frac{A_{j}e^{-\frac{3}{2}(T-T_{*})}}{\sqrt{\left|\phi_{+}\right|}}\right)^{2}.
\end{equation}
After performing the UV integration, we obtain the reduced EoM for
$\phi_{+}$ as follows: 
\begin{equation}
\ddot{\phi}_{+}+3\dot{\phi}_{+}+c_{+}\phi_{+}+\frac{1}{2}A_{j}^{2}e^{-3(T-T_{*})}\approx0\qquad T_{z,j}+\epsilon_{j}\lesssim T\lesssim T_{z,j+1}-\epsilon_{j+1}.
\end{equation}
which has the following approximate solution : 
\begin{align}
\phi_{+}(x>\epsilon_{j}) & \approx\frac{1}{8c_{+}\omega^{2}}\left(-4\omega^{2}A_{j}^{2}e^{-3x}+4\omega^{2}A_{j}^{2}e^{-3x/2}\cos(\omega x)-2\omega(3A_{j}^{2}-4c_{+}f)e^{-3x/2}\sin(\omega x)\right)\label{eq:phip-general}\\
 & \approx\frac{\alpha_{j}F^{2}}{\omega}e^{-3x/2}\sin(\omega x)+\frac{A_{j}^{2}}{4c_{+}\omega}\left(-2\omega e^{-3x}+2\omega e^{-3x/2}\cos(\omega x)-3e^{-3x/2}\sin(\omega x)\right)\\
 & \approx\phi_{+}^{(0)}-\frac{A_{j}^{2}}{2c_{+}}\left(e^{-3x}-e^{-3x/2}\left(\cos(\omega x)-\frac{3}{2\omega}\sin(\omega x)\right)\right)
\end{align}
where $x=T-T_{z,j}$ , $f=\partial_{T}\phi_{+}(T_{z,j})\approx\alpha_{j}F^{2}$,
and $\phi_{+}^{(0)}$ is the zeroth order perturbed solution given
in Eq.~(\ref{eq:approxsol}).

In the limit $A_{j}\rightarrow0$, we find $\phi_{+}\rightarrow\phi_{+}^{(0)}$.
When the amplitude $A_{j}$ is finite and non-negligible, it results
in a deviation from the background solution $\phi_{+}^{(0)}$. Consequently
the next zero-crossing at $T_{z,j+1}$ occurs at 
\begin{equation}
T_{z,j+1}-T_{z,j}=\frac{\pi}{\omega}-\Delta T\left(A_{j}\right)
\end{equation}
where $\pi/\omega$ is the time period between two zero-crossings
for the zeroth order perturbative solution $\phi_{+}^{(0)}$, and
$\Delta T\left(A_{j}\right)$ is a function of $A_{j}$ which can
be obtained by solving the transcendental equation corresponding to
$\phi_{+}(T_{z,j+1}-T_{z,j})=0$ using Eq.~(\ref{eq:phip-general}).

\begin{figure}
\begin{centering}
\includegraphics[scale=0.8]{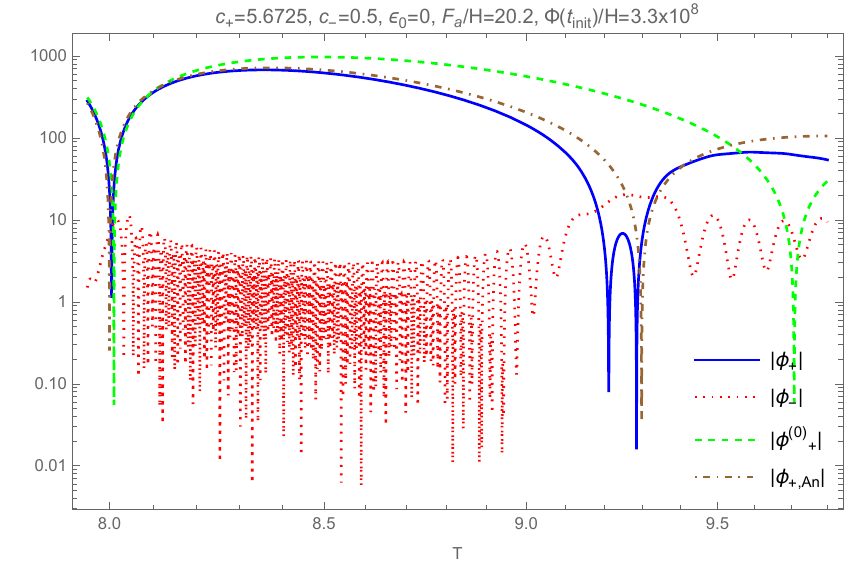} 
\par\end{centering}
\caption{\label{fig:phip-An}In this figure we plot our analytic approximations,
$\phi_{+}^{(0)}$ and $\phi_{+,{\rm An}}$, for the $\phi_{+}$ field
using Eqs.~(\ref{eq:approxsol}) and (\ref{eq:phip-general}) respectively.
For comparison, we also plot the numerical results for $\phi_{\pm}$
fields. The fiducial values used for generating this plot are the
same as the ones used in Fig.~\ref{fig:phim-Tr}. A more accurate
estimation of the amplitude $A_{j}$ of $\phi_{-,{\rm Tr}}$ will
yield a better fit of $\phi_{+,{\rm An}}$ to the numerical results.}
\end{figure}

In Fig.~\ref{fig:phip-An} we plot $\phi_{+}$ using Eq.~(\ref{eq:phip-general})
and compare with the numerical solution for a fiducial case. Since
$f\approx\alpha_{j}F^{2}$ and using Eq.~(\ref{eq:Aj}) we can estimate
the extent of the $A_{j}$-dependent deviation compared to the zeroth
order solution: 
\begin{equation}
\frac{\Delta\phi_{+}(A_{j})}{\phi_{+}^{(0)}}\approx O\left(\frac{F}{\sqrt{c_{+}}\alpha_{j}^{3/2}}\right).
\end{equation}
Hence, the deviation from the $\phi_{+}^{(0)}$ solution is negligible
for values of $\alpha_{j}\gtrsim O(F)$.

\section{\label{sec:Approx_theta}Approximate estimation for $\theta$}

In this Appendix, we give an approximate estimation of the phase,
$\theta$, of the zero-mode solution, $I_{0}$. First, we note that
the zero-mode solution, $I_{0}$, was initialized at $T_{0}$ as 
\begin{equation}
I_{0}(T_{0})=e^{-\left(3/2+i\omega\right)T_{0}}e_{1}
\end{equation}
from Eq.~(\ref{eq:initi0}). For $\phi_{+}\gg F$, the lightest mass
eigenvalue is $m_{1}\approx c_{+}$, and during the time $T<T_{c}$
when $\lambda=F^{2}/\phi_{+}^{2}<1$, the zero-mode solution can be
approximated as 
\begin{equation}
I_{0}(T)=e^{-\left(3/2+i\omega\right)T}\left[\begin{array}{c}
1\\
-\lambda
\end{array}\right].
\end{equation}
The argument of $I_{0}$ during this time is 
\begin{equation}
\arg\left(I_{0}\right)\approx-\omega T.
\end{equation}
As shown in Fig.~\ref{multiplecrossings}, at the transition, $T=T_{c}$,
the background fields cross each other momentarily and subsequently
settle to the minimum of the potential. Since the underdamped background
fields can deviate significantly from the flat direction at $T\sim T_{c}$
and the lightest mass eigenvalue can undergo $O(F_{a}/H)$ oscillations
post-transition. For $c_{-}<c_{+}$, the oscillations do not contribute
a time-averaged mass, thereby freezing the phase angle $\arg\left(I_{0}\right)$
at the transition. For such cases, the final phase angle $\theta$
can be approximated as 
\begin{equation}
\theta\approx-\omega T_{c}.\label{eq:approx_theta}
\end{equation}
The transition time, $T_{c}$, can be estimated using Eq.~(\ref{eq:Tc-emperical}).

\section{\label{sec:Chaotic-structure}Chaos in blue axion system}

In this Appendix, we will show that the axion toy model potential
leads to a dynamical system that can result in chaotic trajectories
due to quartic nonlinear interactions in the $\phi_{\pm}$ phase space.
When infinitesimally separated field trajectories traverse regions
of local instability, they may undergo divergent paths and can indicate
the possible presence of a chaotic system. We begin this analysis
by taking our nonlinear second order system of ordinary differential
equations in Eqs.~(\ref{eq:backgroundeom0}) and (\ref{eq:backgroundeom})
and rewriting them as 
\begin{equation}
\ddot{\phi}+3\dot{\phi}+M^{2}(\phi)\phi=0\label{eq:EoM-1}
\end{equation}
where 
\begin{equation}
M^{2}(\phi)=\left[\begin{array}{cc}
c_{+}+\phi_{-}^{2} & -F^{2}\\
-F^{2} & c_{-}+\phi_{+}^{2}
\end{array}\right]
\end{equation}
and $\phi=(\phi_{+},\phi_{-})$. As shown in Fig.~\ref{fig:zcpalpha},
the above system exhibits chaotic behavior for $\alpha_{c}>\alpha_{{\rm Ch}}$.

To connect with existing literature, consider the following simplified
system of equations\textbf{ 
\begin{equation}
\ddot{\phi}_{\pm}+\phi_{\mp}^{2}\phi_{\pm}=0\label{eq:reduced_eqn}
\end{equation}
}by removing the dissipative and mass terms from the EoM in Eq.~(\ref{eq:EoM-1}).
Note that the above system of equations is a special case of a quartic
coupled oscillator system with the Lagrangian 
\begin{align}
L\left(t,x,y\right) & =\frac{1}{2}\left(\dot{x}^{2}+\dot{y}^{2}\right)-\frac{a}{2}\left(x^{2}+y^{2}\right)\nonumber \\
 & -\left[\frac{b}{4}\left(x^{4}+y^{4}\right)+\frac{1}{2}cx^{2}y^{2}\right]\label{eq:Lagr}
\end{align}
obtained by taking $a=b=0$, and $c=1$. A detailed numerical study
of the Lagrangian in Eq.~(\ref{eq:Lagr}) using the surface of section
technique was performed in \citep{Carnegie_1984} where the authors
concluded that for potential $V=x^{2}y^{2}/2$ where $c=1$, the motion
is completely chaotic. However, \citep{Dahlqvist:1990zz} showed that
there exists a tiny island (occupying $0.005\%$ of the total phase
space) of stable periodic orbits, which was followed by a second set
of stable orbit discovered in \citep{Marcinek:1994} (occupying $4$
orders of magnitude smaller region than found in \citep{Dahlqvist:1990zz}).
Intuitively, chaos occurs due to a nonlinear map between small changes
in the initial conditions and the integrated effects of nonlinear
forces.

\subsection{Effect of dissipation}

Consider the effect of adding the dissipative term and the negative
mass squared term as 
\begin{equation}
\ddot{\phi}_{\pm}+3\dot{\phi}_{\pm}+\left(\phi_{+}\phi_{-}-F^{2}\right)\phi_{\mp}=0
\end{equation}
where the factor of $3$ is the dissipative Hubble term.\footnote{Due to $d=3$ spatial dimensions.}
With the variable changes 
\begin{equation}
x_{1,2}=\phi_{\pm}/F
\end{equation}
and 
\begin{equation}
\tau=TF
\end{equation}
our EoMs ($i=\{1,2\}$) reduce to 
\begin{equation}
\ddot{x}_{i}(\tau)+\gamma\dot{x}_{i}(\tau)+\left(x_{1}(\tau)x_{2}(\tau)-1\right)x_{j\neq i}(\tau)=0\label{eq:x12-dissipative-eqn}
\end{equation}
where the new dissipative constant is $\gamma=3/F.$ For our blue
axion system, the ICs at transition, $T_{c}$, can be approximately
given as 
\begin{equation}
\phi_{+}(T_{c})/F=\phi_{-}(T_{c})/F\approx1-0.2\alpha_{c}
\end{equation}
and 
\begin{equation}
\dot{\phi}_{+}/F^{2}\approx-\alpha_{c}\qquad\dot{\phi}_{-}/F^{2}\approx0.48-0.09/\sqrt{\alpha_{c}}.
\end{equation}
Clearly, we see that the blue axion system with $\gamma=0$ (no Hubble
dissipation/friction) is always chaotic, since the incoming velocities
of the two fields $\phi_{\pm}$ do not lie within the island of stability
(\citep{Dahlqvist:1990zz}) required for a stable periodic solution.

A non-zero positive value of $\gamma$ leads to a constant dissipation
within the system. In the presence of dissipation, a mechanical system
gradually approaches one of its local energy minima. In the context
of the blue axion system, the trajectories in the $\phi_{\pm}$ phase
space are asymptotically approaching the points ($\phi_{\pm}=\phi_{\pm,\min},\dot{\phi}_{\pm}=0$)
known as ``point'' attractors. These point attractors have dimension
zero as they correspond to a stable equilibrium point in the phase
space. Generally, the presence of dissipation tends to make chaotic
motion into a more orderly behavior.\footnote{Non-diagonal dissipative terms can actually enhance chaos in certain
cases \citep{Sheeja:2002}} When a system consists of quartically coupled oscillators and exhibits
chaotic behavior, the introduction of dissipative forces corresponding
to quadratic dissipation functions with diagonal elements ($\gamma\neq0$,
or $F\ne\infty$) causes the system to converge to a fixed point.
If the dissipation is very small, the system can display long-lasting
chaotic transients. Such transient chaos in dissipative systems is
thus essentially a phenomenon that occurs when certain factors, such
as the amplitude of the driving force being above a particular value
or friction constant being below a threshold value.

For our blue axion system, the critical chaos-inducing nonlinear force
is proportional to $\xi=\phi_{+}\phi_{-}-F^{2}$. Hence, if the average
interaction and kinetic energy are substantial during the transition,
we anticipate a temporary phase of chaos until dissipation restores
order in the system. Thus, there must exist a threshold limit on the
value of $\alpha_{c}$ below which chaos does not set in for more
than a few $O(1/F)$ time-period.

\begin{figure}
\begin{centering}
\includegraphics[scale=0.6]{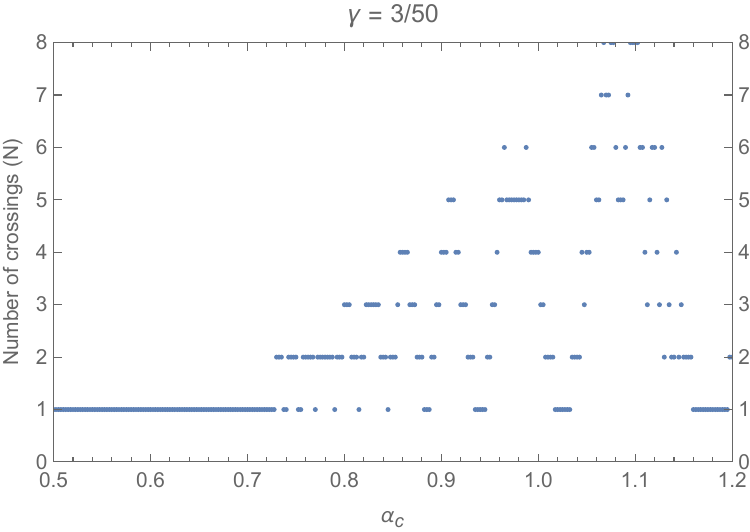}\qquad{}\includegraphics[scale=0.6]{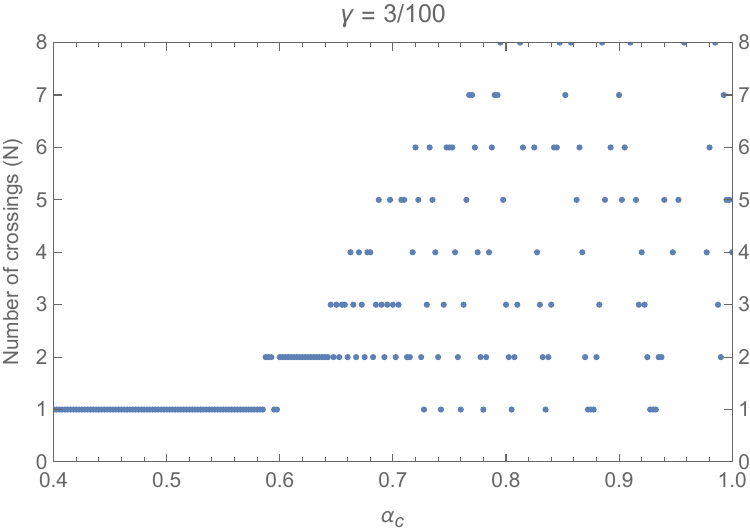} 
\par\end{centering}
\begin{centering}
\includegraphics[scale=0.6]{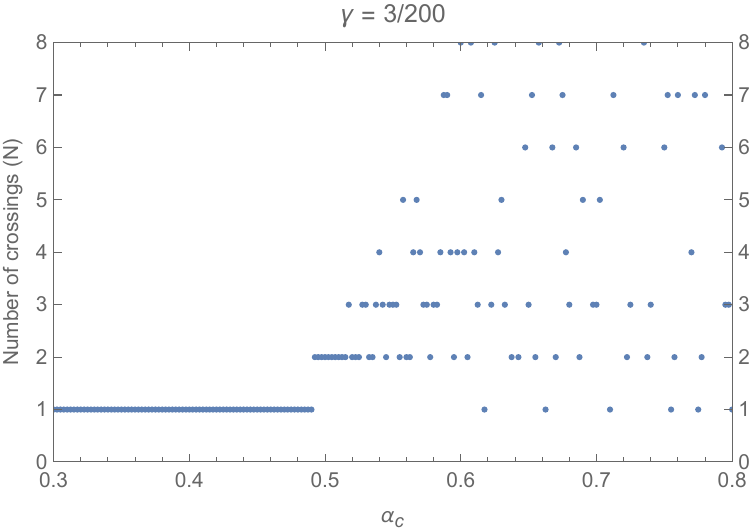}\qquad{}\includegraphics[scale=0.6]{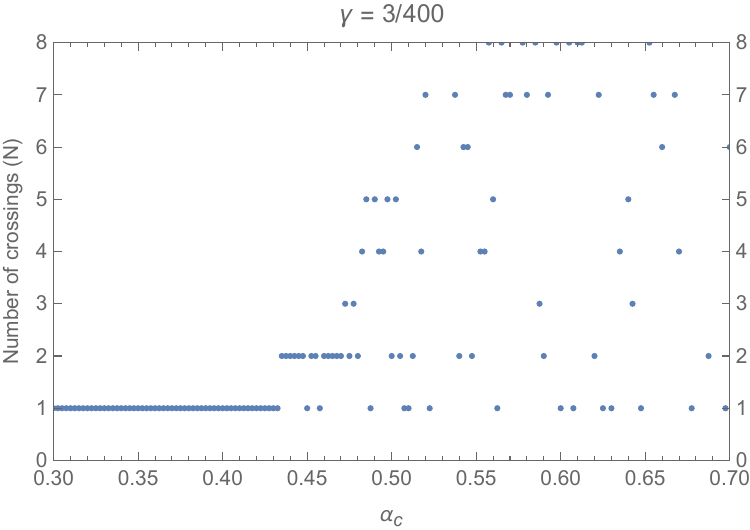} 
\par\end{centering}
\caption{\label{fig:Plot-Nvsac}Plot showing number of crossing (for $\tau>0$)
as a function of parameter $\alpha_{c}$ for different values of dissipative
constant $\gamma=3/F$.}
\end{figure}

\begin{figure}
\begin{centering}
\includegraphics[scale=0.6]{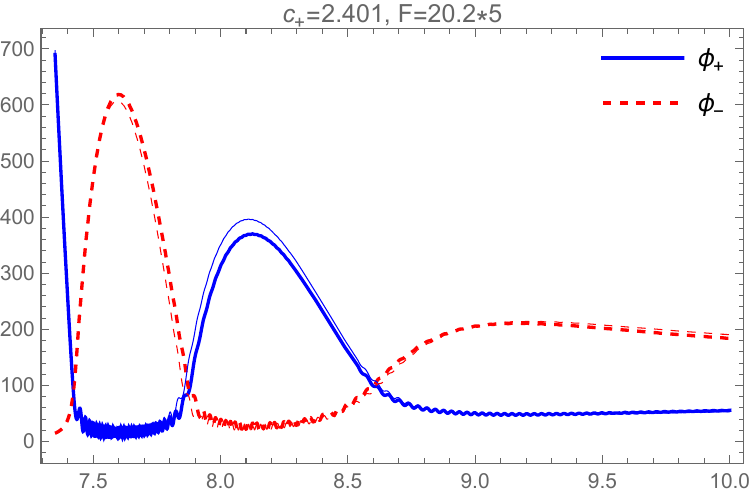}\qquad{}\includegraphics[scale=0.6]{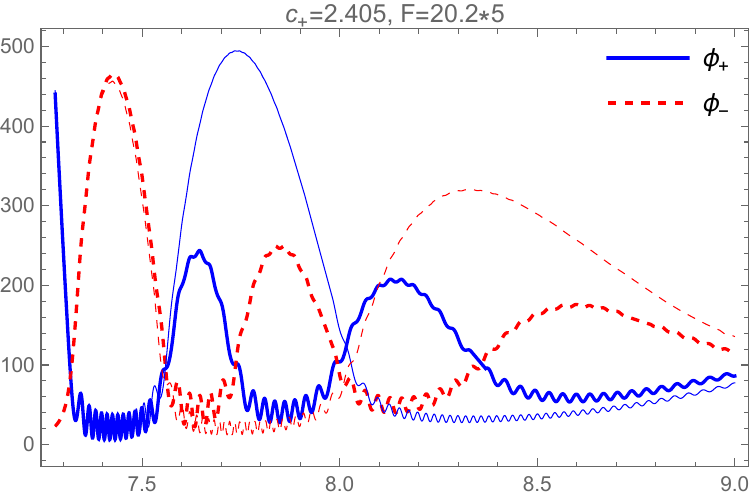} 
\par\end{centering}
\caption{\label{fig:field-trajectory}In each frame, we plot trajectories of
$\phi_{\pm}$ fields (thick curve) and a similar perturbed system
(thin curve) with a $1\%$ deviation in the ICs. The plot on the left
(right) correspond to case where $\alpha_{c}$ is lower (higher) than
$\alpha_{{\rm Ch}}$. We find that tiny changes in the ICs can result
in large deviations in the trajectories when $\alpha_{c}$ is larger
than the threshold value $\alpha_{{\rm Ch}}$.}
\end{figure}

To illustrate the onset of chaos in the presence of dissipation, we
solve the coupled oscillator system in Eq.~(\ref{eq:x12-dissipative-eqn})
by initializing at $\tau=0$ with the initial conditions similar to
that of the blue axionic system at transition. Hence, we set 
\begin{equation}
x_{1}=x_{2}=1-0.2\alpha_{c},
\end{equation}
\begin{equation}
\dot{x}_{1}=-\alpha_{c}\qquad\dot{x}_{2}=0.48-0.09/\sqrt{\alpha_{c}}
\end{equation}
and plot the number of times $x_{1,2}$ cross each other for $\tau>0$
as a function of the parameter $\alpha_{c}$ for four different values
of dissipative constant $\gamma=3/F$ in Fig.~\ref{fig:Plot-Nvsac}.
In each scenario, we observe a phenomenon where the number of crossings,
denoted as $N$, becomes random (unpredictable) once the parameter
$\alpha_{c}$ becomes larger than a threshold value $\alpha_{{\rm Ch}}$.
Across all cases, the onset of chaos seems to occur as $N$ becomes
larger than $1$, indicating a significant kinetic energy level allowing
the oscillators to cross more than once after $\tau=0$.

In Fig.~(\ref{fig:field-trajectory}) we graph the trajectory of
$\phi_{\pm}$ fields for two sample $c_{+}$ values corresponding
to $\alpha_{c}$ lower and higher than the threshold $\alpha_{{\rm Ch}}$
respectively. Within each frame, we also depict a ``similar'' perturbed
$\phi_{\pm}$ system with a $1\%$ deviation in the ICs. We observe
that as $\alpha_{c}$ surpasses the threshold value $\alpha_{{\rm Ch}}$,
the field trajectories can be highly sensitive to the ICs due to the
dominance of the nonlinear quartic interaction.

In a qualitative sense, the chaos tends to become important at the
moment of the first crossing of the oscillators after $\tau=0$. This
may be understood by noticing that as the average interaction energy
exceeds a specific threshold, the subsequent motion (trajectory) following
the crossings can exhibit extensive divergence due to rapid oscillations
($\gg O(\gamma)$) of the interaction term. To obtain a quantitative
limit, we note that in the blue axion system, the interaction term
$\xi$ induces a high frequency component, $\phi_{\pm,f}$, to the
background field solution.\footnote{See Appendix E of \pA} Compared
to the slow varying IR mode, the UV mode has a strength approximately
given by the ratio $\xi/\Omega^{2}$ where $\Omega^{2}=\phi_{+}^{2}+\phi_{-}^{2}$.
Based on our qualitative reasoning, we estimate that to initiate chaos,
the UV mode should be important at the first crossing, $T_{1}$ (after
transition).\footnote{Here, we assume that there are no crossings before the transition,
ensuring that the UV modes are generated mostly after the transition.
See Appendix E of \pA} Hence we require for the lack of chaos the condition 
\begin{align}
\left.\frac{\xi}{\Omega^{2}}\right|_{T_{1}} & <r\ll1
\end{align}
where $r$ is an $O(0.1)$ number.\footnote{This condition is consistent with the condition that the $\phi_{\mp}^{2}\phi_{\pm}$
term is a subdominant force for the equation of motion.} Since $\Omega^{2}=2F^{2}$ at $T_{1}$, we obtain the condition 
\begin{equation}
\left.\xi^{2}\right|_{T_{1}}\ll r^{2}4F^{4}.
\end{equation}
The time average of this quantity over $1/F$ time scale yields 
\begin{equation}
\left\langle \xi^{2}\right\rangle _{T_{1}}\ll2r^{2}F^{4}.\label{eq:xi-condition}
\end{equation}
Numerical analysis reveals that the coupled oscillator system undergoes
transient chaotic motion when the average interaction energy $\left\langle \left(x_{1}x_{2}-1\right)^{2}\right\rangle \approx(1/2)E(0)\exp(-\gamma\tau)$
at the point of first crossing $(\tau_{1}>0)$ is approximately $>0.1$.
In the context of the blue axion system, this condition for the transient
chaotic behavior can be expressed as 
\begin{equation}
\left\langle \xi^{2}\right\rangle _{T_{1}}\approx(1/2)E(T_{c})\exp(-3(T_{1}-T_{c})))\gtrsim\left(0.1\right)F^{4}\label{eq:Fto4th}
\end{equation}
where $E(T_{c})$ is the total energy at transition. This $0.1$ number
was numerically inferred with $c_{+}/F^{2}\ll0.1$ and can be interpreted
as the number related to the Hubble expansion rate and does not reflect
the $c_{+}$ parametric dependence. Presumably, the right-hand side
of Eq.~(\ref{eq:Fto4th}) has contributions that are $c_{+}F^{2}$
which would become important when $c_{+}/F^{2}$ become comparable
to $0.1$. These numerical findings align with the condition described
in Eq.~(\ref{eq:xi-condition}), by setting the $O(0.1)$ parameter
$r\approx0.2$. Hence, while the total energy at the transition is
approximately $O(\alpha_{c}^{2}F^{4})$, the Hubble friction must
cause sufficient damping to achieve a stable trajectory for the background
fields. Since $T_{1}$ is a function of $F$ and $\alpha_{c}$, we
get the following condition for stable trajectories 
\begin{equation}
T_{1}(F,\alpha_{c})\gtrsim T_{c}(F,\alpha_{c})+\frac{1}{3}\ln\left(\frac{f(\alpha_{c})}{0.2}\right)
\end{equation}
where the function $f(\alpha_{c})=E(T_{c})/F^{4}$ and can be approximately
given as 
\begin{equation}
f(\alpha_{c})\approx\alpha_{c}^{2}+(0.48-0.09/\sqrt{\alpha_{c}})^{2}
\end{equation}
where we have neglected mass contributions of order $F^{2}$.

\section{\label{sec:cminus}Exploring $c_{-}$ dependence}

\begin{table}[H]
\centering{}%
\begin{tabular}{|c|c|c|c|}
\hline 
$c_{-}$  & $\alpha_{c}(c_{+})<\alpha_{2}$  & $\alpha_{2}<\alpha_{c}(c_{+})\lesssim\alpha_{{\rm Ch}}$  & $\alpha_{c}(c_{+})>\alpha_{{\rm Ch}}$\tabularnewline
\hline 
\hline 
$<O(0.1)$  & exponential increase/decay  & exponential increase/decay  & Chaotic\tabularnewline
\hline 
$<9/4$  & $(1/c_{-})^{O(0.25)}$  & $(1/c_{-})^{O(0.25)}$  & Chaotic\tabularnewline
\hline 
$F^{2}\gg c_{-}>9/4$  & smooth increase  & Oscillating function of $c_{-}$  & Chaotic\tabularnewline
\hline 
\end{tabular}\caption{\label{tab:Table-c+-}Table summarizing the approximate dependence
of the zero-mode amplitude on the Lagrangian mass parameters $c_{\pm}$,
where we have explicitly considered the $c_{+}$ dependence of the
spectrum using the parameter $\alpha$. The boundaries $\alpha_{2}$
and $\alpha_{{\rm Ch}}$ are approximately independent of $c_{-}$
for $c_{-}\ll F^{2}$. For example, Eq.~(\ref{eq:alpha-Ch}) still
governs $\alpha_{{\rm Ch}}$ in the range of $c_{-}$ here.}
\end{table}

\begin{figure}
\begin{centering}
\includegraphics[scale=0.6]{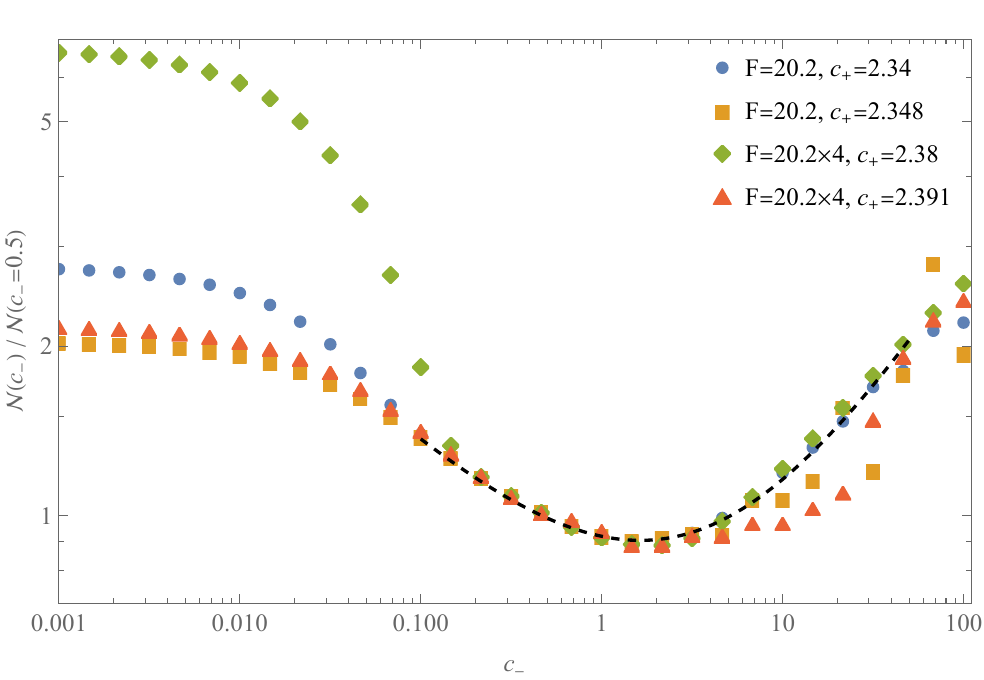} 
\par\end{centering}
\caption{\label{fig:cminus-plot}Plot showing the $c_{-}$ dependence of the
zero-mode amplitude $\mathcal{N}$ for four fiducial instances. The
black-dashed line is our fitting curve as given in Eq.~(\ref{eq:cminus-fitting}).
For each distinct value of $F$, the corresponding $c_{+}$ values
belong to $\alpha_{c}<\alpha_{2}$ and $\alpha_{2}<\alpha_{c}\lesssim\alpha_{{\rm Ch}}$
cases respectively.}
\end{figure}

Post transition, $|\phi_{-}|$ begins to increase due to a positive
velocity of $O(F^{2})$ and becomes dominant compared to a decreasing
$|\phi_{+}|$. During this time, the UV integrated EoM for the IR
component of the dominant $|\phi_{-}|$ field can be given as 
\begin{equation}
\partial_{T}^{2}\phi_{-s}+3\partial_{T}\phi_{-s}+\left(c_{-}+\frac{A^{2}}{2\bar{\Omega}^{2}}e^{-3(T-T_{2})}\right)\phi_{-s}+\sqrt{c_{+}c_{-}}\phi_{+s}\approx0\label{eq:phiminus-IR}
\end{equation}
where the subscript $s$ denotes IR (slow) component, $T_{2}\approx T_{c}+O(1/F/\sqrt{\alpha_{c}})$
, $A/\bar{\Omega}$ is the effective amplitude obtained through the
UV integration of the nonlinear interaction term $\propto\xi$ as
explained in Appendix E of \pA. During this time, $\phi_{+s}\approx F^{2}/\phi_{-s}$.
In the limit $\phi_{-s}>\phi_{+s}$, the general solution to the above
ordinary differential equation is 
\begin{equation}
\phi_{-s}(T)=e^{-\frac{3}{2}(T-T_{2})}\left[c_{1}J_{n}\left(\frac{2}{3}Ae^{-\frac{3}{2}(T-T_{2})}\right)+c_{2}J_{-n}\left(\frac{2}{3}Ae^{-\frac{3}{2}(T-T_{2})}\right)\right]\label{eq:phiminus-IR-soln}
\end{equation}
where 
\begin{equation}
n=\sqrt{1-4c_{-}/9}
\end{equation}
and the coefficients $c_{1,2}$ are obtained by matching with the
incoming solution at $T_{2}$.

At a later time in the evolution of the background fields when the
$A/\bar{\Omega}$ term in Eq.~(\ref{eq:phiminus-IR}) can be neglected
(due to exponential decay), the solution in Eq.~(\ref{eq:phiminus-IR-soln})
reduces to 
\begin{equation}
\phi_{-s}(T)=e^{-\frac{3}{2}(T-T_{2})}\left[c_{1}e^{\frac{3}{2}n(T-T_{2})}+c_{2}e^{-\frac{3}{2}n(T-T_{2})}\right].
\end{equation}
For $c_{-}<9/4$, the exponent $n$ is real, and it was shown in \pA
that the evolution of the background fields towards the minimum of
the potential is governed by an $O(c_{-}/3)$ exponent. Hence, we
can write 
\begin{equation}
\lim_{T\rightarrow T_{\infty}}\phi_{-s}(T)\approx\phi_{-\min}\left(1\pm e^{-\lambda(T-T_{\infty})}\right)
\end{equation}
where $\lambda\approx O(c_{-}/3)$. The $+(-)$ sign indicates the
direction of the field's movement where $\phi_{-}(T)$ is greater
(lesser) than $\phi_{-\min}$ as $T\rightarrow T_{\infty}$.

Due to the duality between the zero-mode and the background fields
(see Eq.~(\ref{eq:dualitymap})), the evolution of the zero-mode
is closely related to that of the background fields, resulting in
an increase/decrease of the zero-mode amplitude.\footnote{The detailed exploration of the effects of slowly varying effective
mass was covered in Appendix I of \pA.} Because the rolling of the $\phi_{-}$ continues until the $\phi_{-}$
field reaches its minimum at $\phi_{-\min}$, the zero mode amplitude
can be written as 
\begin{equation}
I_{0}(T_{\infty})\approx I_{0}(T\gg T_{2})\frac{\phi_{-\min}}{\phi_{-}(T)}.\label{eq:I0-phiminus}
\end{equation}
Hence, for $c_{-}<9/4$ the leading $c_{-}$ dependence of the zero-mode
amplitude is 
\begin{equation}
\mathcal{N}\propto\phi_{-\min}\propto\left(\frac{1}{c_{-}}\right)^{1/4}
\end{equation}
because $\phi_{-\min}\approx F\left(c_{+}/c_{-}\right)^{1/4}$. By
matching with the numerical data, we obtain the following fitting
function for the zero-mode amplitude 
\begin{equation}
\mathcal{N}\left(O(0.1)<c_{-}<9/4,c_{+},F\right)\equiv f_{-}(c_{-})\mathcal{N}\left(c_{-}=0.5,c_{+},F\right)
\end{equation}
where 
\begin{equation}
f_{-}(c_{-})\approx0.26c_{-}^{0.5}+0.66c_{-}^{-0.29}.\label{eq:cminus-fitting}
\end{equation}
Thus, the zero-mode amplitude increases as an inverse power of decreasing
$c_{-}$ for $c_{-}<O(1)$. Interestingly, if $c_{-}\sim O(0.01)$
the exponential slow roll after transition can be extremely gradual,
causing the fields to not fully settle at the minima of the potential
by the end of inflation. This slow roll of the fields is analogous
to inflationary slow roll situations, and in the current axion model
this parametric region is $c_{+}\lesssim O(0.1)$. Consequently, the
mode amplitude at $T_{{\rm end}}$ (number of e-folds to the end of
inflation) appears to be either amplified or attenuated due to the
insufficient number of e-folds for the background fields to settle
down. This amplification (attenuation) of the mode amplitude occurs
when $\phi_{-}(T_{{\rm end}})$ is greater (smaller) than $\phi_{-\min}$,
and the parametric $c_{-}$ dependence is given approximately by $\sim\exp\left(O(c_{-})(T_{\infty}-T_{{\rm end}})\right)$
where $T_{\infty}$ is the hypothetical number of e-folds required
for the background fields to settle to the minima. The value of $T_{\infty}$
is approximately dependent upon the maximum amplitude $\phi_{-\max}$
which, in turn, is controlled by the Lagrangian parameters $c_{\pm}$
and $F$. This subtle yet interesting dynamics for extremely small
$c_{-}$ values was not pointed out in \pA.

For $c_{-}>9/4$, we first consider the case where the $A^{2}$ term
in Eq.~(\ref{eq:phiminus-IR}) is negligible, applicable to scenarios
where $\alpha_{c}<\alpha_{2}$. In these cases, the exponent $n$
in Eq.~(\ref{eq:phiminus-IR-soln}) is imaginary and the solution
for $\phi_{-s}$ becomes oscillatory, leading to subsequent crossings
of the two fields after the transition. At each crossing, the lightest
eigen-mass briefly becomes tachyonic (due to a short, significant
excursion of $-\partial_{T}e_{1}.\partial_{T}e_{1}$) resulting in
an amplification of the zero-mode. The amplitude of the tachyonic
dip depends upon the magnitude of eigenvector rotation $\propto-\partial_{T}e_{1}.\partial_{T}e_{1}$
which decays as $\sim\exp\left(-3\left(T-T_{c}\right)\right)$ where
$T_{c}$ should be distinguished from crossing times other than the
first.\footnote{See Eq.~(98) and Appendix F of P1} Hence, as $c_{-}$
increases beyond $9/4$, because the background fields oscillate with
a short time-scale $\propto1/\sqrt{c_{-}}$, the tachyonic amplitude
is larger at each crossing, consequently making the final mode amplitude
larger where the steepness of the rise in amplitude (which is characterized
by the coefficient $c_{1}$ in Eq.~(\ref{eq:fit-Z-non-chaotic}))
increases as function of $c_{-}$.

When $\alpha_{2}<\alpha_{c}<\alpha_{{\rm Ch}}$, the $A^{2}$ term
in Eq.~(\ref{eq:phiminus-IR}) is significant, and the fields must
cross again after the transition. Close to the crossing, the incoming
velocity of the sub-dominant field $\phi_{+}$ has a significant contribution
from the fast UV oscillations, $\phi_{-f}\sim A/\Omega^{2}\sin(\int_{0}^{T-T_{2}}\Omega(x)dx)\phi_{-s}$
where $\Omega=\sqrt{\phi_{+s}^{2}+\phi_{-s}^{2}}\sim O(F)$ and $A\sim O(\alpha_{c}F^{2})$.
A finite value of $c_{-}$ leads to a marginal increase in the frequency
of $\Omega\sim\phi_{-s}$ causing the fields to cross earlier. As
a result, the phase $\theta=\int_{0}^{T-T_{2}}\Omega(x)dx$ of $\phi_{+f}$
at the crossing reduces. For $c_{-}$ values larger than a threshold,
the change in the phase angle, $\Delta\theta$, can be order $1$,
leading to the oscillatory pattern in the final zero-mode amplitude
as a function of $c_{-}$ as observed in Fig.~\ref{fig:cminus-plot}
for $c_{+}=2.348$ and $2.391$. Up to leading order in $c_{-}$,
we find semi-analytically that we can approximate $\Delta\theta\approx c_{-}\left(0.25+0.2\left(F/20.2\right)\right)$
which implies an almost $F$ independent lower bound of $c_{-}\sim O(2)$
for the oscillations to be prominent.

\section{\label{sec:Fitted-sine-model-parameters}Fitted sine-model parameters}

The sine model presented in Sec.~\ref{sec:sine-model-ftting} can
be most generally written as 
\begin{equation}
\Delta_{{\rm 2-fit}}^{2}\left(\left[c_{1,..,8},k_{{\rm cut}}\right],k\right)=\begin{cases}
c_{1}\left|H_{i\sqrt{c_{2}-9/4}}^{1}(kc_{3})\right|^{2}\left(kc_{3}\right)\left(j_{1}(kc_{3})\right)^{2} & k\lesssim k_{{\rm cut}}\\
\left(c_{4}+c_{5}e^{-c_{6}k}\sin\left(c_{7}\left(k-c_{8}\right)\right)\right)^{2} & k_{{\rm cut}}\lesssim k<O(5)k_{{\rm cut}}
\end{cases}\label{eq:initial sine-model}
\end{equation}
where $j_{1}(x)$ is the spherical Bessel function of order $1$,
and $H_{i\sqrt{c_{2}-9/4}}^{1}(x)$ is the Hankel function of order
$i\sqrt{c_{2}-9/4}$. The functional form of the fitting function
in the blue region ($k<k_{{\rm cut}}$) is jointly motivated from
the analysis presented in Sec.~\ref{subsec:Zero-mode} and from the
results of \pA. Readily one can identify that $c_{2}>9/4$ for underdamped
cases and $<9/4$ for overdamped fields. The first bump in the isocurvature
spectrum (for critical and underdamped cases) lies at the location
$k_{{\rm cut}}$. From Eq.~(\ref{eq:initial sine-model}), we observe
that the first bump of the fitting model is primarily determined by
the function, $j_{1}(kc_{3})$. For the spherical Bessel function,
$j_{1}(x)$, the first bump occurs at $x\approx2.08$. Consequently,
we deduce that $k_{{\rm cut}}c_{3}\approx2.08$, and therefore, we
can eliminate $c_{3}$ as $c_{3}\approx2.08/k_{{\rm cut}}$.

We model the post-cutoff region in Eq.~(\ref{eq:initial sine-model})
using a sine function with an amplitude that exponentially depends
on the mode $k$. During the fitting procedure, we choose a scale
$k_{0}$ as the boundary where we match the two piecewise functions.
The matching allows us to eliminate another model parameter. Hence,
consider the following expression where we match the amplitude of
the two piecewise functions in Eq.~(\ref{eq:initial sine-model})
at an arbitrary scale $k_{0}$: 
\begin{align}
c_{1}\left|H_{i\sqrt{c_{2}-9/4}}^{1}(k_{0}c_{3})\right|^{2}\left(k_{0}c_{3}\right)\left(j_{1}(k_{0}c_{3})\right)^{2} & =\left(c_{4}+c_{5}e^{-c_{6}k_{0}}\sin\left(c_{7}\left(k_{0}-c_{8}\right)\right)\right)^{2}.
\end{align}
Using the above expression we eliminate $c_{4}$ as 
\begin{equation}
c_{4}=\sqrt{c_{1}\left|H_{i\sqrt{c_{2}-9/4}}^{1}(k_{0}c_{3})\right|^{2}\left(k_{0}c_{3}\right)\left(j_{1}(k_{0}c_{3})\right)^{2}}-c_{5}e^{-c_{6}k_{0}}\sin\left(c_{7}\left(k_{0}-c_{8}\right)\right).
\end{equation}
Substituting the above expression for $c_{4}$ into Eq.~(\ref{eq:initial sine-model})
and redefining the remaining parameters yields the form of the model
presented in Eq.~(\ref{eq:final sine-model}).

\begin{figure}
\begin{centering}
\includegraphics[scale=0.7]{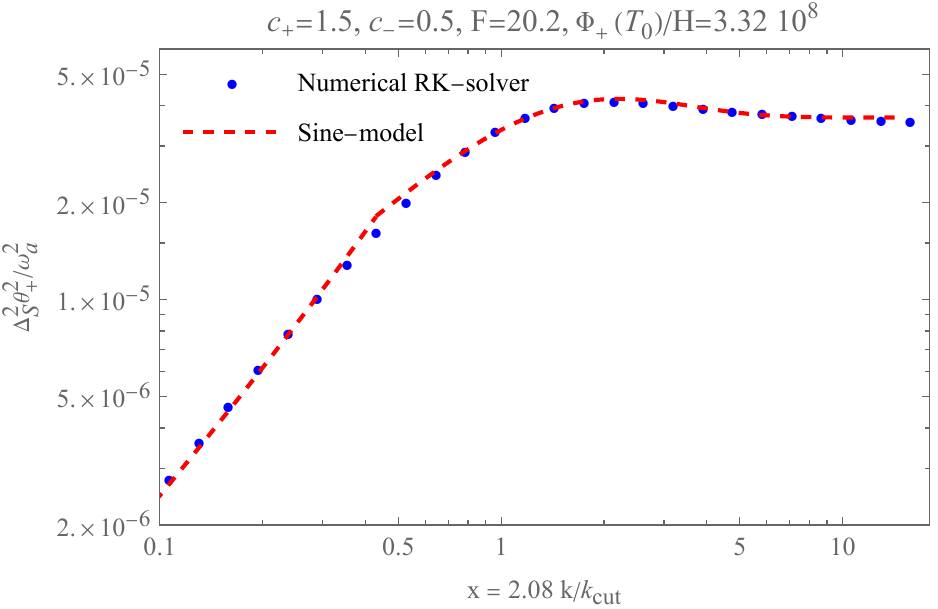} 
\par\end{centering}
\caption{\label{fig:overdamped-sine-model}In this figure we present plot obtained
from fitting the sine-model to a fiducial overdamped scenario.}
\end{figure}

Fig.~\ref{fig:overdamped-sine-model} displays an example where we
fit the sine-model to an overdamped isocurvature power spectrum covered
extensively in \cite{Chung:2016wvv}. The plots in Figs.~\ref{fig:sine-model}
and \ref{fig:overdamped-sine-model} illustrate that the sine-model
can be extended to analyze and fit cases beyond the underdamped scenario
of the axion toy model considered in this work. It may also be applicable
to other Lagrangian models that produce similar shapes of the isocurvature
power spectra.

In the rest of this appendix, we provide the best-fit values of the
model parameters obtained by fitting the piecewise model in Eq.~(\ref{eq:final sine-model})
to the numerical isocurvature power spectra examples shown in Sec.~.\ref{sec:sine-model-ftting}.
In each case mentioned below, we use $x=2.08k/k_{{\rm cut}}$. As
discussed in Sec.~\ref{sec:sine-model-ftting}, for the overdamped
(underdamped) cases, we find $x_{0}\approx0.4$ ($3.0$) is a suitable
choice for fitting: 
\begin{enumerate}
\item $c_{+}=2.30$, $c_{-}=0.5$, $F=20.2$, $\epsilon_{0}=0$, $\phi(T_{0})=3.32\times10^{8}$:
\begin{equation}
c_{1}=0.000361,\,c_{2}=2.3226,\,c_{3}=-0.03957,\,c_{4}=0.88344,\,c_{5}=1.40334,\,c_{6}=3.97287
\end{equation}
and $k_{{\rm cut}}/\left(a_{i}H\right)\approx2.0\times10^{5}$. 
\item $c_{+}=2.34$, $c_{-}=0.5$, $F=20.2$, $\epsilon_{0}=0$, $\phi(T_{0})=3.32\times10^{8}$:
\begin{equation}
c_{1}=0.000816,\,c_{2}=2.40618,\,c_{3}=0.02156,\,c_{4}=0.13236,\,c_{5}=0.89631,\,c_{6}=0.97835
\end{equation}
and $k_{{\rm cut}}/\left(a_{i}H\right)\approx3.2\times10^{4}$. 
\item $c_{+}=2.36718$, $c_{-}=0.5$, $F=20.2$, $\epsilon_{0}=0$, $\phi(T_{0})=3.32\times10^{8}$:
\begin{equation}
c_{1}=307.29,\,c_{2}=2.50251,\,c_{3}=33.571\,c_{4}=0.06304,\,c_{5}=0.61706,\,c_{6}=-0.82489
\end{equation}
and $k_{{\rm cut}}/\left(a_{i}H\right)\approx1.7\times10^{4}$. 
\item $c_{+}=10.0$, $c_{-}=0.5$, $F=20.2$, $\epsilon_{0}=0$, $\phi(T_{0})=3.32\times10^{8}$:
\begin{equation}
c_{1}=1.8431\times10^{-6},\,c_{2}=9.98452,\,c_{3}=0.04836,\,c_{4}=0.000067,\,c_{5}=0.59356,\,c_{6}=-2.4721
\end{equation}
and $k_{{\rm cut}}/\left(a_{i}H\right)\approx1.6\times10^{4}$. 
\item $c_{+}=1.5$, $c_{-}=0.5$, $F=20.2$, $\epsilon_{0}=0$, $\phi(T_{0})=3.32\times10^{8}$:
\begin{equation}
c_{1}=9.155\times10^{-4},\,c_{2}=1.473,\,c_{3}=216.5,\,c_{4}=0.979,\,c_{5}=1.658\times10^{-5},\,c_{6}=-1.1942
\end{equation}
and $k_{{\rm cut}}/\left(a_{i}H\right)\approx1.6\times10^{4}$ and
$x_{0}\approx0.43$. 
\end{enumerate}
The last example corresponds to the overdamped case and the parameter
$c_{5}$ is significant even though its smallness naively might suggest
otherwise. The sinusoidal terms whose frequency is controlled by $c_{5}$
is responsible for describing the small bump near the matching point
of $x_{0}\approx0.43$.

\bibliographystyle{JHEP2}
\bibliography{ref2,blu_iso_axion_cdm,ref,kasuya_citations_and_other_misc_papers2,blueiso_constraints,axion_isocurvature_papers2,file,axionpaper-2,misc}

\end{document}